\documentclass[fleqn,twocolumn,usenatbib,twocolappendix,numberedappendix]{openjournal}

\usepackage{xcolor}
\usepackage{textgreek}
\usepackage[utf8]{inputenc}
\usepackage[english]{babel}
\usepackage{amsmath}
\usepackage{multirow}
\usepackage{placeins}
\usepackage{hyperref}
\hypersetup{
    unicode, 
    colorlinks=true,
    linkcolor=linkcolor,
    citecolor=linkcolor,
    filecolor=linkcolor,
    urlcolor=linkcolor,
}
\usepackage{color,colortbl}
\definecolor{linkcolor}{rgb}{0.0,0.3,0.5}
\usepackage{tensind}
\tensordelimiter{?}
\DeclareGraphicsExtensions{.bmp,.png,.jpg,.pdf}
\usepackage{verbatim}
\usepackage[normalem]{ulem}
\usepackage{orcidlink}
\usepackage{soul}

\DeclareRobustCommand{\VAN}[3]{#2}
\let\VANthebibliography\thebibliography
\def\thebibliography{\DeclareRobustCommand{\VAN}[3]{##3}\VANthebibliography}
\DeclareUnicodeCharacter{2212}{\ensuremath{-}}

\graphicspath{ {./figures/} }

\begin{document}
\title[Unified reanalysis of shear diagnostics in Stage-III]{Reanalysis of Stage-III cosmic shear surveys: A comprehensive study of shear diagnostic tests}

\author{Jazmine Jefferson$^{1,2}$, Yuuki Omori$^{1,2}$, Chihway Chang$^{1,2}$, Shrihan Agarwal$^{1,2}$, Joe Zuntz$^{3}$, Marika Asgari$^{4}$, Marco Gatti$^{1,2}$, Benjamin Giblin$^{3}$, Claire-Alice H\'ebert$^{5}$, Mike Jarvis$^{6}$, Eske M. Pedersen$^{7}$, Judit Prat$^{8}$, Theo Schutt$^{9,10,11}$, Tianqing Zhang$^{12}$, and the LSST Dark Energy Science Collaboration}

\affiliation{$^{1}$Department of Astronomy and Astrophysics, University of Chicago, Chicago, IL 60637, USA\\
$^{2}$Kavli Institute for Cosmological Physics, University of Chicago, Chicago, IL 60637, USA\\
$^{3}$Institute for Astronomy, University of Edinburgh, Royal Observatory, Blackford Hill, Edinburgh, EH9 3HJ, U.K.\\
$^{4}$School of Mathematics, Statistics and Physics, Newcastle University, Herschel Building, NE1 7RU, Newcastle-upon-Tyne, U.K.\\
$^{5}$Physics Department, Brookhaven National Laboratory, Upton, NY 11973, USA.\\
$^{6}$Department of Physics and Astronomy, University of Pennsylvania, Philadelphia, PA 19104, USA\\
$^{7}$Department of Physics, Harvard University, 17 Oxford street, Cambridge, MA 02138, USA\\
$^{8}$Nordita, KTH Royal Institute of Technology and Stockholm University, Hannes Alfvéns väg 12, SE-10691 Stockholm, Sweden\\
$^{9}$Kavli Institute for Particle Astrophysics and Cosmology, P. O. Box 2450, Stanford University, Stanford, CA 94305, USA\\
$^{10}$SLAC National Accelerator Laboratory, Menlo Park, CA 94025, USA\\
$^{11}$Department of Physics, Stanford University, 382 Via Pueblo Mall, Stanford, CA 94305, USA\\
$^{12}$ Department of Physics and Astronomy and PITT PACC, University of Pittsburgh, Pittsburgh, PA 15260, USA\\
}

\begin{abstract}
In recent years, weak lensing shear catalogs have been released by various Stage-III weak lensing surveys including the Kilo-Degree Survey (KiDS), the Dark Energy Survey (DES), and the Hyper Suprime-Cam Subaru Strategic Program (HSC-SSP). These shear catalogs have undergone rigorous validation tests to ensure that the residual shear systematic effects in the catalogs are subdominant relative to the statistical uncertainties, such that the resulting cosmological constraints are unbiased.
While there exists a generic set of diagnostic tests that are designed to probe certain systematic effects, the implementations differ slightly across the individual surveys, making it difficult to make direct comparisons.
In this paper, we use the \textsc{TXPipe} package to conduct a series of predefined diagnostic tests across three public shear catalogs -- the 1,000 deg$^2$ KiDS-1000 shear catalog \citep[KiDS-1000;][]{Giblin_2021}, the Year 3 DES-Y3 shear catalog \citep[DES-Y3;][]{Gatti_2021}, and the Year 3 HSC-Y3 shear catalog \citep[HSC-Y3;][]{Li2022}. We attempt to reproduce the published results when possible and perform key tests uniformly across the surveys. While all surveys pass most of the null tests in this study, we find two tests where some of the surveys fail.
Namely, we find that when measuring the tangential ellipticity around bright and faint star samples, KiDS-1000 fails depending on whether the samples are weighted, with a $\chi^2$/dof of 121.1/16 and 257.7/16 for bins 4 and 5 for faint stars. We also find that DES-Y3 and HSC-Y3 fail the $B$-mode test when estimated with the Hybrid-$E$/$B$ method, with a $\chi^2$/dof of 37.9/10 and 36.0/8 for the fourth and third autocorrelation bins.
For PSF-related systematics, we assess the impacts on the $\Omega_{\rm m}$ - S$_{8}$ parameter space by comparing the posteriors of a simulated data vector with and without PSF contamination -- we find negligible effects in all cases. 
Finally, we propose strategies for performing these tests on future weak lensing surveys such as the Vera C. Rubin Observatory's Legacy Survey of Space and Time.
\end{abstract}


\maketitle

\section{Introduction}
Weak gravitational lensing is a powerful cosmological probe that can be used to map out the distribution of matter in the Universe. This is achieved by measuring the statistical correlations between the shapes of many distant galaxies that appear distorted due to the gravitational lensing effect. For a comprehensive review of  galaxy weak lensing, we refer the readers to e.g., \cite{Bartelmann2001,kilbinger2015}. 

Ideally, a shear catalog used for a cosmological analysis will be free of systematic biases. In practice, however, there are numerous systematic effects that are difficult to characterize and address. It is therefore crucial to have a comprehensive and detailed understanding of all the potential sources of error that may affect the cosmological results, and that those are minimized as much as possible. The galaxy weak lensing community addresses systematics in two ways: 
1) use image simulations to characterize any biases associated with the shear estimation algorithm \citep{Mandelbaum2018a, MacCrann2022, Li2023} and 2) develop diagnostic tests to identify spurious shear signals that are non-cosmological in origin and are not present in the image simulations used in the first approach \citep{Heymans2012, Asgari2019, Jarvis2016, Zuntz_2018, Mandelbaum2018b, Giblin_2021, Li2022}. In this paper, we focus on the classes of tests that fall under the second approach.

Among the many diagnostic tests that are commonly performed, many of them are designed to assess the quality of the point-spread function (PSF), which describes how light from a star or galaxy is distorted as it propagates through the atmosphere, telescope optics, and detectors. These distortions must be modeled and removed precisely to uncover the true shapes of galaxies, from which we measure the weak lensing signal.
A different class of tests, which involves measuring galaxy shapes around random locations, stars, and focal plane properties, are also measured -- all of which are expected to be consistent with zero for an unbiased shear catalog.

Development of diagnostic tests began with the first cosmic shear detection. Initial studies of cosmic shear \citep{Bacon2000,Kaiser2000,Wittman2000} characterized their systematics via methods such as auto- and cross-correlations of PSF shapes, measurement of optics-induced shear, and simulations. 

As weak lensing measurements advanced, researchers began to explore the possibility of using cosmic shear measurements to constrain cosmology. However, realizing this potential demanded substantial improvements in both the robustness and precision of these measurements. This led to the development of the many weak lensing ``challenges'' in the community \citep{Heymans2006, Massey2007, Bridle2010, Kitching2012, Kitching2013, Mandelbaum2015}. The objectives of these challenges included assessing the impact of complex PSF shapes and galaxy morphologies, selection biases, exposure co-addition, and more. 
Through these exercises, the community gained a significant amount of insight into the origin of systematic errors in the shape measurements and converged on a standard way to parametrize and propagate these errors. 

In the last decade, three Stage-III\footnote{The ``Stage-X'' terminology was introduced in \cite{Albrecht2006} to describe the different phases of dark energy experiments. There are currently 4 stages, where Stage-III refers to the dark energy experiments that started in the 2010s and Stage-IV refers to those that start in the 2020s.} weak lensing surveys were conducted, producing  shear catalogs and associated tests. Due to the increased precision that these surveys bring, requirements have become more stringent and tests have become more elaborate \citep{Jarvis2016, Hildebrandt2017, Zuntz_2018, Mandelbaum2018b, Giblin_2021, Gatti_2021, Li2022}.
Since the main goal of these surveys was to constrain cosmological parameters, each survey developed the infrastructure and frameworks to assess how residual systematic effects impact the resulting cosmology independently \citep{Amon2022, Li2022,Zhang23}.

This paper is part of a series of work in the LSST Dark Energy Science Collaboration (DESC) on reanalyzing Stage-III datasets to prepare for LSST data \citep{Chang2019, Longley2023}. Performing reanalysis exercises is critical in both transferring the knowledge and tools from Stage-III to Stage-IV surveys, and in summarizing the lessons learned from combining all the Stage-III surveys in a way that can directly inform the analysis for Stage-IV surveys. The previous papers focused on cosmic shear measurements from previous data-releases of Stage-III surveys. This paper is the first in the series that focuses on the potential shear systematic effects rather than the cosmological signal itself.  

The purpose of this paper is threefold: 1) to expand the functionality in \textsc{TXPipe}\footnote{\url{https://github.com/LSSTDESC/TXPipe}} \citep{Prat2023} -- a DESC software framework developed to measure and test ``3$\times$2pt\footnote{The term 3$\times$2pt, coined by DES-Y3 \citep{DES2018}, refers to combining the three kinds of two-point functions one can form with galaxy density $\delta_{g}$ and weak lensing shear $\gamma$: cosmic shear $\langle \gamma \gamma \rangle$, galaxy-galaxy lensing  $\langle \delta_g \gamma \rangle$ and galaxy clustering $\langle \delta_g \delta_g \rangle$. This combination probes the large-scale structure, while providing a robust self-calibration mechanism to ensure various systematic effects minimally affect the results.}'' data vectors -- to include various shear diagnostic tests carried out by Stage-III surveys and validate the implementation by comparing the outputs with published results from the fourth data release of the Kilo-Degree Survey \citep[KiDS-1000,][]{deJong2015,Giblin_2021}, the first three years of observation for the Dark Energy Survey \citep[DES-Y3,][]{DES:2005,Gatti_2021}, and all three years of observation for the Hyper Suprime-Cam Subaru-Strategic Program \citep[HSC-Y3,][]{Aihara2018b,Li2022};  2) compute a subset of the tests that can be applied to all three surveys and compare the results to gain insights into the difference between the three catalogs; 3) compile the lessons learned from 1) and 2) to formulate a proposal for diagnostic tests for LSST.    

This paper is organized as follows. In Section~\ref{sec:sys_tests}, we provide a concise overview of the different stages in a cosmic shear analysis and how systematic effects can impact the final cosmological inference. We list the diagnostic tests that are performed in Stage-III surveys and detail the purpose of these tests. In Section \ref{sec:stageIII}, we describe the three Stage-III weak lensing surveys and the data products used in this reanalysis. In Section~\ref{sec:results} we present the results for these measurements. In Section~\ref{sec:discussion} we discuss the main lessons learned from the comparisons of tests and provide suggestions on how these can be incorporated in the LSST data pipeline. Finally, we conclude in Section~\ref{sec:conclusion}.

\section{Overview of shear catalog tests}
\label{sec:sys_tests}
Shape catalogs pass through a number of data-level tests before they are used for cosmological studies. Below we briefly outline the formalism used to derive cosmological constraints from galaxy shape catalogs, and highlight the various places where systematic effects can take place.

Under the extended Limber approximation and in a spatially flat universe\footnote{For a non-flat universe, one would replace $\chi$ by $f_{K}(\chi)$ in the following equations, where $K$ is the universe's curvature, $f_{K}(\chi)=K^{-1/2}\sin(K^{1/2}\chi)$ for $K>0$ and $f_{K}(\chi)=(-K)^{-1/2}\sinh((-K)^{1/2}\chi)$ for $K<0$.} \citep{Limber1953, Loeverde2008}, the lensing power spectrum encodes cosmological information through: 
\begin{equation}
C_{\gamma}^{ij}(\ell) = \int_{0}^{\chi_{\rm H}} d\chi \frac{q^{i}(\chi)q^{j}(\chi)}{\chi^2} P_{\rm NL}\left( \frac{\ell + 1/2}{\chi}, \chi \right),
\label{eq:Cl}
\end{equation}
where $\chi$ is the radial comoving distance, $\chi_{\rm H}$ is the comoving distance to the horizon, $P_\text{\rm NL}$ is the nonlinear matter power spectrum, and $q(\chi)$ is the lensing efficiency defined via:
\begin{equation}
q^{i}(\chi) = \frac{3}{2} \Omega_{\rm m} \left( \frac{H_{0}}{c}\right)^{2} \frac{\chi}{a(\chi)} \int_{\chi}^{\chi_{\rm H}}d\chi ' n_{i}(\chi') \frac{\chi' - \chi}{\chi'},
\label{eq:lensing_efficiency}
\end{equation}
where $\Omega_{\rm m}$ is the matter density parameter today, $H_{0}$ is the Hubble parameter today, $a$ is the cosmological scale factor, and $n_{i}(\chi)$ is the redshift distribution of the galaxy sample, computed for the $i$-th tomographic redshift bin. The equivalent position-space shear correlation function can be obtained by applying the transformation:
\begin{equation}
\xi^{ij}_{\pm}(\theta) = \frac{1}{2\pi}\int_{0}^{\infty} d\ell \, \ell J_{0/4}(\theta \ell) \, C^{ij}_{\gamma}(\ell),
\label{eq:xipm}
\end{equation}
where $J_{0/4}$ are the 0th/4th-order Bessel functions of the first kind. 
Equations \ref{eq:Cl} and \ref{eq:xipm} are the signal parts of the model which we compare with observational measurements.
There are a number of mature tools available for one to calculate Equations \ref{eq:Cl} and \ref{eq:xipm} given model parameters \citep{Cosmosis, Chisari2019}.  

To measure the two-point statistics in harmonic space, one can use pseudo-$C_{\ell}$ methods such as \textsc{NaMaster} \citep{Alonso2019} and \textsc{PolSpice} \citep{Szapudi2001, Chon2004}. 
For real-space two-point correlation function (which this work is focused on), we calculate
\begin{equation}
\xi_{\pm}(\theta)=
\frac{1}{N}\underset{ab}{\sum} \left(e_{t,a}\,e_{t,b} (\theta)
\pm
e_{\times,a}\,e_{\times,b} (\theta)\right),
\label{eq:xipm_estimator}
\end{equation} 
\noindent where $e_{a}$ and $e_{b}$ denote the observed shape of two galaxies separated by an angular distance $\theta$ and $t$ and $\times$ refer to the tangential and cross component of the ellipticity with respect to the axis connecting the two galaxies. Before the correlation function can be measured, several calibration steps must be applied to the shear catalogs:

\begin{itemize}

\item Firstly, the quantity \textit{shear} is the distortion of galaxy shapes coming from gravitational lensing (i.e. light rays being perturbed as they travel through the gravitational potential between the galaxy and observer). Shear is a spin-two quantity with two elements (e.g., $\boldsymbol{\gamma} = \gamma_1 + i \gamma_2$). In the weak lensing regime, the observed galaxy shapes ($\boldsymbol{e}_{\rm obs}$) become noisy estimators of shear, where noise here refers to a combination of measurement noise ($\boldsymbol{e}_{\rm n}$) and the intrinsic galaxy shapes ($\boldsymbol{e}_{\rm int}$), both expected to average to zero in the limit of a large number of galaxies (excluding effects from each galaxies' local tidal field). The convergence, defined as the magnification distortions in the observed galaxy shape, is expected to be small in this regime ($\kappa \ll 1$). Therefore, we can write for any single galaxy    
\begin{equation}
\boldsymbol{e}_{\rm obs} = \boldsymbol{\gamma} + \boldsymbol{e}_{\rm int} + \boldsymbol{e}_{\rm n}.
\end{equation}
And for a large enough sample of galaxies in a region of approximately constant shear, we have 
\begin{equation}
\langle \boldsymbol{e}_{\rm obs} \rangle \approx \langle \boldsymbol{\gamma} \rangle.
\end{equation}
Similarly, when using the estimator in Equation ~\ref{eq:xipm_estimator} at $\theta \neq0$, we have 
\begin{equation}
\xi_{\pm, \boldsymbol{e}_{\rm obs} \boldsymbol{e}_{\rm obs}}(\theta) \approx \xi_{\pm, \boldsymbol{\gamma \gamma}}(\theta).
\end{equation}

\item Secondly, the above point is complicated by the fact that the as-is observed galaxy image is a convolution of the true galaxy image with the point-spread function (PSF). The PSF encodes the instrumental and atmospheric response at the time and location where the galaxy image is taken. The PSF cannot be part of the cosmological signal, so an effective {\it deconvolution} step is needed to access the galaxy shapes before being convolved with the PSF. 

Furthermore, for typical ground-based data, we do not know the PSF \textit{a priori} and will need to estimate it using the stars (i.e. point sources) in the same image. The PSF model estimated from stars can be biased due to a number of reasons (e.g., due to contamination of the star sample, too few stars, overlapping sources). Errors in the PSF model, coupled with the presence of noise, can introduce a bias in the shear estimate when averaged over an ensemble of galaxies. We commonly parametrize the bias as 
\begin{equation}
\langle \boldsymbol{e}_{\rm obs} \rangle  = (1+\boldsymbol{m}) \langle \boldsymbol{\gamma} \rangle + \boldsymbol{c},
\label{eq:m_and_c}
\end{equation}
where $\boldsymbol{m}$ and $\boldsymbol{c}$ are referred to as the \textit{multiplicative} and \textit{additive} shear (calibration) bias. The exact way in which $\boldsymbol{m}$ and $\boldsymbol{c}$ depend on the PSF and the galaxy shape depends on the particular algorithm that is used to perform this deconvolution. In \cite{PaulinHenriksson2008}, an analytical form is derived and often used to understand the qualitative behaviors of these biases -- the authors used a simplified scenarios where an unweighted moments-based shape estimator is used.

\item Finally, even with perfect knowledge of the PSF, the estimator used for the shape measurements is often biased. Biases may be incurred from insufficient model parameters, low signal-to-noise, or a non-representative sample resulting from selection effects. Thus, a final calibration step is required. 
This is either done with simulations \citep{Mandelbaum2018a, MacCrann2022} or with the data directly \citep{HuffMandelbaum2017, Sheldon2017}. 

\end{itemize}

\begin{table*}
\begin{center}
\caption{Tests that we will be working on in this paper. We will apply the tests uniformly to the three catalogs: KiDS-1000, DES-Y3, and HSC-Y3. We group the tests into three categories: survey distributions which are useful for general diagnostics, or one- and two-point tests which provide information for more quantitative assessments. The check marks signify whether or not the shear catalog release paper published results for each test. *KiDS-1000- primarily uses PH statistics for their PSF two-point analysis, but additionally published a measurement of $\rho_1$.}
\label{table:tests}
\begin{tabular}{lcccc} \toprule
\multirow{2}{*}{Tests} & KiDS-1000 & DES-Y3 &  HSC-Y3 & \multirow{2}{*}{Section}\\
& \cite{Giblin_2021} & \cite{Gatti_2021}&  \cite{Li2022} & \\
\multicolumn{2}{l}{\textbf{General Diagnostics}}&&&\\
Purity of PSF stars & \checkmark & \checkmark & & \ref{sec:psf_purity}\\
Overall characteristic of the PSF & \checkmark & \checkmark & \checkmark& \ref{sec:1D_PSF}\\
Overall characteristic of the galaxy sample & &\checkmark &\checkmark& \ref{sec:overall_sample}\\[0.2cm]
\multicolumn{2}{l}{\textbf{One-point tests}}&&&\\ 
Galaxy shape vs. PSF size &  &  \checkmark & \checkmark  & \ref{sec:mean_shear} \\
Galaxy shape vs. PSF ellipticity &  & \checkmark &  & \ref{sec:mean_shear}\\
Galaxy shape vs. galaxy signal-to-noise &  & \checkmark & \checkmark  & \ref{sec:mean_shear}\\
Galaxy shape vs. galaxy size  &  & \checkmark &  & \ref{sec:mean_shear}\\
PSF ellipticity vs. focal plane location &  \checkmark &  &  & \ref{sec:fov_psf}\\
PSF (fractional) size vs. focal plane location & \checkmark &  &   &  \ref{sec:fov_psf} \\[0.2cm]
\multicolumn{2}{l}{\textbf{Two-point tests}} &&&\\
$\rho$ statistics & \checkmark * & \checkmark &\checkmark & \ref{sec:rowe}\\
$\tau$ statistics &  & \checkmark & \checkmark & \ref{sec:tau}\\
PH statistics & \checkmark &  &  & \ref{sec:ph} \\
Tangential ellipticity around stars &  &\checkmark & \checkmark  & \ref{sec:tangential} \\
B-modes& \checkmark & \checkmark & \checkmark & \ref{sec:bmode}\\
\hline

\end{tabular}

\end{center}
\end{table*}

\vspace{0.1in}
 The galaxy weak lensing community has devised a collection of tests that are aimed to validate the shear catalogs at different levels to ensure that our final cosmology is unbiased. Most of the tests focus on empirically determining the non-cosmological contamination to the final two-point estimator. For example, a nonzero $\boldsymbol{m}$ or $\boldsymbol{c}$ left uncorrected will bias the final estimator and cosmological parameter inference. Additionally, there are a number of tests that do not focus on how the contamination finally propagates into the estimator, but are instead used to diagnose whether there are peculiar signals present. This second category of tests are most useful in the early stages of the catalog testing as they could reveal more significant issues in the upstream data processing. 
 However, peculiar signals could be a source of bias since data quantities in these tests (e.g., the galaxy shape and PSF quantities) are closely connected to the final cosmological analysis.

In Appendix~\ref{sec:all_tests}, we list all the tests that have been performed in the three Stage-III shear catalogs as described in \cite{Giblin_2021, Gatti_2021, Li2022}. We note that some additional null tests have been investigated in accompanying studies, such as those focusing on PSF quantities as in \citet{JarvisPSFDES2021, Zhang23} or on cosmological analyses as in \citet{Asgari2021, Dalal2023}. From the three shear catalog release papers, we choose to reanalyze a subset of the tests performed, listed in Table~\ref{table:tests}. The subset is chosen such that it can be readily applied to the public catalogs and is general enough that it is relevant to each survey despite the varied survey specifications. In addition to these requirements, we select tests that have the greatest impact on cosmic shear analyses. 

We will first examine the PSF and galaxy samples more generally -- these diagnostics are listed as ``general diagnostics.''
We then divide the remaining tests into ``One-point'' and ``Two-point'' tests. One-point tests are designed to look for trends between shear quantities and other (non-cosmological) quantities that are not expected to correlate with shear. One example of this is examining the relationship between galaxy shape and signal-to-noise. Other one-point tests examine the PSF model itself -- these mostly involve qualitative visual inspections to check that the PSF and its model are well-behaved. Two-point tests involve calculating two-point correlation functions of shear quantities with various other quantities. These tests are more concerned with spurious spatial systematic patterns that are not cosmological but can contaminate the cosmological measurements. There are two classes of tests commonly performed here, targeting either the cosmic shear statistics or the galaxy-galaxy lensing statistics \citep{Heymans2020, Prat2022, More2023}. The main difference is that the latter class looks at tangential ellipticity patterns around certain (spin-0) locations whereas the former concerns statistics where two shear-like (spin-2) quantities are correlated.  

In Table~\ref{table:tests}, we summarize the tests that were carried out in each shear catalog paper. We note that the different surveys do not perform the same tests, and even when the same test is listed, the exact implementation is not identical (e.g., the $B$-mode tests are done differently across surveys) and are somewhat challenging to compare. One of the main goals of this paper is to coherently analyze the three sets of data in a consistent framework so that they can be compared on equal footing. 

In this work, all the tests will be integrated into the LSST DESC measurement pipeline \textsc{TXPipe} \citep{Prat2023}. \textsc{TXPipe} has been extensively tested in \citet{Prat2023} and \citet{Longley2023}. The former performed a 3$\times$2pt analysis on simulated LSST Y1-like data and showed that the input cosmology for the simulation can be recovered. The latter carried out a reanalysis of cosmic shear for  DES-Y3 and HSC-Y3 year-one data and KiDS-1000, demonstrating that the pipeline can be applied to observational data. \textsc{TXPipe} is designed to run efficiently on large data sets and is structured in different ``stages'' connected via \textsc{Ceci}, \footnote{\url{https://github.com/LSSTDESC/Ceci}} a parsl-based framework created to manage DESC workflows and connect computations. In this work, we implement and validate additional stages for diagnostic tests and include a number of input catalogs additional to the shear catalog (e.g., star, PSF catalogs).
Additional catalogs are currently stored at the National Energy Research Scientific Computing Center; further details can be found at the \textsc{TXPipe} repository. 

\section{Shear catalogs from Stage-III surveys}
\label{sec:stageIII}

\begin{figure*}
\centering
    \includegraphics[width=1.5\columnwidth]{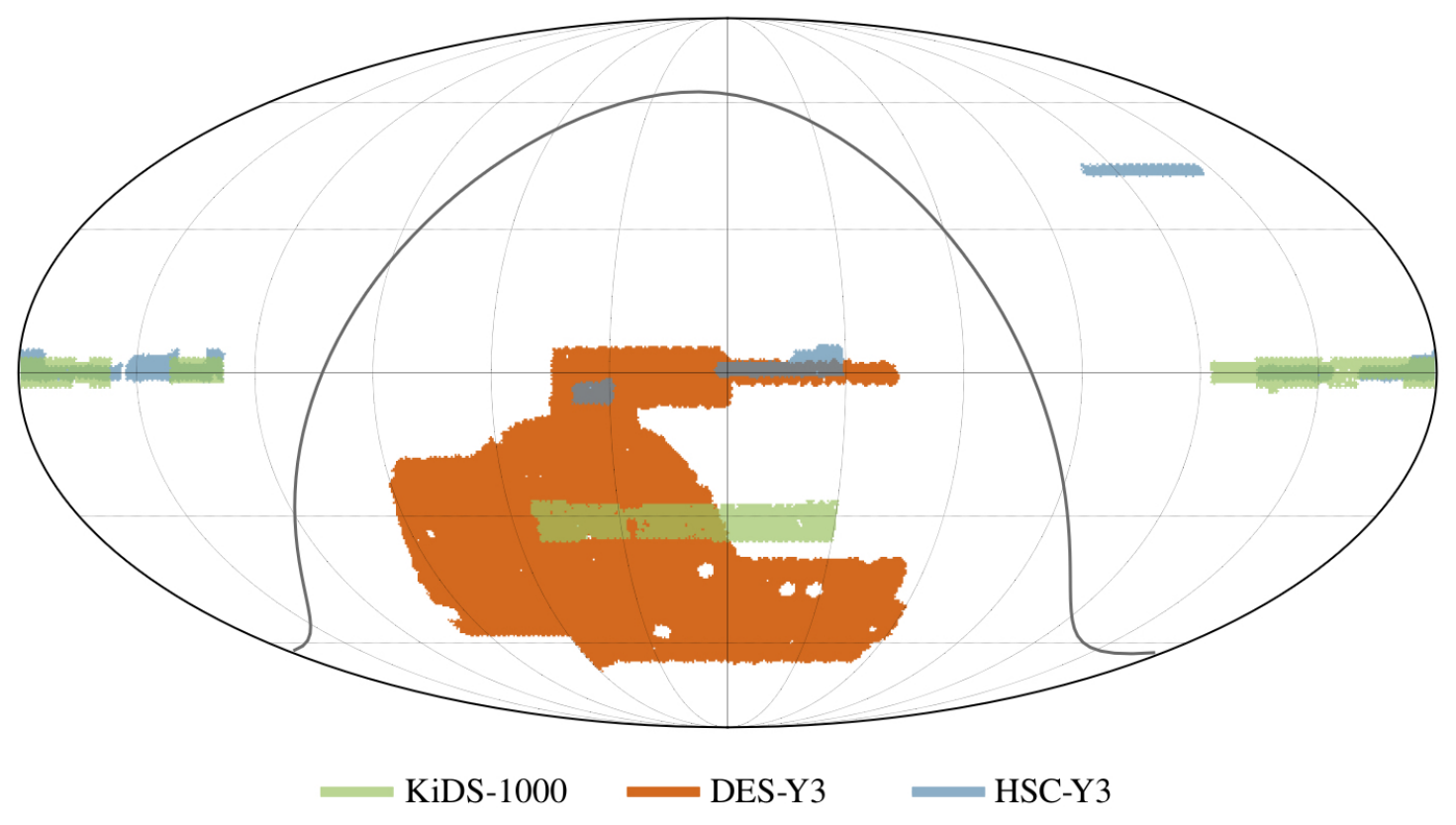}
    \caption{Footprint of the shear catalog from KiDS-1000 (green), DES-Y3 (orange), and HSC-Y3 (blue) that is used in this work. Each survey covers 777, 4,143, and 416 deg$^{2}$ respectively.}
    \label{fig:survey_footprint}
\end{figure*}

In this section we briefly introduce the three Stage-III shear catalogs that we will be testing in this work -- KiDS-1000, DES-Y3, and HSC-Y3. The footprints on the sky for the three catalogs are shown in Figure~\ref{fig:survey_footprint}. The sky area covered by the three surveys are 777, 4,143, and 416 deg$^2$, respectively. Note that there are overlaps between all pairs of the three surveys. We also note that the redshift coverage for the three surveys are slightly different. In particular, HSC-Y3 extends to higher redshift compared to the other two surveys. As a result, when we show results from the highest tomographic bin for each survey, they do not correspond to the same redshift ranges.

\begin{table}
\centering
    \caption{Table of weighted mean $e_1$, $e_2$ and $N_{\rm gal}$. KiDS-1000 and HSC-Y3 values are computed per subfield.}
    \label{table:meanshear}
    \begin{tabular}{lrrr} \toprule
     & $\langle e_{1} \rangle$ & $\langle e_{2} \rangle$ & $N_{\rm gal}$\\
    \textbf{KiDS-1000} &  &   &  \\
    North & \hspace{0.6em}$3.29\times10^{-4}$ & $8.27\times10^{-4}$& $9.86\times10^{6}$\\
    South & $-1.10\times10^{-4}$ & $3.97\times10^{-4}$ & $1.13\times10^{6}$ \\[0.2cm]
    \textbf{DES-Y3} & \hspace{0.6em}$4.97\times10^{-4}$ & $8.40 \times 10^{-5}$  & $1.00\times10^{8}$\\[0.2cm]
    \textbf{HSC-Y3} &  &  &  \\
    GAMA09H & $-1.2\times10^{-3}$ & \hspace{0.6em}$1.01\times10^{-3}$ & $4.87\times10^6$\\
    GAMA15H & $-8.0\times10^{-4}$ & $-1.22\times10^{-3}$ & $2.60\times10^6$\\
    HECTOMAP & $-1.87\times10^{-3}$ & $-7.86\times10^{-4}$ & $2.65\times10^6$\\
    VVDS &  \hspace{0.6em}$3.69\times10^{-5}$ & $-9.75\times10^{-4}$ & $6.02\times10^6$\\
    WIDE12H &\hspace{0.6em}$1.18\times10^{-3}$ & \hspace{0.6em}$2.40\times10^{-4}$ & $7.28\times10^6$\\
    XMM & $-1.54\times10^{-3}$ & $-1.15\times10^{-3}$ & $1.86\times10^6$\\
    \hline
    \end{tabular}
\end{table}

\subsection{The Kilo-Degree Survey 1,000 deg$^2$ shear catalog}

The Kilo-Degree Survey \citep[KiDS; ][]{deJong2015} shear catalog is presented in \citet{Giblin_2021}. KiDS-1000 utilizes observations taken from OmegaCAM, a wide-field optical camera mounted on the 2.6m VLT Survey Telescope at the European Southern Observatory (VST) in Chile. The fourth data release \citep{Kuijken2019} of the survey contains \textit{ugri} band observations, composed of 4-5 dithered exposures per band and is the basis of the shear catalog described in \citet{Giblin_2021}. The shear measurements for galaxies are based off of $r$-band observations. 

KiDS-1000 utilizes \textit{lens}fit \citep{Miller_2007}, a likelihood-based algorithm for shear estimation. KiDS-1000 used the self-calibrating version of lensfit developed for the KiDS-450 data release, described in \citet{FenechConti2017} and validated in \citet{Kannawadi2019}. This method uses a model for surface brightness that combines an exponential disk with a S\'ersic bulge with several additional free parameters. This is convolved with the PSF model and then fitted to each galaxy observation. The final shear estimator is the weighted ellipticity obtained from the likelihood after fitting for each exposure. The PSF is modeled using the Gaussian Aperture and PSF (GAaP) photometry method, presented in \citet{Kuijken2015}. The PSF model is defined on a grid of 32$\times$32 pixels (pixel scale = 0.213 arcsec) with each pixel fitted by a two-dimensional polynomial of order $n$. The model coefficients as well as the flux and centroid of each star are allowed to vary across the field, allowing for PSF discontinuities between CCDs.  There is no explicit selection placed on the catalog, aside from the ``SOM-gold selection'', a selection in the self-organized maps to locate and remove galaxies that were not properly represented in spectroscopic samples.

As shown in Figure~\ref{fig:survey_footprint}, the KiDS-1000 catalog is roughly composed of two contiguous patches around the equator (KiDS-1000 North) and one contiguous patch around DEC$\sim$-25$^{\circ}$ (KiDS-1000 South)
. Note that when using the shear catalog from KiDS-1000, the mean value is subtracted for the North part and the South part separately. We list the mean shape values that were subtracted in Table~\ref{table:meanshear}.

The final catalog contains over 21 million galaxies over approximately 1,006 deg$^2$ of imaging. The final effective area is 777.4 deg$^2$, which includes multi-band masks from the combination of \textit{ugri} data with $ZYJHK_s$ data from the VISTA Kilo-degree INfrared Galaxy survey (VIKING). 
The catalog is split into five tomographic redshift bins within the range $0.1 < z \leq 1.2$. A common metric to quantify the statistical power of a weak lensing catalog is the combination of effective number density per arcmin$^2$, $n_{\rm eff}$, and the standard deviation of the intrinsic ellipticity distribution, $\sigma_{e}$. Here we adopt the definitions described in \citet{Heymans2012}. 
For this shear catalog, we have $n_{\rm eff}=6.17$ arcmin$^{-2}$ and $\sigma_{e}=0.265$. This catalog is used in the KiDS-1000 cosmology analyses in \citet{Asgari2021}. The catalog is publicly available at \url{https://kids.strw.leidenuniv.nl/DR4/KiDS-1000_shearcatalogue.php}. 

\subsection{The Dark Energy Survey Year 3 shear catalog}

The Year 3 shear catalog for the Dark Energy Survey \citep[DES,][]{DES:2005} is presented in \citet{Gatti_2021}. It is based on observations taken from the Dark Energy Camera \citep[DECam,][]{Flaugher2015} mounted on the 4m Victor M. Blanco telescope at the Cerro Tololo Inter-American Observatory (CTIO) in Chile. Observations are taken in five filter bands (\textit{grizY}), with 10 dithered exposures per band. The shear measurement is based on observations in $riz$ only.

DES-Y3 measures galaxy shapes using the \textsc{Metacalibration} \citep{HuffMandelbaum2017} method, an algorithm constructed to estimate shear and shear response directly from an observed image, rather than using simulations \citep[e.g.][]{Zuntz2013, Mandelbaum2018a}. The algorithm functions in three successive layers: the first step is to apply a small artificial shear to a real galaxy image using the package \textsc{GalSim} \citep{galsim} -- typically this is done four times by applying a small shear to the positive or negative direction of $\gamma_1$ or $\gamma_2$. Next, the galaxy shape is estimated on the modified and original images. This is done using \textsc{ngmix} \citep{Sheldon:2014}, which applies a maximum-likelihood fit of a simplified Gaussian model convolved with the PSF to the galaxy shapes across multiple observations for all three bands. The PSF model is derived using the \textsc{PiFF} \citep[PSFs In the Full Field-of-view,][]{JarvisPSFDES2021} algorithm. The PSF model is defined on a grid of 17$\times$17 pixels (pixel scale = 0.30 arcsec), using sky coordinates rather than pixel coordinates to mitigate optical and atmospheric distortions. The PSF is fitted for each CCD, with two-dimensional polynomial of third order. From the shape measurements on the modified images, one can derive the per-object shear response. Per-object response estimates are noisy, so the final layer is to estimate the ensemble shear response $\boldsymbol{R}$ from a sample of galaxies.
\begin{equation}
\boldsymbol{R} = \frac{\partial \boldsymbol{e}_{\rm obs}}{\partial \boldsymbol{\gamma}}.
\label{eq:R}
\end{equation}
There are then a large number of selection cuts placed on the catalog to select objects with well-measured shear estimates. The cuts that remove most of the objects are a signal-to-noise cut (SNR$>$10) and a cut in the ratio of the size, $T$, between the galaxy and the PSF ($T_{\rm gal}/T_{\rm PSF}>0.5$). The shear catalog possesses a non-zero mean shape (see Table \ref{table:meanshear}) that is subtracted from each galaxy prior to any science analysis.

The final shear catalog contains $\sim$100 million objects over an area of approximately 4,143 deg$^2$, split into four redshift bins between the ranges of $0 < z \leq 3$. For the full DES-Y3 shear catalog, we have $n_{\rm eff}=5.592$ and $\sigma_{e}=0.265$. This catalog was used in the DES-Y3 Y3 cosmic shear analysis in \citet{Secco2022, Amon2022}. The catalog is publicly available at \url{https://des.ncsa.illinois.edu/releases/y3a2/Y3key-catalogs}.

\subsection{The Hyper Suprime-Cam Year 3 shear catalog}

The Year 3 shear catalog for the Hyper Suprime-Cam Subaru Strategic Program \citep[HSC-SSP,][]{Aihara2018b} is presented in \citet{Li2022}. HSC-SSP is carried out with the Hyper Suprime-Cam on the 8.2m Subaru telescope located at the Mauna Kea Observatory in Hawaii. In this work we use the wide survey in the Data Release S19A of HSC-SSP \citep{Aihara2022}, which is comprised of six sub-fields: GAMA09H, GAMA15H, HECTOMAP, VVDS, WIDE12H, and XMM. These fields were chosen to overlap with CMB, X-ray and spectroscopic surveys. The survey is carried out in $grizy$ with 4-5 exposures per band; the shape catalog was derived using the $i$-band coadd images. The mean shape was measured as a function of various galaxy properties, but was found to be null within 2$\sigma$; the mean shape of the sample was proposed to be largely due to cosmic variance as estimated from mock galaxy catalogs \citep{Mandelbaum2018a}. For consistency of presentation, we quote the mean shape for each subfield in Table \ref{table:meanshear}, although it was not subtracted in their analysis.

HSC-Y3 measures galaxy shear using the re-Gaussianization PSF correction method \citep{Hirata2003} that has been incorporated into the package \textsc{GalSim}. This method first makes corrections to both the galaxy shape and PSF to account for non-Gaussianities. A `re-Gaussianized' image of the galaxy is then constructed, approximating what the galaxy would have looked like had the PSF been perfectly Gaussian. 
The final shear estimation is normalized with the response factor, $\boldsymbol{R}$ (Equation \ref{eq:R}) to account for the average galaxy shape's response to a small shear distortion.
HSC-Y3 uses a coadd PSF in their analysis; as a result, all tests in this work regarding the PSF model use the coadd PSF instead of single-exposure PSFs (which are used by other surveys). The PSF is first modeled for single exposures via an updated version of \textsc{PSFEx}, which was run on objects brighter than magnitude $\approx$22.3 in the $i$-band. \citep{Bertin2011, Bosch2018, Aihara2022}. The PSF model is defined on a 20$\times$20 pixel grid (pixel scale = 0.168 arcsec) and is fitted for each CCD, using a two-dimensional polynomial. To transform these into a coadd PSF model, each exposure is convolved with a warping kernel and then resampled with a common grid. Shear calibration is done via image simulations similar to that done in \citet{Mandelbaum2018a}. A large number of selection cuts are applied to the catalog to ensure that the uncertainty in shear calibration is reliable. The main selection cuts include a signal-to-noise cut (\texttt{i\_cmodel\_flux/i\_cmodel\_fluxerr} $>$ 10), a size cut (\texttt{i\_ hsmshaperegauss\_resolution} $>0.3$), and a magnitude cut (\texttt{i\_cmodel\_mag} $-$ \texttt{a\_i} $<$ 24.5). We refer the reader to Table 2 of \citet{Li2022} for the full list of selection cuts applied to HSC data. 

The final shear catalog contains $\sim$25 million objects for an effective area of 416 deg$^2$. The data is split into four redshift bins between $0 < z \leq 3$. For the HSC-Y3 shear catalog, we have an $n_{\rm eff}=16.2$ and $\sigma_{e}=0.264$. This catalog was used in the HSC-Y3 Y3 cosmic shear analysis in \citet{Li2pt_2023}, \citet{Dalal2023}. The wide field data is publicly available at \url{https://hsc-release.mtk.nao.ac.jp/doc/index.php/data-access__pdr3/}.

\section{Results}
\label{sec:results}

In this section we carry out a series of diagnostic tests across the three datasets in a uniform fashion and discuss the results. Before carrying out the unified analysis, to ensure we have implemented the tests the same way as the Stage-III surveys, we first cross-checked that we can reproduce any published results -- this is done for most of the tests below, and we summarize what has been done in Appendix~\ref{sec:reproduce}. We begin our analysis focusing on the overall characteristics of the PSF and shear samples for each survey to understand the basic properties of the data. We then turn to analyzing a number of two-point statistics, which are designed to identify contamination in the cosmic shear measurement. Some of the two-point tests were presented only for the full galaxy sample in the Stage-III papers, however because cosmic shear analysis is performed tomographically, we also compute the two-point statistics in redshift bins when possible.

As discussed in Section~\ref{sec:sys_tests}, the goal of this paper is not to conduct all possible tests; therefore, some effects will be missed. However, the tests performed here should cover the most important diagnostic tests that will allow us to judge to first order whether a shear catalog is ready for cosmological analysis. 

\begin{figure*}
\centering
\includegraphics[width=1.7\columnwidth]{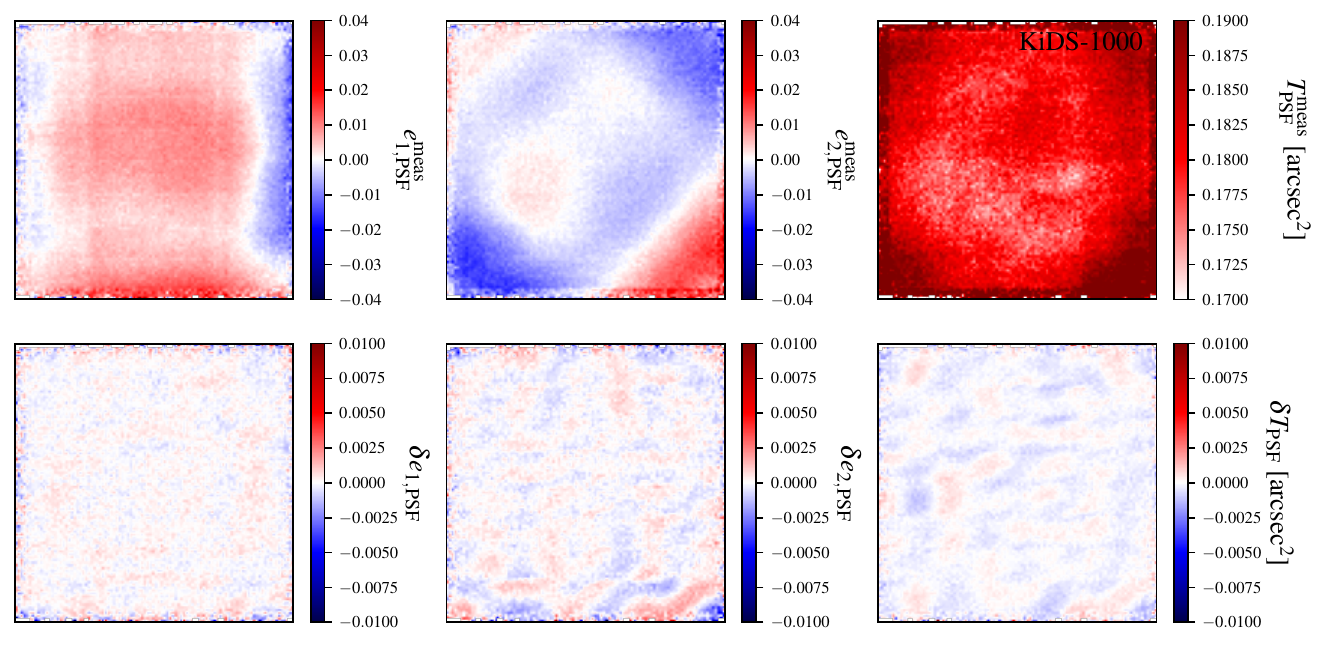}
\includegraphics[width=1.7\columnwidth]{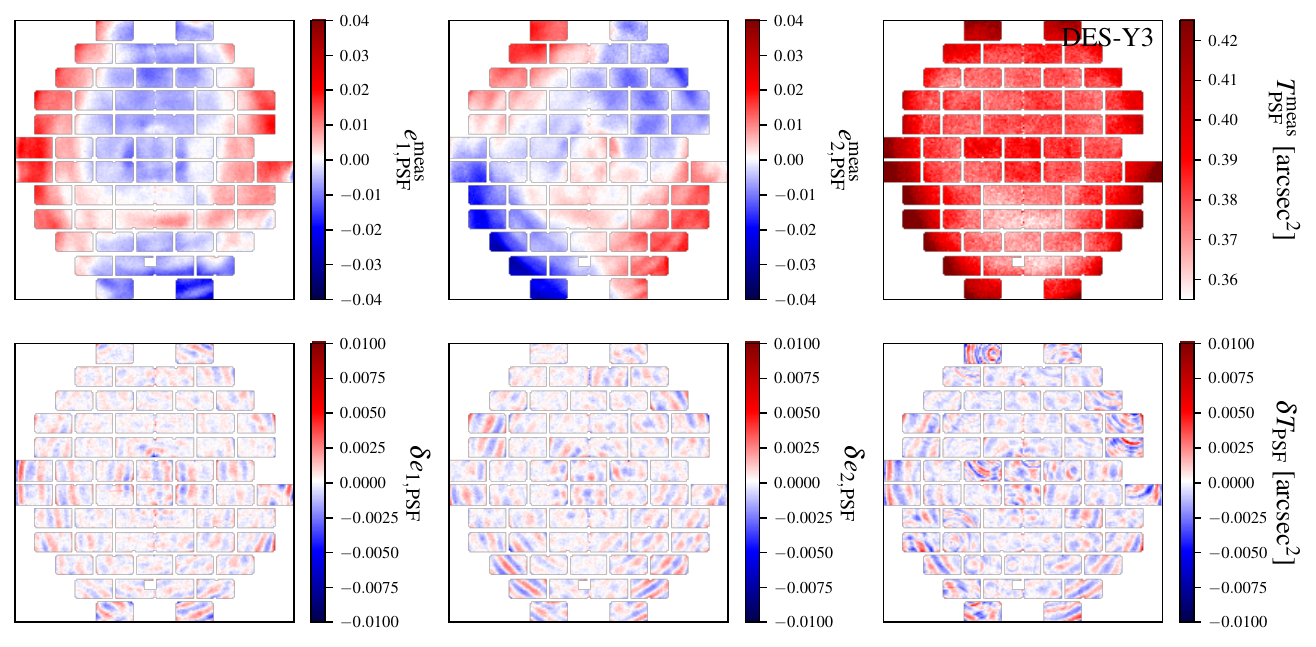}
\includegraphics[width=1.7\columnwidth]{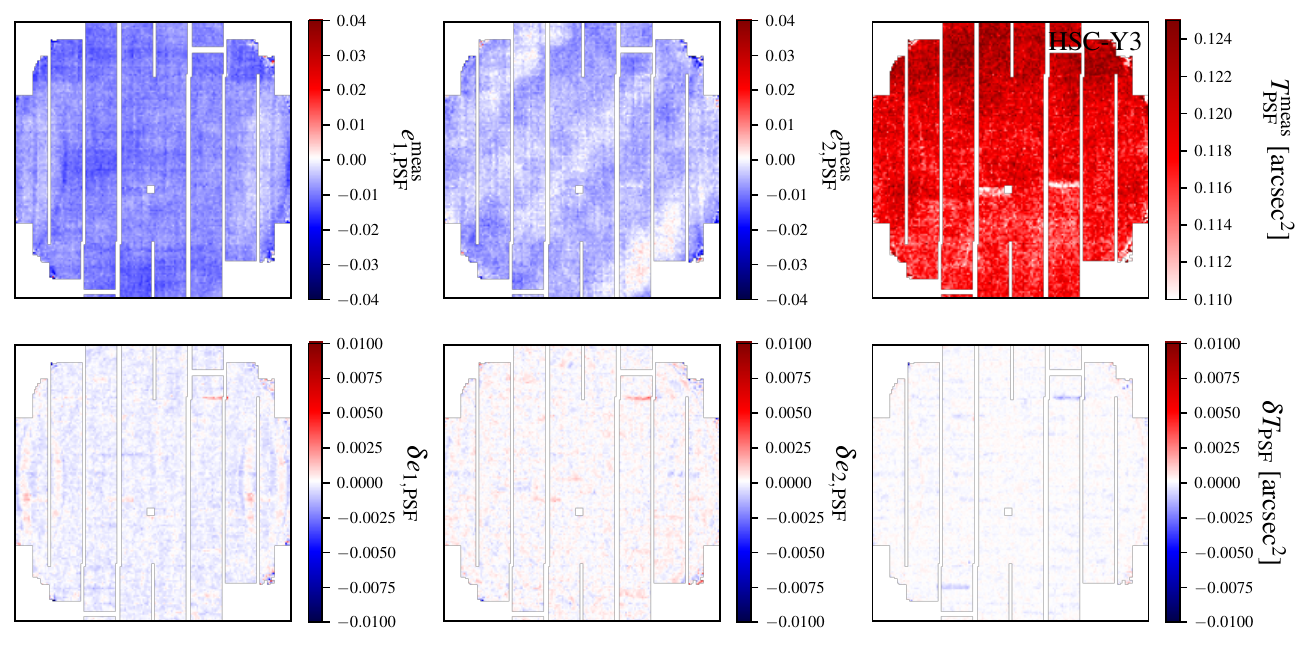}
\caption{PSF ellipticity and size distribution over the focal plane for the three surveys -- KiDS-1000, DES-Y3 and HSC-Y3 from top to bottom. We show here maps of the ellipticity and size (upper row) as well as the model residuals (lower row). 
For DES-Y3 and HSC, these are measured for a "reserved" sample of stars only used for validating the PSF model. For KiDS-1000, the model and validation star samples are the same, as they found this maintained their PSF quality. ($\boldsymbol{e}_{\rm PSF}^{\rm meas}$, $T_{\rm PSF}^{\rm meas}$) is the shape and size measured for the star. We also show the residuals for the measured values and values inferred from the model ($\boldsymbol{e}_{\rm PSF}^{\rm mod}$, $T_{\rm PSF}^{\rm mod}$): $\delta \boldsymbol{e}_{\rm PSF}= \boldsymbol{e}_{\rm PSF}^{\rm meas} - \boldsymbol{e}_{\rm PSF}^{\rm mod}$ and $\delta T_{\rm PSF} = T_{\rm PSF}^{\rm meas} - T_{\rm PSF}^{\rm mod}$.}
\label{fig:focalplane}
\end{figure*}

\subsection{Overall PSF characteristics}
\label{sec:overall_psf}

The PSF affects weak lensing measurements in two ways. First, the PSF is one of the key measures of the image quality in a survey -- a poor PSF results in lower signal-to-noise and lower resolution in the images, which in turn lowers the statistical power of a survey. Second, we need a model of the PSF in each individual image in order to ``deconvolve'' it from the observed galaxy image and recover the galaxy shape before the PSF convolution. An inaccurate PSF model would result in a biased shear estimate.     

Following \citet{Seitz1991}, a source can be defined in terms of the second moments of the surface brightness profile, $Q_{ij}$. 

\begin{align}
& Q_{ij} = \frac{\int d^2\beta q(I(\boldsymbol{\beta}))(\beta_i - \Bar{\beta}_i)(\beta_j-\Bar{\beta}_j)}{\int d^2 \beta q(I(\boldsymbol{\beta}))},    
\end{align}
where $q$ is a model for the surface brightness profile and $\Bar{\beta}_{ij}$ represents the centroid of the object.  

The complex ellipticity $\boldsymbol{e}$ and size $T$ are defined as:
\begin{align}
&\boldsymbol{e} = \frac{Q_{11}+2iQ_{12}-Q_{22}}{Q_{11} + Q_{22} +2\sqrt{Q_{11}Q_{22}-Q_{12}^2}}, \notag\\
&T = Q_{11} + Q_{22}.
\label{eq:eT_def}
\end{align}

\subsubsection{PSF as a function of focal plane position}
\label{sec:fov_psf}

We first examine qualitatively the distribution of PSF quantities across the focal plane -- the PSF ellipticity $\boldsymbol{e}_{\rm PSF}$, the PSF ellipticity error $\delta \boldsymbol{e}_{\rm PSF}\equiv \boldsymbol{e}_{\rm PSF}^{\rm meas}-\boldsymbol{e}_{\rm PSF}^{\rm mod}$, the PSF size $T_{\rm PSF}^{\rm meas}$ and the PSF size error $\delta T_{\rm PSF} \equiv T_{\rm PSF}^{\rm meas} - T_{\rm PSF}^{\rm mod}$. Here $\boldsymbol{e}_{\rm PSF}^{\rm meas}$ and $T_{\rm PSF}^{\rm meas}$ are the ellipticity and size of stars as measured directly on the star images -- these quantities represent a measure of the "true" PSF at the star's location. The $\boldsymbol{e}_{\rm PSF}^{\rm mod}$ and $T_{\rm PSF}^{\rm mod}$ quantities are inferred from the PSF model at the location of the stars. When averaged over many exposures, the PSF pattern in the focal plane is dominated by the optics and camera hardware. Figure~\ref{fig:focalplane} shows the PSF ellipticity and size across the focal plane for the three surveys (upper row) as well as the model residuals (lower row). We bin each focal plane in a square grid of length 100$\times$D, where D is the size of each survey's field of view in degrees -- approximately 1$^{\circ}$ for KiDS-1000, 2.2$^{\circ}$ for DES-Y3, and 1.5$^{\circ}$ for HSC-Y3. To obtain focal plane positions for HSC-Y3, we use individual exposures and cross-match the sources with the PSF catalog (RA,Dec $\textless$ one arcsec) to conserve the coadd PSF shape and size.   

Positive and negative values for $e_1$, correspond to an object being sheared in a horizontal or vertical direction while positive and negative $e_2$ values correspond to the object being sheared positively/negatively in the $45^{\circ}$ direction. The $e_1$ and $e_2$ maps together are equivalent to a ``whisker plot'' showing the distortion pattern in the focal plane.
We note that the PSF size ranges are somewhat different, which we will also discuss in later sections. 

For KiDS-1000, we see the strength of the measured PSF ellipticity as well as size is at its highest at the edges of the focal plane. Although a high-order polynomial model for the PSF interpolation ($n=4$) is employed, a residual oscillatory pattern with no obvious direction remains of effects unable to be captured. The residual is somewhat weaker for the first component of the PSF shape compared to the second.

For DES-Y3, we also find that the 
the PSF ellipticity is strongest at the edges of the focal plane, corresponding to more distortion at the edge of the field. The PSF is the smallest close to the middle of the field, though some asymmetric patterns can be seen. The model residual reveals a concentric pattern of alternating values for both the ellipticity and the size -- these residuals primarily come from the optics and the imperfection in the model is due to the particular choice of the functional form used for the PSF interpolation \citep{JarvisPSFDES2021}. DES-Y3 used a third-order polynomial for the interpolation, while the residual shape arises from fourth- and higher-order effects not able to be captured by this modeling choice.

For HSC-Y3, we find a strong uniformity of PSF shapes across the focal plane, where $e_1$ and $e_2$ measure predominantly negative values. The PSF size shows a diverging pattern, in the upper and lower portions of the focal plane. The residuals are particularly weak for all three quantities, showing little error in the PSF modeling.

\subsubsection{1D distributions of PSF quantities}
\label{sec:1D_PSF}

\begin{figure*}
\centering
\includegraphics[width=0.95 \linewidth]{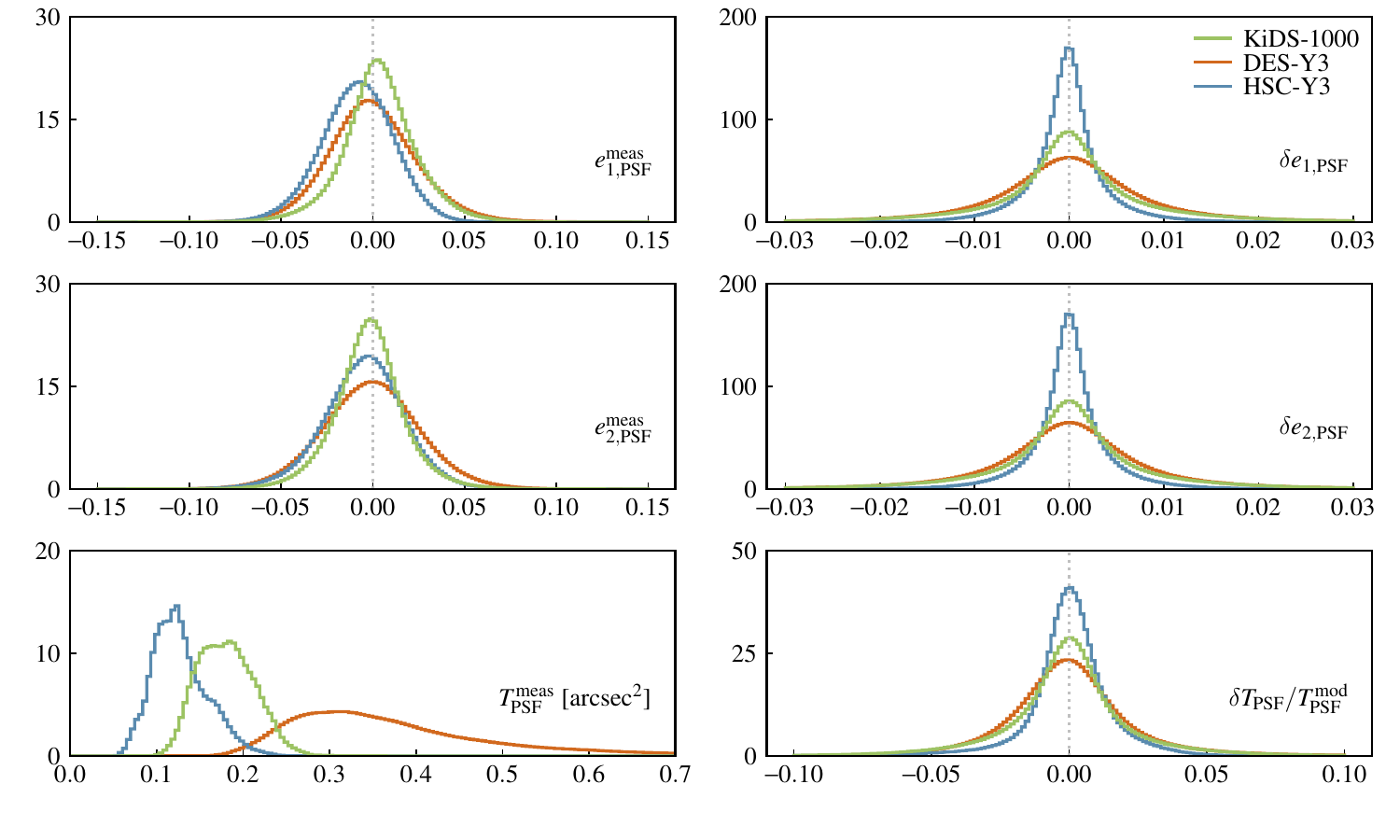}
\caption{Normalized distributions of PSF ellipticities (top, middle) and PSF size (bottom) from each survey's PSF catalog. Left columns display raw measured values and right columns display residuals. These are measured from reserved PSF stars with shape both measured from the star itself ($\boldsymbol{e}_{\rm PSF}^{\rm meas}$, $T_{\rm PSF}^{\rm meas}$) and inferred from the PSF model ($\boldsymbol{e}_{\rm PSF}^{\rm mod}$, $T_{\rm PSF}^{\rm mod}$).
} 
\label{fig:psfhistograms}
\end{figure*}

Next we look at the 1D distribution of the same PSF quantities in Section~\ref{sec:fov_psf} in Figure~\ref{fig:psfhistograms}. For the size residuals, we show the fractional size error instead of the size error.

For $e_1$ and $e_2$, all survey distributions display some level of asymmetry around zero. For $e_1$, DES-Y3 and HSC-Y3 have more negative values while KiDS-1000 displays a tendency towards positive values. For $e_2$, all surveys favor negative values. One can gain insight comparing these distributions with Figure~\ref{fig:focalplane} to see where the negative/positive PSF ellipticities are located in the focal plane.

For the size of the PSF, it is interesting to observe that the three surveys are quite different. As expected, HSC-Y3 has the smallest PSF size at around 0.11 arcsec$^2$ -- the Subaru telescope's 8-meter aperture and exquisite site condition at Mauna Kea results in a much better seeing. DES-Y3 on the other hand shows a wider spread in the PSF size distribution, stretching beyond $T = 0.5$ arcsec$^2$. 

Next we turn to the model residuals. We find that all three surveys similarly display fairly symmetric error distributions around zero, with long tails. For PSF ellipticity, HSC-Y3 yields a kurtosis of 48 and 56 for $\delta e_{1,2\rm PSF}$, respectively. HSC-Y3 measures the strongest skewness, with a score of 6.5 and 6.6 for $\delta e_{1,2\rm PSF}$, indicating a tendency to underestimate the PSF ellipticity. Both DES-Y3 and KiDS-1000 measure similar values for the residual distributions: KiDS-1000 measures a kurtosis of 5.8, 5.1 with a skewness of 1.9 for $\delta e_{1,2\rm PSF}$ whereas DES-Y3 measures a kurtosis of 4.8, 4.9 with a skewness of 1.9, 1.7. 

For $\delta T/T$, DES-Y3 measures the strongest kurtosis score of 4 with a skewness of approximately 2, indicating that on average they underestimate the PSF size. Both HSC-Y3 and KiDS-1000 yield scores near zero. 

\begin{figure*}
\centering
\includegraphics[width=1.9\columnwidth]{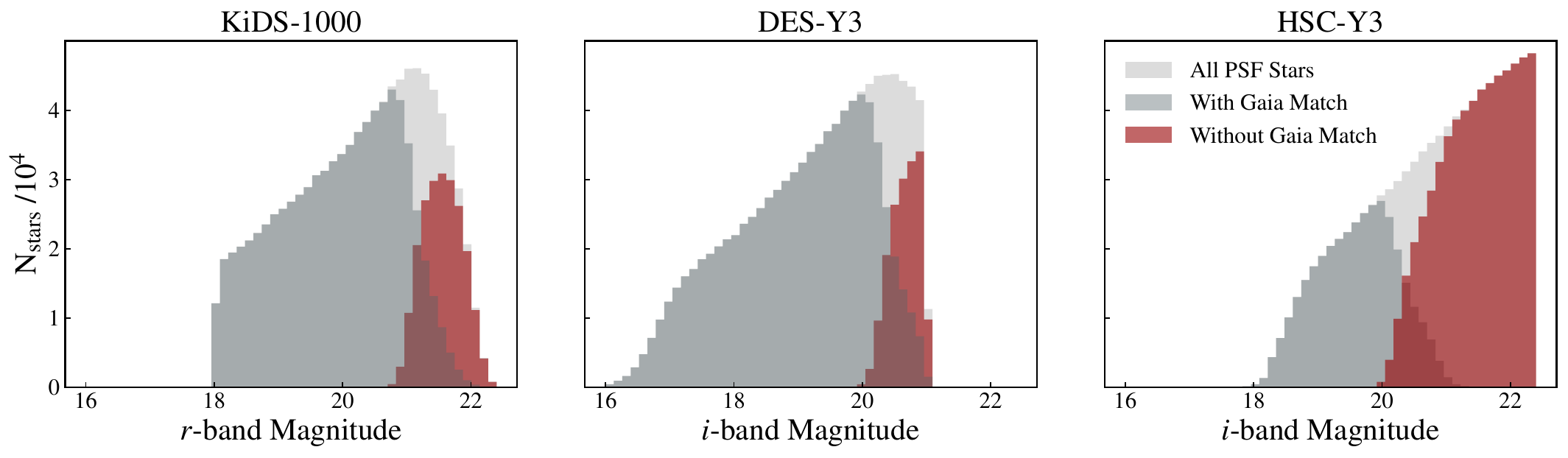}      
\caption{
Histograms of match completeness between Gaia sources and PSF stars for each survey. 
In gray, we have a magnitude histogram of all successfully matched PSF-stars to Gaia sources, while in red, we bin all PSF-stars that go unmatched. 
In light gray, we show the magnitude histogram of the entire subsample of PSF-stars.}
\label{fig:des_crossmatch}
\end{figure*}
\subsubsection{Purity of PSF stars}
\label{sec:psf_purity}

We finally look at the purity of the PSF stars, a slightly less examined aspect of the PSF. The procedure to model the PSF relies on selecting stars in the single-exposure images that are sufficiently high signal-to-noise to have a good shape measurement but not so bright that detector nonlinearities set in. A bad star selection could result in errors in the PSF model. For example, if the PSF star sample is contaminated by extended sources such as galaxies, it can lead to systematic biases on shear measurements since the modeled PSF will be larger than its true size. 

We use a test similar to that of \citet{amon2018, JarvisPSFDES2021}, to check if any galaxies are incorrectly included in the stellar sample used to model the PSF. 
We cross-match PSF stars from the three surveys with the \textit{Gaia} Data Release 3 (DR3) dataset. DR3 contains robust object classifications for 1.59 billion sources with a limiting magnitude of $G \approx 21$; 4.8 million of those sources are classified as galaxies and over 7 million classified as non-single stars or quasi-stellar objects (QSOs). 
For the star-galaxy separation in \textit{Gaia}, we use the following separator based on \verb\astrometric_excess_noise\:
\begin{align*}
    & \log_{10}(\text{\texttt{astrometric\_excess\_noise}}) < \notag \\
   & \text{max}((\text{\texttt{phot\_g\_mean\_mag}} - 18.2) \times 0.3 + 0.2, 0.3),
\end{align*}
which conservatively ensures that all selected objects are confirmed stars. As such, this separator may imply the inferred amount of galaxy contamination is higher than the truth.

In Figure~\ref{fig:des_crossmatch}, we show the magnitude distribution of the PSF stars for different matching scenarios with \textit{Gaia}. 
For KiDS-1000 and DES-Y3, the match remains very close to complete at brighter magnitudes whereas for dimmer magnitudes, \textit{Gaia} does not offer a good constraint on the completeness due to its magnitude limit.
HSC-Y3, on the other hand, appears to use much fainter stars for PSF modeling given its depth, which means that much of the star sample fainter than $m=21$ is not matched to \textit{Gaia}. The overall contamination in all three surveys is below 1\% within the \textit{Gaia} magnitude limit, estimated using the conservative star-galaxy separator above. More robust methods to estimate stellar purity such as cross-matching using catalogs from existing infrared/near-infrared surveys e.g., the Wide-field Infrared Survey Explorer (WISE), the VISTA Hemisphere Survey (VHS) \citep{Wright2010,McMahon2013} is not feasible in this analysis since no one catalog has a sufficient depth and sky footprint that accommodates all three surveys.

\begin{figure*}
\centering
\includegraphics[width=0.95\linewidth]{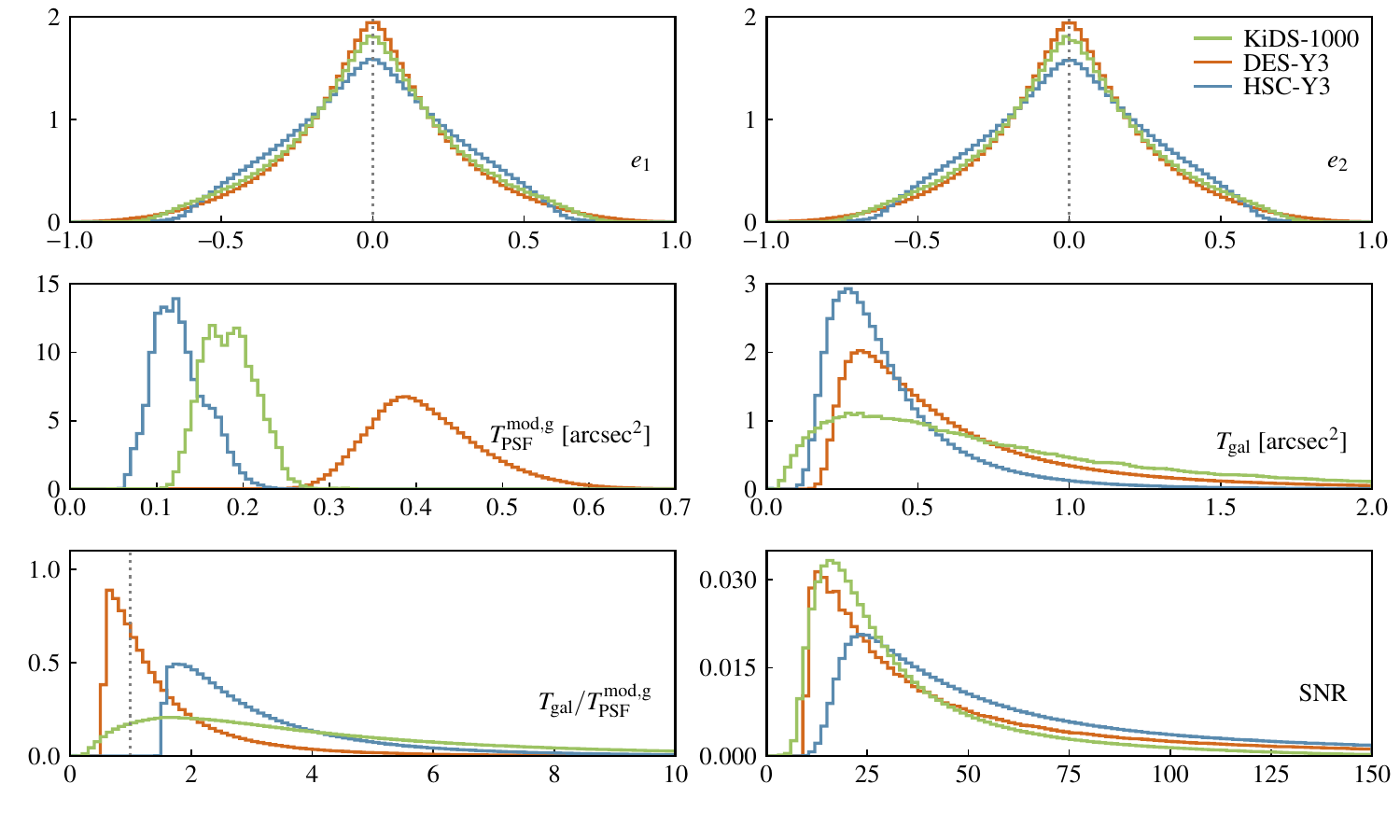}
\caption{Normalized distributions of galaxy properties in the three shear catalogs -- these are the final galaxies that are in the cosmology sample. We show the galaxy shape (top row), the PSF and the galaxy size (middle row), and the ratio of the sizes (bottom left) and the signal-to-noise ratio (bottom right). Note that the $T_{\rm PSF}^{\rm mod,g}$ shown here is the size of the PSF model interpolated to the galaxy positions, rather than the size of the PSF model evaluated at the reserved star positions as shown in Figure \ref{fig:psfhistograms}. One can see that there are clear selections placed in $T_{\rm gal}/T_{\rm PSF}^{\rm mod,g}$ and SNR for some of the samples.}
\label{fig:histograms}
\end{figure*}

\subsection{Overall source sample characteristics}
\label{sec:overall_sample}

Considering that each survey employed different selection and methods for estimating the galaxy shapes, we examine the distribution of various shape-related quantities: galaxy ellipticities ($e_1$ and $e_2$), galaxy size ($T_{\rm gal}$), the ratio of galaxy size to PSF ($T_{\rm gal}/T_{\rm PSF}^{\rm mod,g}$), and the signal-to-noise ratio (SNR). The results are shown in Figure~\ref{fig:histograms}. 

For $e_1$ and $e_2$, we observe that the overall distributions are similar for all the three surveys. All of them are fairly non-Gaussian with a more peaked distribution at low shear values. HSC-Y3 displays a visible drop of the distribution at $|e_{1}|, |e_{2}| \sim 0.6-0.7$, which may be a result from a selection cut based on shear values. KiDS-1000 also displays a subtle drop in $|e_{1,2}|$ at this range.   

For the SNR distributions,DES-Y3 and HSC-Y3 remove objects with $\rm{SNR} \leq 10$ and $<10$, respectively. DES-Y3 employed this cut to remove objects impacted by detection biases whereas HSC-Y3 employed this cut to avoid complications from blending, as the HSC-Y3 data is significantly deeper than the other two surveys. KiDS-1000 did not specifically employ selections for the SNR of each object, rather they placed selections on objects where their properties in color-magnitude space were represented well by the spectroscopy used for their SOM calibration for photometric redshifts. Their SOM-Gold selection includes objects down to $\rm {SNR} \approx 3$.
For the ratios of the galaxy size to the PSF size, we find that the KiDS-1000 and DES-Y3 distributions begin before one, indicating that there are a non-negligible amount of objects whose PSF is of equal or greater size than the galaxy.
DES-Y3 in particular displays a strong peak in the distribution between 0.5 and 1. 
On the other hand, all HSC-Y3 galaxies are larger than the PSFs. This again showcases that the HSC-Y3 selection is rather conservative and only uses well-measured shapes. 
For the measured galaxy sizes, KiDS-1000's galaxy size distribution begins near 0; this is consistent with the fact that their shear catalog does not contain a size cut. HSC-Y3 has a slightly narrower distribution of galaxy size measurements, likely due to the fact that they are capable of including smaller galaxies due to the exceptionally small PSF. Furthermore, the catalog is deeper and extends to higher redshifts, with smaller galaxies.

\subsection{Trends in mean shape}
\label{sec:mean_shear}
 
On average, galaxies are oriented randomly and thus the mean of $e_{1}$ and $e_{2}$ for a large collection of galaxies is expected to be small, and this mean is often subtracted in cosmic shear analyses (as well as other tests in this paper). Significant deviations from zero in the mean hint at systematics arising from either an error in the PSF modeling or the shear calibration. 

\begin{figure*}
    \centering
    \includegraphics[width=0.68\linewidth]{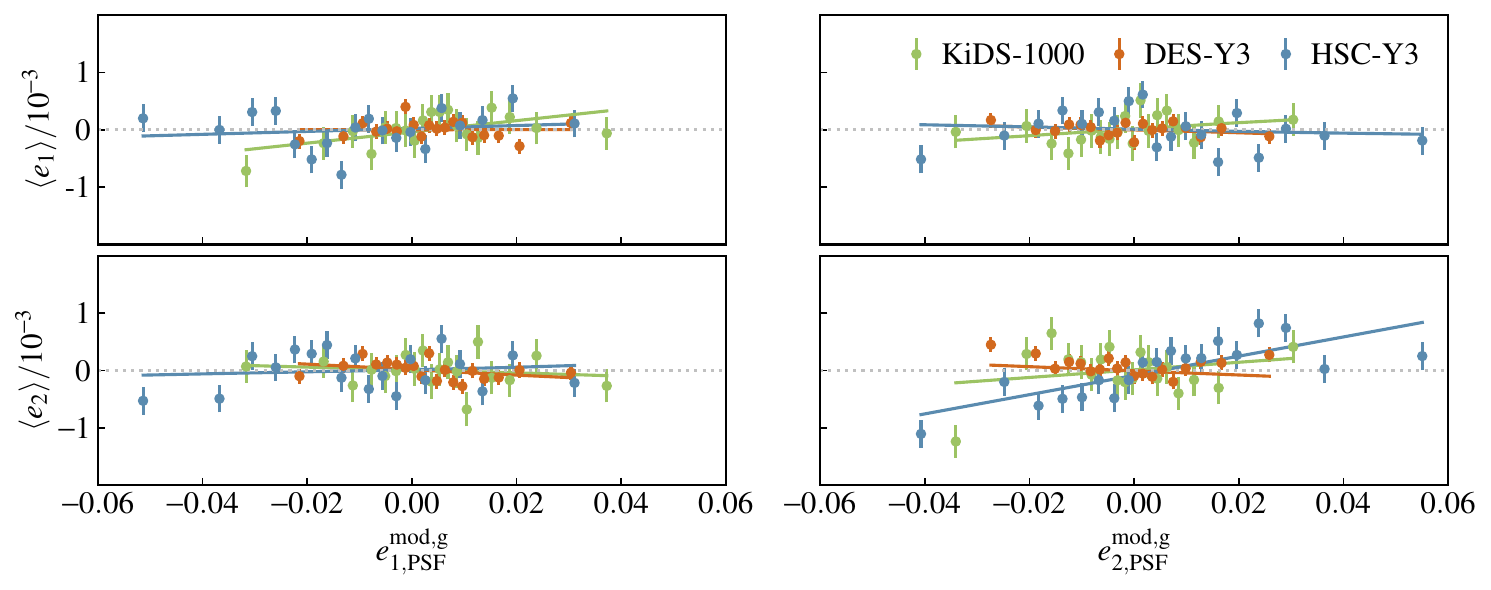}
    
    \vspace{0.1in}
     \includegraphics[width=\linewidth]{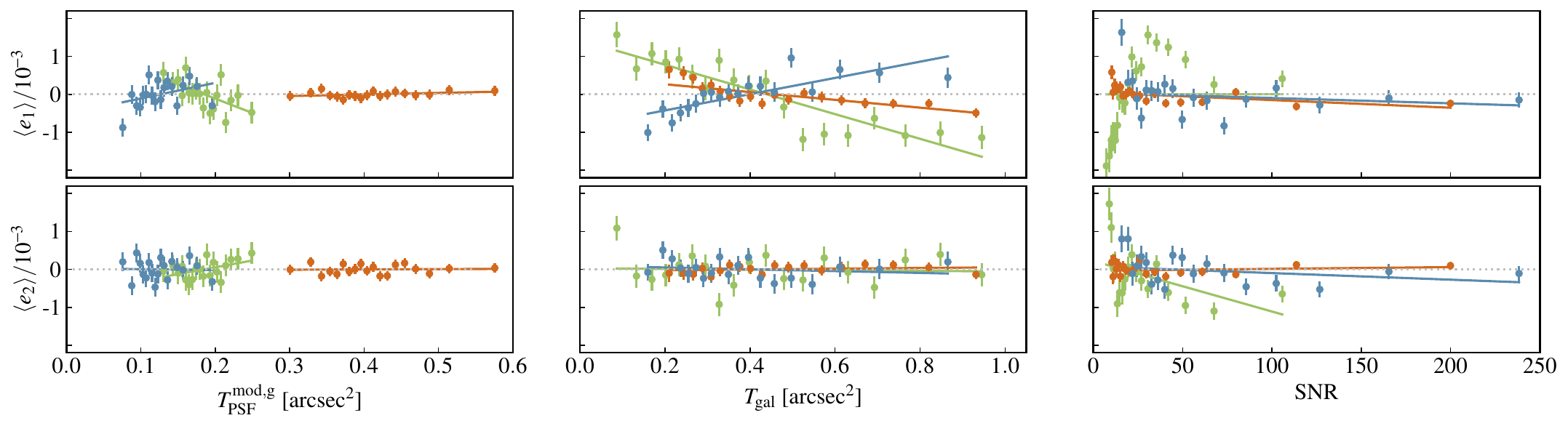}
       
    \caption{Components of the mean galaxy shape $\langle e_{1,2} \rangle$ as a function of the input PSF ellipticities (top row), PSF size (bottom left), galaxy size (bottom center), and galaxy SNR (bottom right). Each measurement is evaluated for a total of 20 bins within the range [-1, 1] for $e_{1,2,\rm{PSF}}$, [0, 1] for $T_{\rm{PSF}}$, $T_{\rm{gal}}$ and [0, 300] for SNR. The widths of the bins for each survey are chosen to achieve an equal number of galaxies per bin.  The linear best-fit to each panel is shown on the plot with the slopes listed in Table~\ref{table:linear_fit}.}
    \label{fig:meanshear}
\end{figure*}

We show the overall mean shape in each survey's sub-areas in Table~\ref{table:meanshear}. We find that the mean shape values range from below $10^{-5}$ to 10$^{-3}$, spanning two orders of magnitude. 
Overall, the amplitude of the mean shape in the HSC-Y3 fields is generally higher;
the exception is the VVDS field, for which we find $\langle e_{1} \rangle$  value of $10^{-5}$, comparable to the $\langle e_{2} \rangle$ obtained by DES-Y3.
In \citet{Gatti_2021}, the authors noted that the $\langle e_{1} \rangle$ value is larger than expectations from cosmic variance, and seems to be correlated with the ratio between the galaxy size and the PSF size (also see Figure~\ref{fig:meanshear}). 
Across the different HSC-Y3 fields and across the north and south KiDS-1000 fields, the mean shape values have similar orders of magnitude, indicating that there are no single fields that stand out as having a large systematic residual.

We next look at trends of the mean shape as a function of different quantities that are not expected to correlate with shape. We measure the two components of the mean galaxy shape $\langle e_{1,2}\rangle$ as functions of several PSF and galaxy quantities: the components of the PSF ellipticity and the size of the PSF interpolated to the galaxy positions, the size of the galaxy, and the signal-to-noise ratio of the galaxy.

These tests are designed to empirically reveal unexpected systematic effects associated with the shear measurement process as in theory, the shear measurement should be independent of these quantities. We show the measurements in Figure~\ref{fig:meanshear}.

For each quantity, we first subtract the overall mean shape of the catalog, and then bin the objects according to a selection of PSF and galaxy quantities into 20 bins with equal numbers of galaxies. We then calculate the mean shape and the error on the mean in each bin (estimated via the standard deviation divided by $\sqrt{N}$ where $N$ is the number of objects in each bin). Each survey does not appear in the same range as the different datasets have different observing conditions as well as selection criteria for the samples (see also Figure~\ref{fig:histograms}). The equal-number binning ensures that we have good statistics in each bin. 
We note that for DES-Y3, calibration needs to be done for each of these bins according to the \textsc{Metacalibration} algorithm, since the binning introduces selection bias that needs to be calibrated.   
A linear fit is performed to each set of data points and shown in the figure, with the slope of the linear best-fit listed in Table~\ref{table:linear_fit}.

\begin{table*}
\center
    \caption{Slope of the linear fit for all the quantities in Figure~\ref{fig:meanshear}. For each survey, the first row corresponds to $\langle e_1 \rangle$ and the second row corresponds to $\langle e_2 \rangle$.}
    \label{table:linear_fit}
    \begin{tabular}{rrrrr} \toprule
    $e_{\rm 1,PSF}^{\rm mod,g}$&$e_{\rm 2, PSF}^{\rm mod,g}$&$T_{\rm PSF}^{\rm mod,g}$&$T_{\rm gal}$&SNR \\
    \multicolumn{1}{l}{\textbf{KiDS-1000}}&&&&\\
     $(9.81\pm03.59)\times 10^{-3}$& $(5.72\pm3.62)\times 10^{-3}$& $(-7.58\pm2.38)\times 10^{-3}$ & $(-3.26\pm0.39)\times 10^{-3}$ & $(1.08\pm0.75)\times 10^{-5}$ \\
     $(-2.49\pm4.04)\times 10^{-3}$ & $(6.62\pm6.77)\times 10^{-3}$ & $(3.61\pm1.65)\times 10^{-3}$ & $(-0.83\pm4.13)\times 10^{-4}$ & $(-1.33\pm0.57)\times10^{-5}$\\[0.2cm]
     \multicolumn{1}{l}{\textbf{DES-Y3}}&&&&\\
     $(0.04\pm3.00)\times 10^{-3}$ & $(-2.42\pm2.00)\times 10^{-3}$ & $(4.06\pm2.48)\times 10^{-4}$ & $(-1.03\pm0.18)\times 10^{-3}$ & $(-2.02\pm0.72)\times10^{-6}$ \\
     $(-4.65\pm2.86)\times 10^{-3}$ & $(-3.63\pm2.80)\times 10^{-3}$ & $(0.77\pm4.26)\times 10^{-4}$ & $(0.95\pm1.20)\times 10^{-4}$ & $(4.63\pm5.97)\times10^{-7}$ \\[0.2cm]
     \multicolumn{1}{l}{\textbf{HSC-Y3}}&&&&\\
     $(2.56\pm3.36)\times 10^{-3}$& $(-1.95\pm3.52)\times 10^{-3}$ & $(4.12\pm2.45)\times 10^{-3}$ & $(2.14\pm0.40)\times 10^{-3}$ & $(-1.41\pm1.58)\times10^{-6}$ \\
    $(2.24\pm3.98)\times 10^{-3}$ & $(1.68\pm0.33)\times 10^{-2}$ & $(-0.22\pm2.05)\times 10^{-3}$ & $(-2.38\pm3.31)\times 10^{-4}$ & $(-1.74\pm1.29)\times10^{-6}$ \\
    \hline
    \end{tabular}
\end{table*}

\begin{table*}
\center
    \caption{$\chi^2$/dof and $p$-value in parenthesis for all quantities in Figure~\ref{fig:meanshear}. For each survey, the first row is corresponds to $\langle e_1 \rangle$ and the second row corresponds to $\langle e_2 \rangle$.}
    \label{table:nullhyp_fit}
    \begin{tabular}{rrrrr} \toprule
    $e_{\rm 1,PSF}^{\rm mod,g}$&$e_{\rm 2, PSF}^{\rm mod,g}$&$T_{\rm PSF}^{\rm mod,g}$&$T_{\rm gal}$&SNR \\
    \multicolumn{1}{l}{\textbf{KiDS-1000}}&&&&\\
     16.185/19 (0.64) & 14.553/19 (0.75)& 33.082/19 (0.02)& 156.121/19 (--) & 217.567/19 (--)\\
     11.158/19 (0.92) & 35.788/19 (0.01)& 12.667/19 (0.86)& 34.48/19 (0.02) & 144.767/19 (--)\\
     \multicolumn{1}{l}{\textbf{DES-Y3}}&&&&\\
     24.567/19 (0.18) & 25.638/19 (0.14) & 5.837/19 (2.5e-4) & 62.704/19 (1.4e-6) & 43.702/19 (1.0e-3)\\
     12.984/19 (0.84) & 25.725/19 (0.14) & 15.022/19 (4.3e-4) & 9.454/19 (0.97) & 19.392/19 (0.43)\\
     \multicolumn{1}{l}{\textbf{HSC-Y3}}&&&&\\
     32.619/19 (2.7e-2)& 34.417/19 (1.6e-2) & 36.567/19 (9.0e-3)& 71.922/19 (4.4e-8)& 55.123/19 (2.2e-5)\\
     33.230/19 (2.3e-2)& 72.416/19 (3.6e-8) & 21.375/19 (0.32)& 18.472/19 (0.49)& 38.889/19 (4.6e-3)\\
    \hline
    \end{tabular}
\end{table*}

We find that overall, DES-Y3 has the smallest scatter in the measured data points. This is mostly driven by the fact that DES-Y3 has more galaxies. We note that for SNR, the linear fit does not pick up the trend in the data well, which is especially apparent for the $e_1$ component of KiDS-1000, but is also present in the other surveys. With this in mind, we also compute the reduced $\chi^2$ for each survey quantity and associated $p$-value in Table ~\ref{table:nullhyp_fit} for a null signal. We  set a threshold of $p = 3 \times 10^{-3}$. Measurements that result in $p$-values below this threshold are $\gtrsim 3\sigma$ from a null signal and therefore highly unlikely to arise from noise alone if there is no underlying trend. We will assume this threshold value throughout the paper for other tests where we evaluate a $p$-value.

Some $p$-values are estimated to be excessively small i.e., beyond machine-level precision; these values are not considered meaningful and thus are replaced with hyphens in the table. We maintain this notation throughout the paper. For KiDS-1000, $\langle e_{1} \rangle$ vs. $T_{\rm gal}$, $\langle e_{1} \rangle$ vs. SNR, and $\langle e_{2} \rangle$ vs. SNR are significant; for DES-Y3, $\langle e_{1} \rangle$ vs. $T_{\rm PSF}^{\rm mod,g}$, $\langle e_{1} \rangle$ vs. $T_{\rm gal}$, $\langle e_{1} \rangle$ vs. SNR, and $\langle e_{2} \rangle$ vs. $T_{\rm gal}$ are significant; for HSC-Y3, $\langle e_{1} \rangle$ vs. $T_{\rm gal}$, $\langle e_{1} \rangle$ vs. SNR, and $\langle e_{2} \rangle$ vs. $e_{\rm 2, PSF}^{\rm mod,g}$ are significant.

Identifying the origin and quantifying the potential influence of these trends is key to building a robust understanding of the systematics present in a survey. Some of the trends that are associated with PSF quantities will later also be captured in the tau statistics (Section~\ref{sec:tau}). The impact of correlations with galaxy size and SNR on biases in cosmological constraints remains unclear and could serve as motivation for further inquiry.

\subsection{PSF and shear two-point statistics}
\label{sec:2pt}

Two-point diagnostic tests involving the PSF and shear quantities were developed to estimate how errors in the PSF directly  propagate into a bias in the cosmic shear measurement.
A spatially random PSF error will not introduce a bias in the cosmic shear signal; whereas if the PSF error contains spatially coherent structures, it may result in a spurious shear signal.   

There are three related statistics in the literature, all of which are based on the formalism derived in \cite{PaulinHenriksson2008}: 1) rho statistics (Section~\ref{sec:rowe}) look at the two-point correlation of PSF quantities without propagating it into an error in cosmic shear; 2) tau statistics (Section~\ref{sec:tau}) were introduced to connect rho statistics more directly to cosmological signals by assuming a model of how the shear errors are related to the PSF errors and empirically fitting the model parameters; 3) Paulin-Henriksson (PH) statistics (Section~\ref{sec:ph}) rely on a more analytical approach to assume a simple fixed form in which PSF errors propagate into shear errors, and directly calculate the expected error in cosmic shear from the PSF errors.    

We repeat previous studies' analysis choices so that our measurements reproduce the surveys' published results when possible. In the case where surveys conducted the same diagnostic test with small variations between the two (e.g., rho statistics), we match the analysis choices for one of the surveys. Each of these statistics comes with different pass or fail criteria, and comparing between statistics is not entirely straightforward. We use the criteria defined in each survey and examine all three datasets in a unified way.

\subsubsection{Rho statistics}
\label{sec:rowe}

\begin{figure*}
    \includegraphics[width=0.95\textwidth]{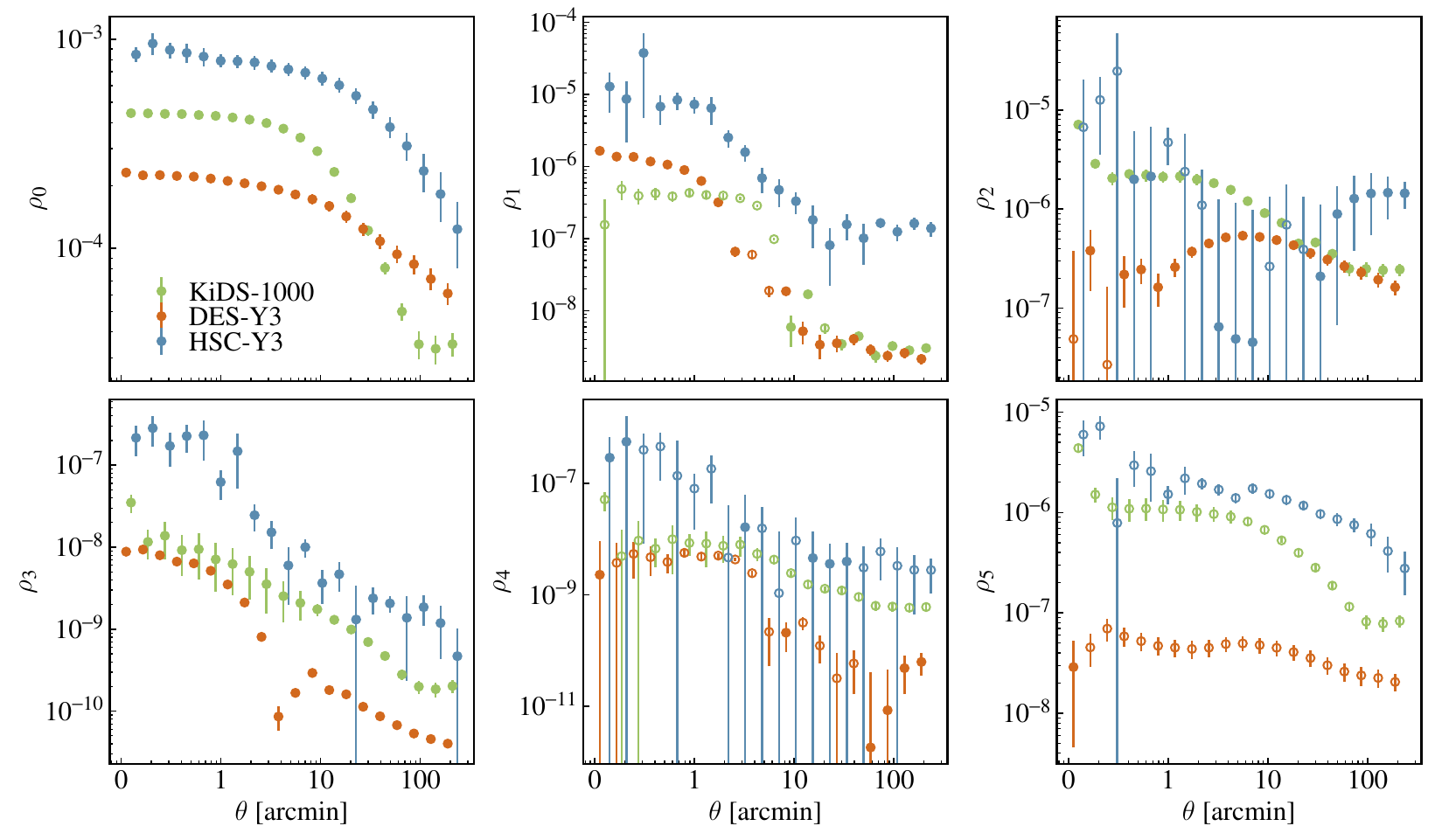}
    \caption{Rho statistics as described in Equation \ref{eq:rho} for each survey, displaying only the components corresponding to $\xi_{+}$. Negative correlations are shown in absolute value with an open circle. Theta ranges are plotted with slight offsets for clearer visibility. The $\rho$s are evaluated for 20 angular bins between 0.1 to 250 arcminutes and error bars are estimated using jackknife resamplings for 250, 1000, and 150 patches for KiDS-1000, DES-Y3, and HSC-Y3.
    }
    \label{fig:rowes}
\end{figure*}

The rho statistics \citep{rowe2010, Jarvis2016} are a set of six two-point correlation functions formed by the auto- and cross-correlation of PSF quantities.
It allows us to isolate the most important sources of error in the PSF modeling. We note that while the rho statistics are informative in understanding the quality of the PSF model, we cannot use them to directly infer how the PSF errors contribute to the galaxy shape measurements. We would need to combine the rho statistics with the tau statistics described in the next section. The six rho statistics are
\begin{align}
\rho_{0}(\theta)&=\langle \boldsymbol{e}_{\rm PSF}^{\rm mod}(x) \boldsymbol{e}_{\rm PSF}^{\rm mod}(x + \theta)\rangle, \notag\\
\rho_{1}(\theta)&=\langle\delta \boldsymbol{e}_{\rm PSF}(x)\delta \boldsymbol{e}_{\rm PSF}(x + \theta)\rangle, \notag\\
\rho_{2}(\theta)&=\langle \boldsymbol{e}_{\rm PSF}^{\rm mod}(x)\delta \boldsymbol{e}_{\rm PSF}(x + \theta)\rangle, \notag\\
\rho_{3}(\theta)&=\left\langle \left( \boldsymbol{e}_{\rm PSF}^{\rm meas}\frac{ \delta T_{\rm PSF} }{ T_{\rm PSF}^{\rm meas} }\right)(x) \left( \boldsymbol{e}_{\rm PSF}^{\rm meas}\frac{ \delta T_{\rm PSF} }{ T_{\rm PSF}^{\rm meas} }\right)(x+\theta)
\right\rangle, \notag\\
\rho_{4}(\theta)&=\left\langle  \delta \boldsymbol{e}_{\rm PSF}^{\rm mod}(x) \left( \boldsymbol{e}_{\rm PSF}^{\rm meas}\frac{ \delta T_{\rm PSF} }{ T_{\rm PSF}^{\rm meas} }\right)(x+\theta)
\right\rangle, \notag\\
\rho_{5}(\theta)&=\left\langle \boldsymbol{e}_{\rm PSF}^{\rm mod}(x) \left( \boldsymbol{e}_{\rm PSF}^{\rm meas}\frac{ \delta T_{\rm PSF} }{ T_{\rm PSF}^{\rm meas} }\right)(x+\theta)
\right\rangle,
\label{eq:rho}
\end{align}

where $x$ and $x+\theta$ represent coordinates of pairs of stars separated by an angle $\theta$. We follow the definitions as presented in \citet{Gatti_2021}, which differ slightly from \citet{Jarvis2016,JarvisPSFDES2021}. We note that formally each of the rho statistics have a plus and a minus component analogous to  $\xi_{+}$ and $\xi_{-}$. We only show here the component corresponding to $\xi_{+}$, though we do use the component corresponding to $\xi_{-}$ later when computing the tau statistics (see next section).

To measure the rho statistics, \textsc{TXPipe} calls the software package \textsc{TreeCorr} \citep{Jarvis2004,Jarvis2015} to compute rho statistics, as shown in Figure~\ref{fig:rowes}. We choose to calculate the correlations in the range of 0.1 to 250 arcminutes with with a \texttt{binslop} parameter set to 0.0. Jackknife error estimations are computed with a different number of spatial patches for each survey to achieve a similar area per patch. We choose 250 patches for KiDS-1000, 1000 for DES-Y3, and 150 for HSC-Y3.

We now examine each of the rho statistics and compare across the different surveys. We note that the correlation function amplitudes are broadly expected to scale inversely with the number of exposures per area, since individual structures captured within each exposure will average down. 

$\rho_0$ is the auto-correlation of the PSF  and quantifies the intrinsic pattern of the PSF. We find that HSC-Y3 consistently measures the largest overall values for each rho correlation aside from $\rho_0$. However, the ellipticity of the PSF scales inversely with its size and since the HSC-Y3 PSFs are much smaller, this trend is to be expected. This does not affect how we infer its impact on cosmic shear when combined with tau in the next section. We also find that each survey's measured correlations drop at large separation, showing that shear signals measured for galaxies at closer separation will be subject to stronger additive biases.  For all surveys we see higher power on small scales between $10^{-4}$ and $10^{-3}$ and lower power on large scales at several times $10^{-5}$, with DES-Y3 having an overall lower amplitude. There is a sharp drop in $\rho_0$ for KiDS-1000 and HSC, which corresponds to the size of the focal plane. For DES-Y3 the drop is less steep, corresponding to a larger focal plane and less vignetted field.  
$\rho_{1}$ is the auto-correlation of the PSF errors, which gives an estimate of the quality of the model. We see that the PSF ellipticity model errors are approximately two orders of magnitude smaller than the PSF ellipticity itself as represented by the amplitudes of $\rho_1$ and $\rho_0$. This shows that all surveys have achieved percent-level precision modeling. 
$\rho_{3}$ is the auto-correlation of the fractional error of the PSF model size, scaled by the PSF ellipticity to make it a spin-2 quantity. Comparing $\rho_3$ and $\rho_0$ illustrates again that the PSF model error is much smaller than the PSF itself (approximately two-to-three orders of magnitude lower). The overall behavior of $\rho_3$ is similar to $\rho_1$, where HSC-Y3 is higher than the other two surveys. 
$\rho_{2}, \rho_{4}, \rho_{5}$ are generally harder to interpret since they involve cross-correlation between the different PSF components, but we observe that generally DES-Y3 yields the lowest amplitudes. 
Interestingly $\rho_5$ stands out visually, in that the DES-Y3 measurements are much lower than the other two surveys. This indicates that for DES-Y3 specifically, the PSF model size errors are uncorrelated with the PSF shapes themselves. 

We finally note that in previous work, some form of metric was used to judge whether the PSF model is sufficiently good for cosmic shear analyses. These metrics are often arbitrary, such as 10\% of the cosmic shear signal in the lowest redshift bins. These guides are useful in catching evident problems, but are insufficient for placing quantitative statements.    

\subsubsection{Tau statistics}
\label{sec:tau}

We can expand from the notation in Equation~\ref{eq:m_and_c} and realize that the additive bias in shear, or the term $\boldsymbol{c}$, often is associated with the PSF. One common model that is used in the literature is to define  
\begin{equation}
\boldsymbol{c} \equiv \alpha \boldsymbol{e}_{\rm PSF}^{\rm mod} + \beta \delta \boldsymbol{e}_{\rm PSF} + \eta \boldsymbol{e}_{\rm PSF}^{\rm meas}\frac{\delta T_{\rm PSF}}{T_{\rm PSF}^{\rm meas}}, 
\label{eq:define_c}
\end{equation}
where the assumption here is that any additive shear error can be decomposed into three terms. The first term scales with the PSF ellipticity itself, the second term scales with the PSF ellipticity error and the final scales with the PSF size error. This model is motivated by \citet{PaulinHenriksson2008}, who performed a Taylor expansion on a moment-based shear measurement algorithm, keeping the first-order terms. 

One can attempt to extract the parameters $\alpha, \beta, \eta$ via the data, which will tell us how the PSF deconvolution from galaxy images and PSF modeling errors are contributing to an additive bias in the shear via Equation~\ref{eq:define_c}. This can be done using the tau statistics first introduced in \citet{Gatti_2021}.
These are cross-correlation of galaxy shape measurements with PSF quantities, or 
\begin{align}
\tau_{0}(\theta)&=\langle \boldsymbol{e}(x)\boldsymbol{e}_{\rm PSF}^{\rm mod}(x + \theta)\rangle, \notag\\
\tau_{2}(\theta)&=\langle \boldsymbol{e}(x)\delta \boldsymbol{e}_{\rm PSF}(x + \theta)\rangle, \notag\\
\tau_{5}(\theta)&= \left\langle \boldsymbol{e}(x) \left( \boldsymbol{e}_{\rm PSF}^{\rm meas}\frac{\delta T_{\rm PSF}}{T_{\rm PSF}^{\rm meas}} (x + \theta) \right) \right\rangle.  
\label{eq:tau}
\end{align}
Substituting $\boldsymbol{e}$ with $\boldsymbol{e}+\boldsymbol{c}$ and using Equation~\ref{eq:define_c}, we can rewrite the tau statistics as a linear combination of rho statistics. 

\begin{align}
\tau_{0}(\theta)=\alpha\rho_{0}(\theta)+\beta\rho_{2}(\theta)+\eta\rho_{5}(\theta), \notag \\
\tau_{2}(\theta)=\alpha\rho_{2}(\theta)+\beta\rho_{1}(\theta)+\eta\rho_{4}(\theta), \notag\\
\tau_{5}(\theta)=\alpha\rho_{5}(\theta)+\beta\rho_{4}(\theta)+\eta\rho_{3}(\theta).
\label{eq:tau_rho}
\end{align}
The above assumes that there is no correlation between the true shear and any of the PSF quantities.

\begin{figure}
    \centering
    \includegraphics[width=\columnwidth]{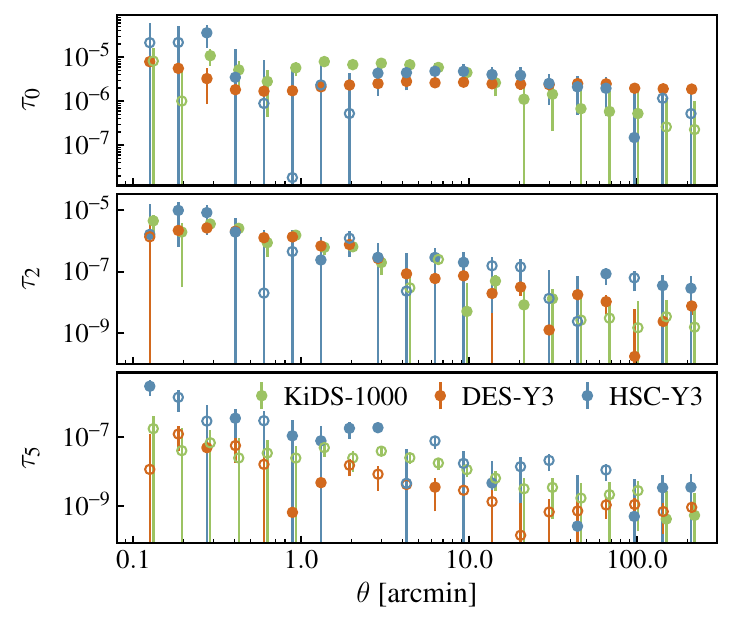}
    \caption{Non-tomographic $\tau_{0}$ (top), $\tau_{2}$ (middle), and $\tau_{5}$ (bottom) from Equation \ref{eq:tau}, displaying only the components corresponding to $\xi_+$. Theta ranges are plotted with slight offsets for clearer visibility. Negative correlations are shown in absolute value with an open circle. Tau statistics are evaluated for 20 angular bins between 0.1 and 250 arcminutes for KiDS-1000, DES-Y3, and HSC-Y3. Error bars are estimated with jackknife resamplings for 250, 1000, and 150 patches in the respective survey footprints.}
    \label{fig:tau_stats}
\end{figure}

We measure the tau statistics for 20 angular bins within the range of 0.1 to 250 arcminutes for all surveys, using a \texttt{binslop} of 0.0. All the shear values and PSF quantities are mean-subtracted. The covariance is estimated using jackknife resampling with the same patch configuration as detailed in Section~\ref{sec:rowe}.
The positive components of the three tau statistics are shown in Figure~\ref{fig:tau_stats}.  We find that HSC-Y3 measures larger signals compared to DES-Y3 and KiDS-1000 in all tau measurements. The most significant correlations are reflected in $\tau_0$ and $\tau_2$, showing that the PSF ellipticity and modeling errors are correlated with the shear. 

In the last step, we combine the measured rho and tau statistics to constrain $\alpha, \beta, \eta$ using Equation~\ref{eq:tau_rho}. To do this, we first use \textsc{SciPy}'s \texttt{optimize.minimize} method, which uses the Nelder-Mead  algorithm, to find the optimal initial values for each parameter \citep{SciPy2020, NelderMead}. We then use the package \textsc{EMCEE} \citep{Foreman-Mackey2013} to create Markov Chain Monte Carlo (MCMC) samples to determine the final best-fit values for the coefficients. We choose to use 100 walkers and 5000 steps with a burn-in of 2000 steps in our inference. In both steps, we use the Gaussian likelihood $\mathcal{L}$ defined via
\begin{equation}
\label{eq:Likelihood}
	\ln \mathcal{L}(\boldsymbol{\theta}) = -\frac{1}{2} (\mathbf{d}-\mathbf{m})^T \mathcal{C}_{\rm debias}^{-1} (\mathbf{d}-\mathbf{m}) \equiv -\frac{1}{2} \chi^2 \; ,
\end{equation}
where $\boldsymbol{\theta} = (\alpha, \beta, \eta)$ is the model parameters
, $\mathbf{d}$ is the concatenated data vector formed out of all the $\tau$ statistics and $\mathbf{m}$ is the model for the $\tau$ statistics which takes the form of Equation~\ref{eq:tau_rho} and uses fixed templates of $\rho$'s from the measurements. We do not account for the error bars in the $\rho$ statistics in our analysis since including the errors degrades the quality of the fit. $\mathcal{C}_{\rm debias}^{-1}$ is the corrected inverse covariance, where we apply two correction factors to the jackknife covariance 
$\mathcal{C}$,
\begin{align}
\label{eq:HartlapCovariance}
	&\mathcal{C}_{\rm debias}^{-1} = a b \mathcal{C}^{-1}, \notag \\
 &a=\frac{N_{\rm JK} - (N_{\rm dv} - N_{\theta}) - 2}{N_{\rm JK}-1}, \notag \\
 &b= (N_{\rm dv} - N_{\theta})\frac{N_{\rm JK}-N_{\rm dv}-2}{(N_{\rm JK}-N_{\rm dv}-1)(N_{\rm JK}-N_{\rm dv}-4)}.
\end{align}
 $a$ is the correction factor described in \citet{Hartlap2006}, which accounts for the fact that there are finite number of resampled instances, making the covariance inherently noisy and inverting a noisy matrix tends to bias the covariance low. $b$ is the correction factor described in \citet{Dodelson2013}, which accounts for the fact that when inverting a noisy covariance matrix one not only underestimates the error bars, the best-fit could also be biased. 
 These correction methods were originally formulated for simulated data vectors, but here we adopt these methods by substituting jackknife samples of the data for simulations, where $N_{\rm JK}= 250, 1000$, and $150$ for KiDS-1000, DES-Y3 and HSC-Y3.
$N_{\theta}=3$ is the number of free parameters in the model and $N_{\rm dv} = 120$ is the total number of elements in the data vector: 20 measurements per $\tau_{\pm 0,2,5}$. The values and resulting reduced $\chi^2$ can be found in Table~\ref{table:taustats}. 
Across all surveys and tomographic bins, the reduced $\chi^2$ remains less than 1.
While $\alpha$ and $\beta$ values are near zero, $\eta$ can measure to a significantly larger value. This, however, does not appear to affect the fit drastically. 
We note that overall, KiDS-1000's best fitting $\alpha$ values are both larger and greater than one standard deviation from DES-Y3 and HSC-Y3 best-fitting $\alpha$s; this trend does not occur for $\beta$ or $\eta$. A possible cause is the $\tau_0$ for KiDS-1000 may correlate more strongly to $\rho_0$, since $\rho_0$ is not used to estimate the latter variables as described in Equation \ref{eq:tau_rho}. 
We explore how these best-fit values affect cosmological constraints in Section~\ref{sec:cosmology}.

\begin{table*}
\begin{center}
\caption{Best fitting $\alpha$,$\beta$,$\eta$ parameters to characterize the relationship between rho and tau statistics according to Equation \ref{eq:tau_rho}. Best-fits are determined by performing an MCMC analysis with a Gaussian likelihood. We include best-fit values and an estimated $\hat{\chi}^2\equiv \chi^2/{\rm dof}$ for the non-tomographic and tomographic autocorrelation bins.}
\label{table:taustats}
\begin{tabular}{lrrrr} \toprule
 &  $\alpha$ & $\beta$ & $\eta$ & $\hat{\chi}^2$\\
\multicolumn{2}{l}{\textbf{Bin 1}}&&&\\
KiDS-1000& $(6.73 \pm 2.53)\times 10^{-2}$ & $-1.11 \pm 0.543$ & $14.2 \pm 7.00$ & 0.49\\
DES-Y3& $(9.10 \pm 5.01)\times 10^{-3}$ & $0.491 \pm 0.282$ & $-2.31 \pm 2.85$ & 0.75\\
HSC-Y3& $(1.40 \pm 0.798)\times 10^{-2}$ & $-0.0503 \pm 0.228$ & $1.76 \pm 2.10$ & 0.35\\[0.2cm]
\multicolumn{2}{l}{\textbf{Bin 2}}&&&\\
KiDS-1000& $(-1.26 \pm 1.49)\times 10^{-3}$ & $-0.493 \pm 0.322$ & $-11.8 \pm 4.43$ & 0.57\\
DES-Y3& $(4.31 \pm 5.27)\times 10^{-3}$ & $1.84 \pm 0.301$ & $-5.55 \pm 3.32$ & 0.87\\
HSC-Y3& $(1.00 \pm 0.648)\times 10^{-2}$ & $0.414 \pm 0.199$ & $0.444 \pm 1.52$ & 0.22\\[0.2cm]
\multicolumn{2}{l}{\textbf{Bin 3}}&&&\\
KiDS-1000& $(2.32 \pm 1.42)\times 10^{-2}$ & $-0.121 \pm 0.305$ & $2.08 \pm 3.91$ &0.68\\
DES-Y3& $(3.50 \pm 5.83)\times 10^{-3}$ & $2.40 \pm 0.285$ & $1.54 \pm 3.21$ & 0.73\\
HSC-Y3& $(-7.49 \pm 11.5)\times 10^{-3}$ & $0.0102 \pm 0.230$ & $-0.275 \pm 1.90$ & 0.15\\[0.2cm]
\multicolumn{2}{l}{\textbf{Bin 4}}&&&\\
KiDS-1000& $(4.05 \pm 1.45)\times 10^{-2}$ & $-0.222 \pm 0.355$ & $-1.31 \pm 4.82$ &0.65\\
DES-Y3& $(1.23 \pm 0.692)\times 10^{-2}$ & $1.42 \pm 0.357$ & $-2.86 \pm 3.85$ & 0.88\\
HSC-Y3& $(-2.01 \pm 13.7)\times 10^{-3}$ & $-0.514 \pm 0.424$ & $-6.03 \pm 2.59$ & 0.29\\[0.2cm]
\multicolumn{2}{l}{\textbf{Bin 5}}&&&\\
KiDS-1000 &$(-5.98 \pm 1.64)\times 10^{-2}$ & $0.402 \pm 0.392$ & $-10.3 \pm 4.04$ &0.61\\[0.2cm]
\multicolumn{2}{l}{\textbf{Non-tomographic}}&&&\\
KiDS-1000 &$(1.47 \pm 0.685)\times 10^{-2}$ & $-0.254 \pm 0.172$ & $1.05 \pm 1.89$& 0.78 \\
DES-Y3 & $(4.85 \pm 3.03)\times 10^{-3}$ & $1.40 \pm 0.158$ & $-2.72 \pm 1.73$ & 0.97\\
HSC-Y3 & $(1.90 \pm 0.468)\times 10^{-2}$ & $0.197 \pm 0.111$ & $1.50 \pm 0.811$ & 0.40\\
\hline
\end{tabular}
\end{center}
\end{table*}

\subsubsection{PH statistics}
\label{sec:ph}

KiDS-1000 derived a slightly different statistic to estimate the PSF modeling error called Paulin-Henriksson (PH) statistics. \citet{PaulinHenriksson2008} define an estimator to quantify the impact of the residual PSF shape and size on the observed galaxy shape. They perform a Taylor expansion of this estimator which takes the form:
\begin{equation}
\boldsymbol{e}_{\rm obs}\simeq \boldsymbol{e}_{\rm obs}^{\rm perfect} +  ( \boldsymbol{e}_{\rm obs}^{\rm perfect}-\boldsymbol{e}_{\rm PSF}^{\rm mod}) \frac{\delta T_{\rm PSF}}{T_{\rm gal}}-\frac{T_{\rm PSF}^{\rm mod}}{T_{\rm gal}}\delta \boldsymbol{e}_{\rm PSF},
\end{equation}
where $\boldsymbol{e}_{\rm obs}^{\rm perfect}$ is the true galaxy ellipticity, obtained from theoretical modeling. $T_{\rm gal}$ is the true size of the galaxy in absence of PSF convolution and  $\delta e_{\rm PSF}$, $\delta T_{\rm PSF}$ are the differences between the true and model PSF ellipticities and sizes. 

This systematics model is then incorporated in the shear two-point correlation as:
\begin{align}
\langle \boldsymbol{e}_{\rm obs}\boldsymbol{e}_{\rm obs}\rangle &\simeq  \langle \boldsymbol{e}_{\rm obs}^{\rm perfect}\boldsymbol{e}_{\rm obs}^{\rm perfect}\rangle \nonumber \\
&+ 2\left[\frac{\delta T_{\rm PSF}}{T_{\rm gal}}\right]\langle \boldsymbol{e}_{\rm obs}^{\rm perfect}\boldsymbol{e}_{\rm obs}^{\rm perfect}\rangle \nonumber \nonumber\\
&+\hspace{0.22cm}\left[\frac{1}{T_{\rm gal}}\right]^{2}\langle(\boldsymbol{e}_{\rm PSF}^{\rm mod}\delta T_{\rm PSF})(\boldsymbol{e}_{\rm PSF}^{\rm mod}\delta T_{\rm PSF})\rangle\nonumber\\
&+2\left[\frac{1}{T_{\rm gal}}\right]^{2}\langle(\boldsymbol{e}_{\rm PSF}^{\rm mod}\delta T_{\rm PSF})(\delta \boldsymbol{e}_{\rm PSF} T_{\rm PSF}^{\rm mod})\rangle\nonumber\\
&+\hspace{0.22cm}\left[\frac{1}{T_{\rm gal}}\right]^{2}\langle(\delta \boldsymbol{e}_{\rm PSF} T_{\rm PSF}^{\rm mod})(\delta \boldsymbol{e}_{\rm PSF} T_{\rm PSF}^{\rm mod})\rangle.
\label{eq:PH}
\end{align}

\begin{figure}
    \centering
    \includegraphics[width=\columnwidth]{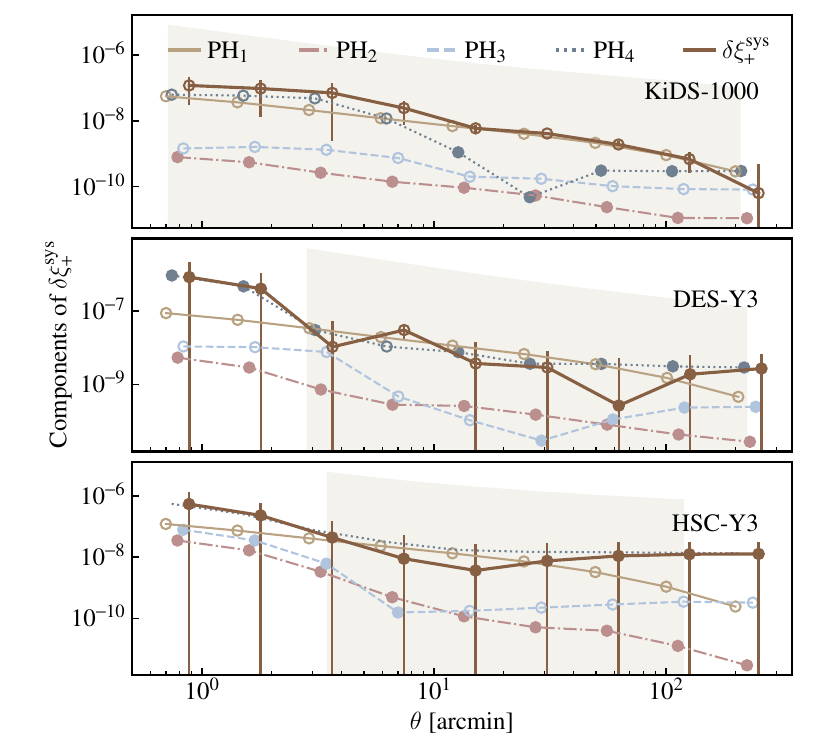}
    \caption{Values for $\delta \xi_{+}^{\rm sys}$ and terms $\rm PH_{1-4}$ of Equation \ref{eq:PH} for the highest tomographic autocorrelation bin for each survey. Theta ranges are plotted with slight offsets for clearer visibility. Negative correlations are shown in absolute value with an open circle. Beige band represents half the uncertainty for the respective tomographic cosmic shear signal $\xi_+$.}
    \label{fig:PHStats}
\end{figure}

We can derive $\langle \boldsymbol{e}_{\rm obs}^{\rm perfect} \boldsymbol{e}_{\rm obs}^{\rm perfect}\rangle$ in the first two terms (terms in Equation \ref{eq:PH} are hereafter notated as $\rm PH_{0-4}$) from theoretical calculations in \textsc{TXPipe} assuming a fiducial cosmology and each survey's official redshift distribution. We use each survey's best-fit parameters for their respective fiducial cosmology.
The estimated systematic bias is defined as $\delta \xi_{+}^{\rm sys} = \langle\boldsymbol{e}_{\rm obs} \boldsymbol{e}_{\rm obs} \rangle - \rm PH_0$.
We compute $\delta \xi_{+}^{\rm sys}$ tomographically for each survey. $\rm PH_{2-4}$ are computed between the ranges of 0.5 to 300 arcminutes with a \texttt{binslop} of 0.05. We show $\delta \xi_+^{\rm sys}$ and individual terms $\rm PH_{1-4}$ for the highest tomographic bin pair in Figure \ref{fig:PHStats}. We follow \citet{Giblin_2021} and compute an error allowance for the PH statistics, which we define as half the uncertainty of each survey's tomographic cosmic shear signal, 0.5$\sigma_{\xi+}$. 

We find that the amplitudes  of $\delta \xi_+^{\rm sys}$ fit within the error allowances across the three surveys with a maximum absolute amplitude of $\approx 10^{-6}$.
We also find that the amplitude for $\rm PH_1$, which accounts for a multiplicative bias induced from PSF modeling errors, is nearly equivalent across all surveys. Systematics captured in terms $\rm PH_{2-4}$ vary in their level of contribution to $\delta \xi_{+}^{\rm sys}$.

\subsection{Tangential ellipticity}
\label{sec:tangential}
We measure the tangential ellipticity of galaxies around the positions of stars as the final null test. We measure this quantity for two samples of stars: bright stars queried from \textit{Gaia} matched to the footprint of each survey with g-band magnitudes $\leq 16.5$ and faint stars belonging to each survey's PSF catalog.
A non-zero galaxy shape correlation around the former may indicate contaminants related to the stars brightness such as extended light halos. A non-zero signal around the latter may serve as another indication of contaminants from PSF modeling.
 
\begin {figure}
\centering
\includegraphics[width=\columnwidth]{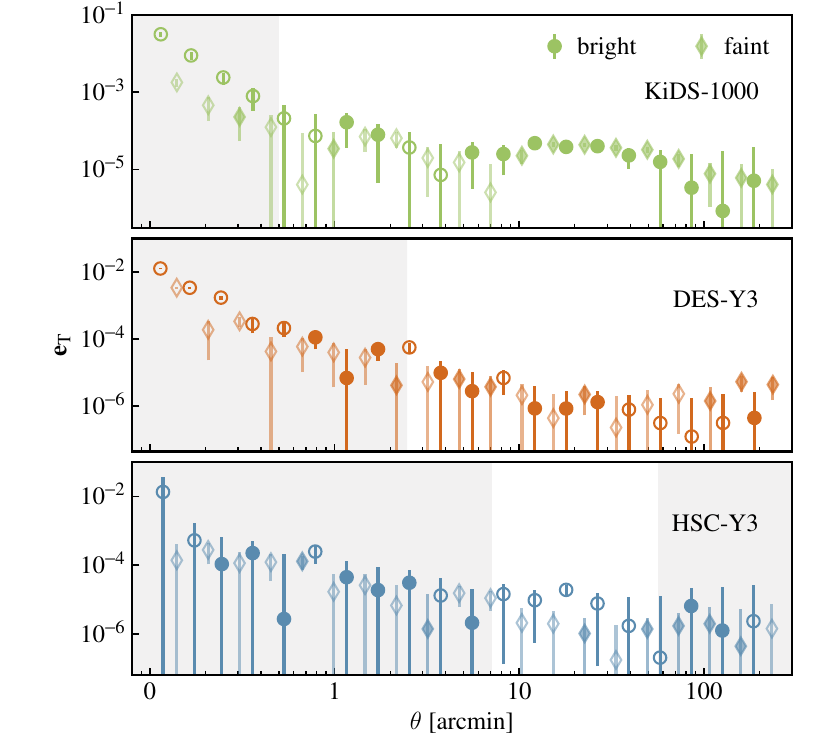}
\caption{Tangential component of the observed galaxy shape measured around two samples of stars. Measurements around bright stars are marked by opaque circles while measurements around faint stars are marked by semitransparent diamonds. Negative correlations are shown in absolute value with open markers. Gray regions denote the scale cuts for each survey, quoted in Table \ref{table:cosmology}. Error bars are estimated with jackknife resamplings for 250, 1000, and 150 patches for KiDS-1000, DES-Y3, and HSC-Y3.}
\label{fig:GTStars}
\end{figure}

We make these measurements with a \texttt{binslop} of 0.0 from the range of 0.1 to 250 arcminutes. 
We follow the procedure in \citet{Gatti_2021} to mitigate potential anti-correlations induced from under-dense star regions by weighting the stars with $w_{s} = \left(1 + \frac{n_s - \langle n_s\rangle} {\langle n_s \rangle}\right)^{-1}$, where $n_s$ is the star density.
We choose to also subtract the galaxy tangential shape around the positions of random points in the survey footprint to mitigate masking effects. We find this reduces the error estimation. The non-tomographic results are displayed in Figure \ref{fig:GTStars}. 

The amplitude of the signals are approximately the same between the bright and faint samples. For smaller separations, each survey displays relatively high signals around 10$^{-2}$. This may be due to smaller density variations in the galaxy sample, which would not have been corrected for in the weighting for the stars. The small-scales signal is unlikely to be cause for concern, since all surveys employ cuts to avoid uncertainties associated with these scales. 
We record the reduced $\chi^2$ and $p$-value nonetheless in Table \ref{table:GTStars}, in Appendix~\ref{sec: GTStarsUW} alongside the reduced $\chi^2$, $p$-value for scales that are deemed science-ready for each survey. Within the scales approved for cosmology inference, both DES-Y3 and HSC-Y3 yield signals that are consistent with zero. In contrast, KiDS-1000 measures a significantly non-zero signal for redshift bins 4 and 5. 
We show the results when we do not assign weights to the stars in Table \ref{sec: GTStarsUW} in Appendix~\ref{sec: GTStarsUW}. We find that this choice produces passing $p$-values for KiDS-1000 and HSC-Y3 within the scale cuts, however DES-Y3 fails for the signal around faint stars with a $p$-value of $6.2\times 10^{-6}$. 

\subsection{B-modes}
\label{sec:bmode}

The effect of weak gravitational lensing to first order is dominated by curl free $E$-mode, and the amplitude of divergence free $B$-mode is much lower. On the other hand, the amplitude of $B$-modes  is typically more comparable to $E$-modes  for systematic effects. Therefore, to first approximation, the presence of $B$-modes  is a powerful diagnostics for the presence of systematic effects \citep{Kaiser1992}.

The challenge in measuring $B$-modes  is the mixing of modes  between $E$ and $B$, which comes from the fact that some of the modes  are intrinsically ambiguous in the observations -- they cannot be uniquely classified as $E$ or $B$ mode. This ambiguity can come from two sources: one is on large scales where the wavelengths of the modes  are comparable to the size of the footprint. Here there is insufficient information to determine whether the mode is $E$ or $B$. Another comes from the incomplete $E$/$B$ decomposition due to the pixel mask that is used in galaxy surveys. Both scenarios introduce variance which can obscure a $B$-mode signal, depending on the level of noise. Extracting ``pure'' $B$-mode signal without the contamination of $E$-mode or ambiguous modes is important to make this a conclusive diagnostic for lensing. 

All three surveys carry out the $B$-mode test (as well as earlier shear catalog papers of the same surveys), but the specific estimator used differs. KiDS-1000 and DES-Y3 used both the angular band powers and the Complete Orthogonal Sets of $E$/$B$-Integrals \citep[COSEBIs,][]{Schneider2010}. Meanwhile, HSC-Y3 focused on testing $B$-mode using pseudo-$C_{\ell}$ alone. 

In this work, we adopt two $B$-mode estimators: the pseudo-$C_{\ell}$  and the \textsc{HybridEB} method described in \cite{Becker2016}. For the former, we call \textsc{NaMaster} internally in \textsc{TXPipe} and  run our calculation setting the \texttt{purify\_b} flag to False.\footnote{For $E/B$-mode purification to work effectively, a heavily apodized mask is required, which is not optimal for galaxy weak lensing surveys because of the complicated mask geometry.} The latter approach involves making shear correlation measurements in pixel-space and then converting to harmonic-space. More specifically, one defines some $E$ and $B$-mode estimator $X_{E,B}$ that is an integral of the linear combination of some form of the shear correlation functions multiplied by a window function $F$. 
\begin{align}
X_{E} &= \frac{1}{2} \int d\theta \theta [F_{+} \xi_{+} + F_{-} \xi_{-}], \notag \\
X_{B} &= \frac{1}{2} \int d\theta \theta [F_{+} \xi_{+} - F_{-} \xi_{-}].
\label{eq:cosebi}
\end{align}
All configuration-space $E$/$B$-mode estimators described above have similar forms as Equation~\ref{eq:cosebi} but differ in the precise way the binning of $\xi_{\pm}$ is accounted for and the exact form of $F_{\pm}$. This estimator has an advantage that the measurement is carried out in configuration space, which avoids $E$-$B$ mixing as in the case of directly computing $B$-modes  applying masks with complicated geometry, which is often the case for galaxy weak lensing surveys. 

For \textsc{HybridEB}, we use the $\xi_{\pm}$ in the real-space angular scales used for the cosmic shear analysis (see also Table~\ref{table:cosmology}), and choose to look at $25<\ell<1600$, which is determined by the shape of the filter used. Errorbars are estimated by converting the individual $\xi_{\pm}$ jackknife subsamples into \textsc{HybridEB} $B$-mode measurements, and computing the jackknife covariance from those samples. For pseudo-$C_{\ell}$, we evaluate the scales that correspond to the harmonic space cosmic shear analysis performed in each survey \citep{Asgari2021, Doux2022, Dalal2023}, and the error bars are derived by measuring $C_{\ell}^{BB}$ from maps of randomly rotated shear by applying:
\begin{align}
\gamma_{1}^{\rm rot} &= \hspace{0.2cm}\gamma_{1}  \cos(2\varphi) + \gamma_{2}  \sin(2\varphi), \nonumber \\
\gamma_{2}^{\rm rot} &= -\gamma_{1}  \sin(2 \varphi) + \gamma_{2}  \cos(2\varphi).
\label{eq:bmode_rot}
\end{align}

The measurement of $B$-modes  for the highest tomographic redshift bin for each survey and the two estimators are shown in Figure~\ref{fig:bmodes} as an example while the $\chi^2$ for all bin pairs is listed in Table~\ref{table:bmodes} in Appendix~\ref{sec:tomoBmodes}. We find that the two $B$-mode estimators have different amplitudes  and are uncorrelated -- it is challenging to compare between the estimators since the scales used are different and the pseudo-$C_{\ell}$ approach may contain some non-pure modes  ($E$-to-$B$ leakage and ambiguous modes). All of the $B$-mode measurements in the plot are consistent with null as also seen in Table~\ref{table:bmodes}.

Table~\ref{table:bmodes} also reveals that it is possible for one of the estimators to pass while the other fails. In particular, in our analysis, some of the bins show non-null signal for \textsc{HybridEB}: bin 4-4 for DES-Y3 and bins 2-4, 3-3, and 3-4 for HSC-Y3. We note that this is not inconsistent with previous work, as \textsc{HybridEB} has not been used in any surveys so far, and some of the analyses only quoted the overall $\chi^2$ values instead of the bin-by-bin values. 

We note that while we assume that the detection of a $B$-mode signal is indicative of the presence of systematic effects, there other effects that are known to generate $B$-modes at higher-order. For example, \citet{Krause2021} derived the impact of higher-order lensing contributions arising from galaxy clustering and reduced shear, which may be detectable in LSST data. Additionally, \citet{Zhang23} derived how the higher order moments for the PSF model contributes to the estimated galaxy shapes which may in turn introduce spurious $B$-modes. Extending this analysis is therefore needed to achieve a more comprehensive understanding of how each source contributes to the overall lensing signal. 

\begin{figure*}
\centering
\includegraphics[width= 0.9\linewidth]{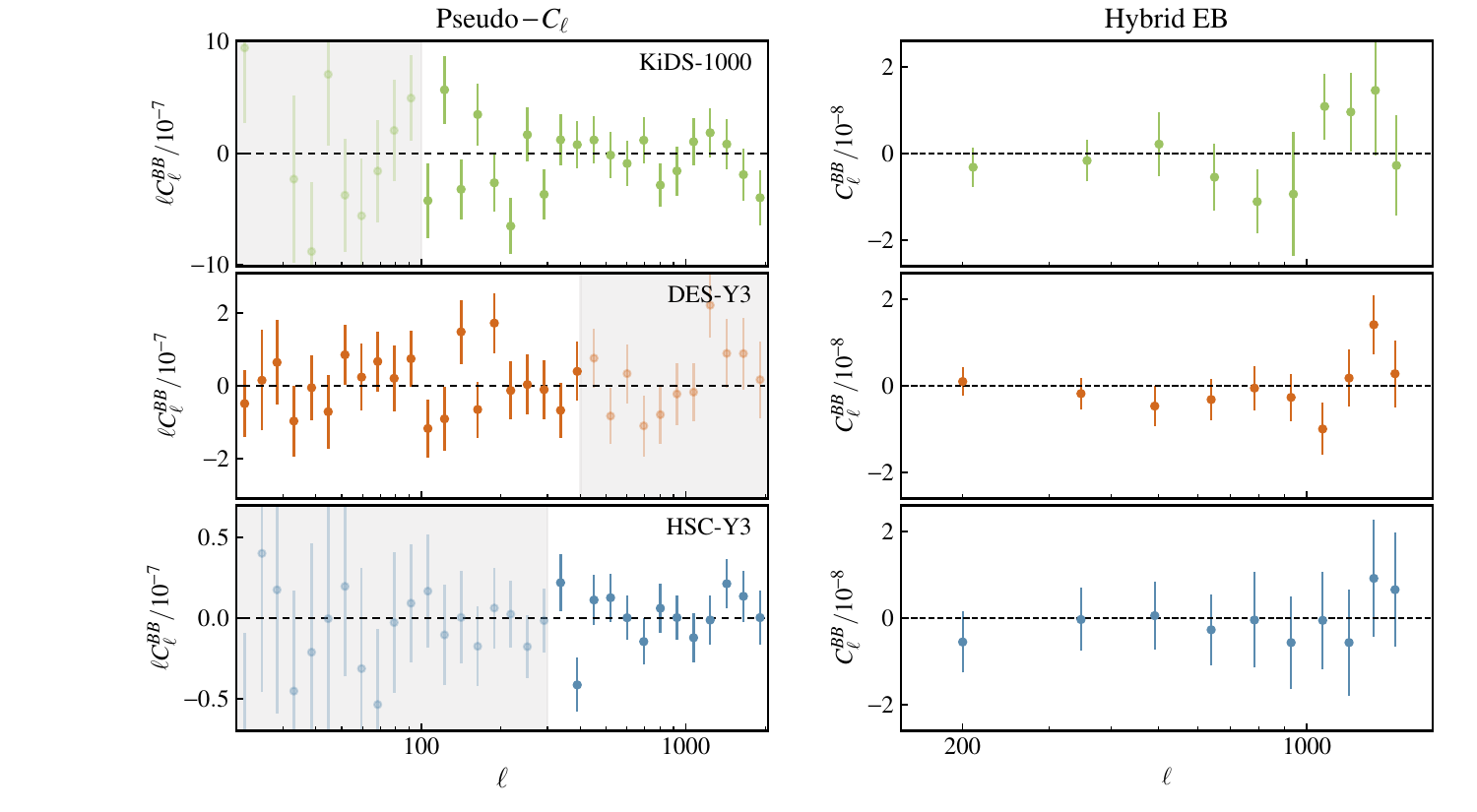}
\caption{$B$-modes  for the highest tomographic bin for KiDS-1000 (upper), DES-Y3 (middle), and HSC-Y3 (lower), estimated with two methods: pseudo-$C_{\ell}$ (left) and \textsc{HybridEB} (right) from \protect\cite{Becker2016}. Pseudo-$C_{\ell}$ estimated $B$-modes  are evaluated according to each survey's official harmonic space cosmic shear analysis \citep{Asgari2021,Doux2022,Dalal2023}, and gray regions denote scales that were excluded. Error bars are estimated by randomly rotated shear maps according to Equation \ref{eq:bmode_rot}.\textsc{HybridEB} estimated $B$-modes  are estimated within the scales used for real space cosmic shear analysis (Table \ref{table:cosmology}) for modes  $25 < \ell < 1600$. Error bars are estimated by converting the individual $\xi_{\pm}$ jackknife subsamples into \textsc{HybridEB} $B$-mode measurements, and computing the jackknife covariance from those samples.}
\label{fig:bmodes}
\end{figure*}

\subsection{Implication of PSF contamination on cosmological constraints}
\label{sec:cosmology}

All the tests presented in this paper are useful for diagnosing issues in our catalogs, but most lack a straightforward way to quantify the effect of a particular systematic on the final cosmological parameter constraints. The tau statistics, which parameterize the relationship between PSF modeling errors and galaxy shape estimation,  (Section~\ref{sec:tau}) are the only test for which the community has developed a method to do so. We perform a simulated likelihood analysis to determine the effects on cosmological constraints from each survey's measured tau statistics.

 For our simulated analysis, we unify the cosmological and astrophysical model parameters and use the shear and redshift calibration parameters derived in each survey's fiducial analysis. Table~\ref{table:cosmology} summarizes the modeling choices and priors on the model parameters.

To quantify the effect of the PSF contamination on cosmological parameters, we first generate a theoretical data vector of $\xi_{\pm}$ without PSF contamination and then one with the contamination.
The theoretical data vector without PSF contamination is generated with the fiducial model parameters listed in Table~\ref{table:cosmology}, the redshift distribution from each survey, and angular binning that is matched to the published data vectors from each survey.
We note that HSC-Y3 official analysis uses two different angular binnings for  $\xi_+$ and $\xi_-$, which is not accommodated for in the \textsc{CosmoSIS} standard library. Our inference for HSC-Y3 therefore only uses $\xi_+$ to constrain the cosmology, which we find to be more constraining than the $\xi_-$ component. The simulated likelihood analysis is run using the official covariance from each survey for the theoretical and contaminated data vectors.

To add the PSF contamination as parametrized by Equation~\ref{eq:define_c}, we add to the theoretical data vector $\xi$ a contamination in the form of $\delta\xi^{\rm PSF} = \langle \boldsymbol{c}\boldsymbol{c}\rangle$. For each bin pair $i,j$, combining Equation~\ref{eq:define_c} and Equation~\ref{eq:rho}, this expression simplifies to: 
\begin{align}
    \delta\xi_{ij}^{\rm PSF} = \alpha_i\alpha_j\rho_{0}  
    +\, \beta_i\beta_j\rho_{1}   
    + (\alpha_i\beta_j + \alpha_j\beta_i)\rho_{2} \nonumber\\ 
    +\,\,\,\eta_i\eta_j\rho_{3}
    +\, (\beta_i\eta_j+\beta_j\eta_i)\rho_{4} \nonumber\\
    +\,\,\, (\alpha_i\eta_j+\,\alpha_j\eta_i)\rho_{5}.
\label{eq:contaminant}
\end{align}

We evaluate Equation~\ref{eq:contaminant} for each survey using the best-fit $\alpha$, $\beta$, $\eta$ values listed in Table~\ref{table:taustats}. The $\rho$s are re-evaluated to match the scale cuts listed in Table \ref{table:cosmology}. This step is done twice for $\delta\xi_+$ and $\delta\xi_-$. The covariance of the contaminated data vector is estimated by plugging in the $\alpha, \beta, \eta$ samples in the MCMC chain generated in Section~\ref{sec:tau} into Equation~\ref{eq:contaminant} and calculating the covariance between the sample data vectors. Note that the steps to compute the $\alpha$, $\beta$, $\eta$ best fits as detailed in Section \ref{sec:tau} are not repeated to match each survey's scale-cuts. This choice results in a slightly more conservative estimate of the PSF contamination. 

\begin{figure*}
\center
    \includegraphics[width=\textwidth]{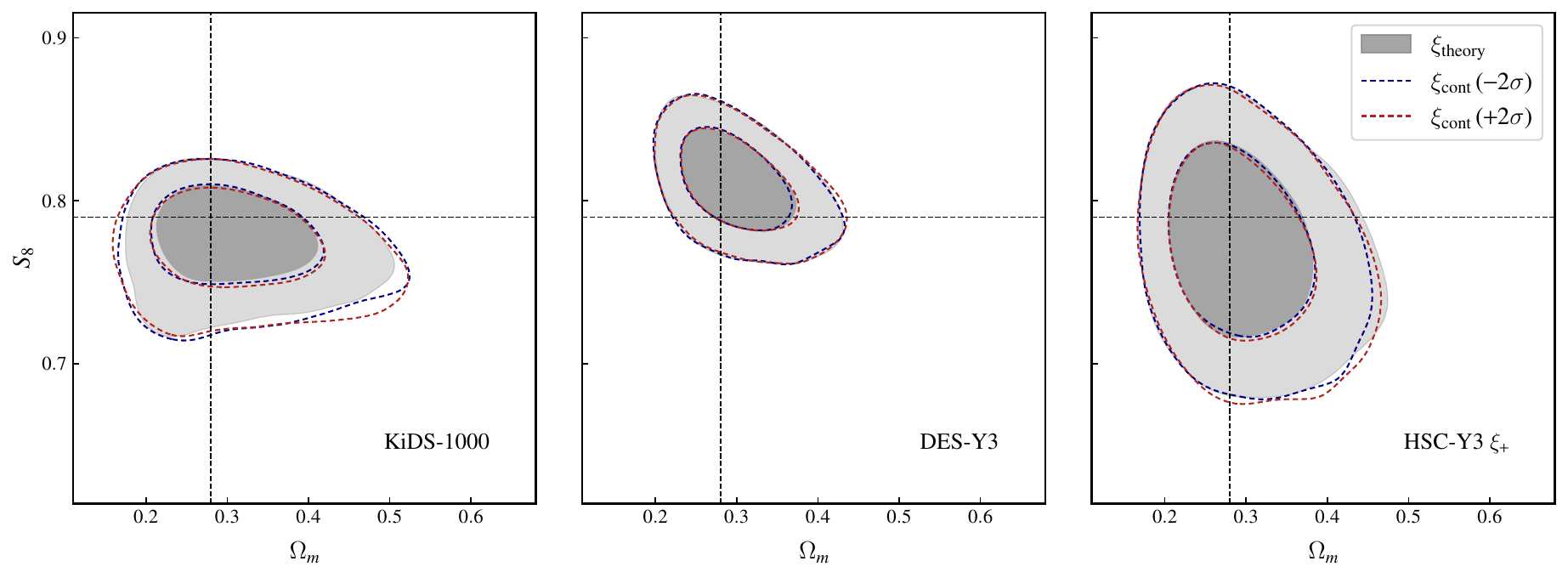}
    \caption{2D posteriors for the $S_8 - \Omega_m$ plane from our baseline simulated cosmic shear data vector (gray) and data vectors constructed with our expected PSF contamination with a $\pm 2\sigma$ uncertainty from Equation \ref{eq:contaminant}. Black horizontal and vertical dashed lines signify the fiducial values for $S_8$ and $\Omega_m$ recorded in Table \ref{table:cosmology}. Inner and outer contours depict a 95\% and 68\% CI, showing that all surveys are very weakly impacted by the PSF contamination model.}
    \label{fig:cosmo}
\end{figure*}

Following \cite{Amon2022}, we check the weak and strong limits of PSF-related systematics by adding a $\pm 2\sigma$ shift, computed from the covariance. 
The results can be found in Figure~\ref{fig:cosmo}. We find minimal changes on the constraints for the $\Omega_{\rm m}$-$S_8$ parameter space for all three surveys, where the exceptionally small shift for DES-Y3 is likely due to both having the smallest $\rho$ amplitudes  and small best-fit $\alpha$ parameters. 

\begin{table*}
\begin{center}
\caption{Priors used in the cosmological inference in Section~\ref{sec:cosmology}. We use a set of unified priors for the cosmological and astrophysical model parameters, and the survey-dependent shear and redshift calibration uncertainty depending on each of the survey's cosmic shear analysis.  For all priors, the $[X_{\rm min}, X_{\rm max}]$ indicates a tophat prior with the lower bound $X_{\rm min}$ and upper bound $X_{\rm max}$; $(\mu, \sigma)$ indicates a Gaussian prior with mean $\mu$ and standard deviation $\sigma$. KiDS-1000 in addition correlates the calibration and redshift uncertainties.  
The scale cuts for each survey are also listed.}
\label{table:cosmology}
\begin{tabular}{lcc|cccc} \toprule
\multicolumn{3}{c|}{\textbf{Unified model parameters}} & \multicolumn{4}{c}{\textbf{Survey-specific model parameters}} \\[0.2cm]
& & Fiducial & & KiDS-1000  & DES-Y3 & HSC-Y3 \\
S$_8$ & [0.5, 0.9]  & 0.790 & $\Delta m_1$ & (−0.009,0.019) & (-0.0063,0.0091) & (0,0.01) \\
$h$ &  [0.55, 0.91] & 0.627 & $\Delta m_2$ &(−0.011,0.020) & (-0.0198,0.0078) &(0,0.01)\\
$\Omega_m$ & [0.1, 0.9] & 0.280 & $\Delta m_3$ &(−0.015,0.017) & (-0.0241,0.0076) & (0,0.01)\\
$\Omega_b$ & [0.03, 0.07] & 0.031& $\Delta m_4$ &(0.002,0.012) & (-0.0369,0.0076) & (0,0.01) \\
$n_s$ & [0.87, 1.07] & 0.894& $\Delta m_5$ & (0.007,0.010) & & \\
$\sum m_{\nu}$ & [0.06, 0.6]eV &  0.077 & $\Delta z_1$ & (0.000,0.011) & (0,0.018) & (0,0.024) \\
& & & $\Delta z_2$& (0.002,0.011) & (0,0.015) & (0,0.022)\\
NLA\_z A$_1$, $\alpha_1$ & [-5, 5] & 0 & $\Delta z_3$ & (0.013,0.012) & (0,0.011) & [-1,1]\\
Nonlinear $P(k)$ & \textsc{HMCode2020} & & $\Delta z_4$ & (0.011,0.009) & (0;0.017) & [-1,1]\\
 & & & $\Delta z_5$ & (−0.006;0.010) && \\
&&& Scale cuts & $ 0.5' < \theta_{\xi+} < 300' $ & $2.475' \leq \theta_{\xi+} <250'$ & $7.1' < \theta_{\xi+} < 56.6'$  \\
&&&&  $ 0.5' < \theta_{\xi-} < 300'$ & $24.75' \leq \theta_{\xi-} <250'$ & $31.2' <\theta_{\xi-} < 248' $  \\
\hline
\end{tabular}
\end{center}
\end{table*}

\section{Discussion}
\label{sec:discussion}

In Section~\ref{sec:results} we performed uniformly a large number of diagnostic tests. We discuss below the overall lessons learned.

\begin{itemize}
\item \textbf{The power of simple tests.} When first given a catalog, examining the most basic quantities may be the most useful exercise to gain an understanding of the catalog rather than complex tests that require further operations on the data. These include, for example: distribution of PSF and galaxy quantities, number counts of stars and galaxies, mean shape for the full sample.
\item \textbf{Documentation.} Our analysis highlights that there are numerous analysis parameters and choices in all of the diagnostic tests that may not be included in the papers (e.g. number of bins, treatment of covariance matrix, tolerance of errors in two-point measurements), yet changing some of them could have an impact on the results. It is therefore critically important to clearly document the exact operation of the tests and make the code publicly available. This is essential for reproducibility and is helpful for comparison with previous work.
\item \textbf{Criteria for pass/fail.} Many of the tests presented as well as in previous papers are somewhat qualitative and do not have well-defined criteria for pass/fail. Even when there is some nominal criteria (such as $\chi^{2}$ or $p$-values), small changes in the analysis choices could affect the outcome, and the failed tests do not always translate directly into biased cosmology. This suggests that instead of treating the results of these diagnostic tests as binary pass/fails, we may take a holistic view and examine the collective suite of tests to determine whether the catalog is ready for science. That is, if many of the tests show signs of unexpected systematics in a certain redshift bin, pausing the analysis and carefully checking the various aspects of the data would be the responsible thing to do. On the other hand, if one of the tests show slightly low $p$-values but all aspects of the data otherwise appear reasonable, one may decide to move on to the next steps.   
\item \textbf{Preservation of metadata.} There are a number of tests that we were not able to reproduce given that the data is not public and otherwise hard to reconstruct. Some examples include location of galaxy on focal plane and colors of PSF stars. If these tests are critical for the science, it would be prudent to preserve the metadata from the upstream processing. Recent analyses from \citet{Zhang23} and \citet{Schutt2025} investigate higher-order moments of the PSF and color-dependence and show they may have significant contributions to the overall modeling error.  Such work is outside the scope of our analysis due to a lack of data.
\item \textbf{Connecting tests to cosmology.} The two-point level tests translate more directly to its impact on cosmological two-point functions. In particular, there is a prescription to test the impact of PSF modeling error on cosmology through the tau statistics. Producing more tests like this would be valuable, particularly for quantities that yield low $p$-values in one-point tests. However, it is also important to remember that there are many assumptions taken in these tests (assuming the form of the PSF model and several analysis choices). That is, \textit{only} passing certain tests (e.g., tau statistics) and failing many others does not constitute a good shear catalog.     
\end{itemize}

With these findings, we formulate the following recommendations for the diagnostic tests for the LSST Y1 shear catalog.
\begin{itemize}
\item With a given catalog, start from the simple tests and once they are understood, move on to more complex ones. We found that the distribution of PSF quantities (Section~\ref{sec:1D_PSF}) to be particularly useful in interpreting the results for the rho statistics (Section~\ref{sec:rowe}). The galaxy property distributions (Section~\ref{sec:overall_sample}) were useful in understanding the quality of the sample. These tests should be performed early on to ensure high quality data. 
\item Some of the metadata needed for the tests are not trivial to compile (e.g. exposure-level quantities for shear measurement on the coadd image), even though they can be in principle reconstructed from the public data. Whenever possible, store these intermediate files and document the process to construct them. 
\item Tests which can be directly connected to cosmology i.e., tau statistics (Section~\ref{sec:tau}), PH statistics (Section~\ref{sec:ph}), $B$-modes  (Section~\ref{sec:bmode}), should be done carefully once the earlier mistakes in the pipeline are caught and fixed. Understand the sensitivity of the results to various analysis choices. 
\item We should continue to develop \textsc{TXPipe} so that we have a uniform framework to perform all of these shear diagnostic tests. Noticeably, the tests in this paper focus on one and two-point level shear tests, and do not include tests associated with e.g. galaxy clustering, chromatic effects, or higher-order statistics. 
\end{itemize}

As Rubin commissioning begins and we learn more about the data we are expecting, there are a number of extensions to this work that will be useful to pursue next: 
\begin{itemize}
\item Check the performance of the commissioning data for the PSF-related tests and understand the basic characteristics. Ensure the metadata related to these products are accessible and easily parsable. 
\item Develop more tests that can probe effects that are not captured with rho and tau statistics. Ensure these tests can straightforwardly propagate to cosmology. In addition, develop tests that may be unique to a specific shear-estimation algorithm. 
\item When checking for $B$-mode signal, one should use a method that is analogous to the estimator for cosmic shear. That is, if the cosmic shear measurement is in harmonic space one may choose to use \textsc{NaMaster} to estimate the $B$-mode, whereas for real-space one may consider estimators such as \textsc{HybridEB} or \textsc{COSEBIs}. The latter two have an additional benefit of being able to cleanly separate $E$ and $B$ modes. Additionally, it is useful to perform a tomographic analysis when looking at $B$-modes as that directly connects to the data vector that is used for cosmology inference.
\end{itemize}

\section{Summary}
\label{sec:conclusion}

In this paper, we set out to uniformly perform a number of diagnostic tests across three Stage-III shear catalogs from KiDS-1000 \citep{Giblin_2021}, DES-Y3 \citep{Gatti_2021} and HSC-Y3 \citep{Li2022}, using \textsc{TXPipe}, a software package developed within the Rubin LSST DESC. The main goals of this exercise is to 1) build a robust pipeline to perform these tests and validate the pipeline by comparing with previous literature; 2) apply the tests uniformly across all 3 datasets and compare their results; 3) derive recommendations for the LSST Y1 cosmic shear analysis regarding what to do for the diagnostic tests.  We have the following findings from the reanalysis
\begin{itemize}
\item We find that when averaged over many exposures, the PSF for all three surveys has prominent patterns in the focal plane that reflect the optics of each of the telescopes. The PSF models used in all three surveys perform well in modeling the pattern. The distributions of the PSF model residuals also show that the models are fairly accurate, with DES-Y3 and KiDS-1000 possessing very slight tendencies to both underestimate the PSF ellipticity and overestimate the PSF size. We also find that there is very little ($\textless$1\%) contamination of non-star objects in the PSF source samples, however this estimate is not valid for objects fainter than \textit{Gaia}'s magnitude limit and another verification method is needed. 
\item The galaxy sample distributions reveal differences in each survey's source selection methods. We find that the selection choices for HSC-Y3 are the most conservative, resulting in a galaxy sample whose shapes can be measured particularly well. One-point tests of the mean galaxy shape as a function of galaxy and PSF properties reveal that DES-Y3 performs the best, with the least amount of scatter and the smallest linear fits for the measured quantities, which may be mostly due to having the largest sample of galaxies. We also find that in some cases, a linear fit does not properly capture the trend in the data.
\item HSC-Y3 generally measures the largest amplitudes  for the $\rho$ statistics, all of which are at least one order of magnitude greater than the weakest $\rho$s as measured by DES-Y3. HSC-Y3 also generally measures the largest signals for the tau statistics, where $\tau_0$ is the most dominant correlation between the galaxy shear and PSF ellipticity. We also find that the fitting parameters for the tau measurements can vary significantly between surveys and between tomographic bins. For PH statistics, all surveys measure a similar amplitude for $\delta \xi^{\rm sys}$, but the signal for KiDS-1000 has greater significance. 
\item For the $B$-mode measurements, we find that while each survey overall measures null \textit{B}-modes  with the pseudo-$C_{\ell}$ estimator, the signal may be significantly different when estimated with the \textsc{HybridEB} method. In the latter estimator, DES-Y3 and HSC-Y3 fail for some of the higher tomographic bins. 
\item The measurements of tangential ellipticity around stars were significantly impacted by the inclusion of scale-cuts for DES-Y3 and KiDS-1000. We also find a non-null signal of unknown origin for KiDS-1000. 
\item The PSF contamination model when propagated to cosmology showed little impact for all surveys.
\end{itemize}

Following our analysis, we discuss overall lessons learned and recommendations for both Rubin commissioning and LSST Y1 diagnostics. 
We find that robust testing is needed as there is a wealth of analysis choices that may lead to the data presenting differently in diagnostic testing. Another challenge is ensuring one is able to estimate the systematic effect's impact on cosmology. Determining whether a catalog is science-ready relies thus on the analysis team to holistically assess the catalog and ensure the analysis choices are well-motivated. In addition, providing metadata and documentation for published results is crucial to ensure the science products are verifiable.

\section*{Acknowledgments}
This paper has undergone internal review in the LSST Dark Energy Science Collaboration. The internal reviewers were Theo Schutt, Claire-Alice H\'ebert and Marika Asgari. 

JJ, YO and CC were supported by DOE grant DESC0021949. JJ was also supported by the National Science Foundation Graduate Research Fellowship under Grant No.  2140001 for much of the duration of this work. JJ also thanks the LSST Discovery Alliance Data Science Fellowship Program, which is funded by LSST Discovery Alliance, NSF Cybertraining Grant No. 1829740, the Brinson Foundation, and the Moore Foundation; her participation in the program has benefited this work. This work has been supported by STFC funding for UK participation in LSST, through grant ST/X001334/1. BG acknowledges support from the UKRI Stephen Hawking Fellowship (grant reference EP/Y017137/1). EMP is grateful to both Harvard University and the Department of Energy for their support of this work, the latter through the Cosmic Frontier grant DE-SC0007881. MA is supported by the UK Science and Technology Facilities Council (STFC) under grant number ST/Y002652/1 and the Royal Society under grant numbers RGSR2222268 and ICAR1231094. TZ is supported by Schmidt Sciences. We thank Angus Wright, Cyrille Doux, Dhayaa Anbajagane, Jack Elvin-Poole, Rachel Mandelbaum, and Xiangchong Li for helpful discussions.

The DESC acknowledges ongoing support from the Institut National de 
Physique Nucl\'eaire et de Physique des Particules in France; the 
Science \& Technology Facilities Council in the United Kingdom; and the
Department of Energy and the LSST Discovery Alliance
in the United States.  DESC uses resources of the IN2P3 
Computing Center (CC-IN2P3--Lyon/Villeurbanne - France) funded by the 
Centre National de la Recherche Scientifique; the National Energy 
Research Scientific Computing Center, a DOE Office of Science User 
Facility supported by the Office of Science of the U.S.\ Department of
Energy under Contract No.\ DE-AC02-05CH11231; STFC DiRAC HPC Facilities, 
funded by UK BEIS National E-infrastructure capital grants; and the UK 
particle physics grid, supported by the GridPP Collaboration.  This 
work was performed in part under DOE Contract DE-AC02-76SF00515. Resources provided by the Research Computing Center at the University of Chicago were also used to support this work. This work made extensive use of tools provided by various open-source packages: healpy, HEALPix, Matplotlib, NumPy, and Astropy \citep{Healpy,Healpix, Matplotlib2007, Numpy2020, astropy:2013,astropy:2018, astropy:2022}. 

JJ performed the main analysis of diagnostic tests, with the exception of B-modes and stellar purity, and the cosmological inference analysis. YO led the B-mode analysis, produced the plots related to B-modes, and provided vital discussion and feedback throughout the analysis. SA led the stellar purity analysis and produced the plots related to stellar purity.
CC developed the premise of the project and provided essential discussion and guiding feedback throughout the analysis. 
JJ and YO contributed work towards developing the diagnostic tests in TXPipe. JZ developed and managed the TXPipe pipeline software and provided valuable discussion and feedback which resulted in improved developments. MJ developed TreeCorr which was used to produce many of the diagnostic tests. EMP and JP assisted with debugging code and resolving issues within TXPipe. BG, MG, and TZ assisted with collecting and interpreting KiDS, DES, and HSC data. 
JJ led the paper writing and YO, CC, and SA contributed writing and polishing of the text. CAH, TS, and MA first reviewed the paper and provided extensive feedback resulting in vital improvements to the paper, especially regarding interpreting PSF test results and analyzing KiDS data. MJ and EMP reviewed the paper and provided final feedback resulting in improvements to the paper.

\bibliographystyle{mnras}

\bibliography{referencelist}

\begin{thebibliography}{}
\makeatletter
\relax
\def\mn@urlcharsother{\let\do\@makeother \do\$\do\&\do\#\do\^\do\_\do\%\do\~}
\def\mn@doi{\begingroup\mn@urlcharsother \@ifnextchar [ {\mn@doi@} {\mn@doi@[]}}
\def\mn@doi@[#1]#2{\def\@tempa{#1}\ifx\@tempa\@empty \href {http://dx.doi.org/#2} {doi:#2}\else \href {http://dx.doi.org/#2} {#1}\fi \endgroup}
\def\mn@eprint#1#2{\mn@eprint@#1:#2::\@nil}
\def\mn@eprint@arXiv#1{\href {http://arxiv.org/abs/#1} {{\tt arXiv:#1}}}
\def\mn@eprint@dblp#1{\href {http://dblp.uni-trier.de/rec/bibtex/#1.xml} {dblp:#1}}
\def\mn@eprint@#1:#2:#3:#4\@nil{\def\@tempa {#1}\def\@tempb {#2}\def\@tempc {#3}\ifx \@tempc \@empty \let \@tempc \@tempb \let \@tempb \@tempa \fi \ifx \@tempb \@empty \def\@tempb {arXiv}\fi \@ifundefined {mn@eprint@\@tempb}{\@tempb:\@tempc}{\expandafter \expandafter \csname mn@eprint@\@tempb\endcsname \expandafter{\@tempc}}}

\bibitem[\protect\citeauthoryear{{Abbott} et~al.,}{{Abbott} et~al.}{2018}]{DES2018}
{Abbott} T.~M.~C.,  et~al., 2018, \mn@doi [\prd] {10.1103/PhysRevD.98.043526}, \href {https://ui.adsabs.harvard.edu/abs/2018PhRvD..98d3526A} {98, 043526}

\bibitem[\protect\citeauthoryear{{Aihara} et~al.,}{{Aihara} et~al.}{2018}]{Aihara2018b}
{Aihara} H.,  et~al., 2018, \mn@doi [\pasj] {10.1093/pasj/psx066}, \href {https://ui.adsabs.harvard.edu/abs/2018PASJ...70S...4A} {70, S4}

\bibitem[\protect\citeauthoryear{{Aihara} et~al.,}{{Aihara} et~al.}{2022}]{Aihara2022}
{Aihara} H.,  et~al., 2022, \mn@doi [\pasj] {10.1093/pasj/psab122}, \href {https://ui.adsabs.harvard.edu/abs/2022PASJ...74..247A} {74, 247}

\bibitem[\protect\citeauthoryear{{Albrecht} et~al.,}{{Albrecht} et~al.}{2006}]{Albrecht2006}
{Albrecht} A.,  et~al., 2006, \mn@doi [arXiv e-prints] {10.48550/arXiv.astro-ph/0609591}, \href {https://ui.adsabs.harvard.edu/abs/2006astro.ph..9591A} {pp astro--ph/0609591}

\bibitem[\protect\citeauthoryear{{Alonso}, {Sanchez}, {Slosar}  \& {LSST Dark Energy Science Collaboration}}{{Alonso} et~al.}{2019}]{Alonso2019}
{Alonso} D.,  {Sanchez} J.,  {Slosar} A.,   {LSST Dark Energy Science Collaboration} 2019, \mn@doi [\mnras] {10.1093/mnras/stz093}, \href {https://ui.adsabs.harvard.edu/abs/2019MNRAS.484.4127A} {484, 4127}

\bibitem[\protect\citeauthoryear{Amon et~al.,}{Amon et~al.}{2018}]{amon2018}
Amon A.,  et~al., 2018, \mn@doi [Monthly Notices of the Royal Astronomical Society] {10.1093/mnras/sty859}, 477, 4285

\bibitem[\protect\citeauthoryear{{Amon} et~al.,}{{Amon} et~al.}{2022}]{Amon2022}
{Amon} A.,  et~al., 2022, \mn@doi [\prd] {10.1103/PhysRevD.105.023514}, \href {https://ui.adsabs.harvard.edu/abs/2022PhRvD.105b3514A} {105, 023514}

\bibitem[\protect\citeauthoryear{{Asgari} \& {Heymans}}{{Asgari} \& {Heymans}}{2019}]{Asgari2019}
{Asgari} M.,  {Heymans} C.,  2019, \mn@doi [Monthly Notices of the Royal Astronomical Society: Letters] {10.1093/mnrasl/slz006}, 484, L59

\bibitem[\protect\citeauthoryear{{Asgari} et~al.,}{{Asgari} et~al.}{2021}]{Asgari2021}
{Asgari} M.,  et~al., 2021, \mn@doi [\aap] {10.1051/0004-6361/202039070}, \href {https://ui.adsabs.harvard.edu/abs/2021A&A...645A.104A} {645, A104}

\bibitem[\protect\citeauthoryear{{Astropy Collaboration} et~al.,}{{Astropy Collaboration} et~al.}{2013}]{astropy:2013}
{Astropy Collaboration} et~al., 2013, \mn@doi [\aap] {10.1051/0004-6361/201322068}, \href {https://ui.adsabs.harvard.edu/abs/2013A&A...558A..33A} {558, A33}

\bibitem[\protect\citeauthoryear{{Astropy Collaboration} et~al.,}{{Astropy Collaboration} et~al.}{2018}]{astropy:2018}
{Astropy Collaboration} et~al., 2018, \mn@doi [\aj] {10.3847/1538-3881/aabc4f}, \href {https://ui.adsabs.harvard.edu/abs/2018AJ....156..123A} {156, 123}

\bibitem[\protect\citeauthoryear{{Astropy Collaboration} et~al.,}{{Astropy Collaboration} et~al.}{2022}]{astropy:2022}
{Astropy Collaboration} et~al., 2022, \mn@doi [\apj] {10.3847/1538-4357/ac7c74}, \href {https://ui.adsabs.harvard.edu/abs/2022ApJ...935..167A} {935, 167}

\bibitem[\protect\citeauthoryear{{Bacon}, {Refregier}  \& {Ellis}}{{Bacon} et~al.}{2000}]{Bacon2000}
{Bacon} D.~J.,  {Refregier} A.~R.,   {Ellis} R.~S.,  2000, \mn@doi [\mnras] {10.1046/j.1365-8711.2000.03851.x}, \href {http://adsabs.harvard.edu/abs/2000MNRAS.318..625B} {318, 625}

\bibitem[\protect\citeauthoryear{{Bartelmann} \& {Schneider}}{{Bartelmann} \& {Schneider}}{2001}]{Bartelmann2001}
{Bartelmann} M.,  {Schneider} P.,  2001, \mn@doi [\physrep] {10.1016/S0370-1573(00)00082-X}, \href {https://ui.adsabs.harvard.edu/abs/2001PhR...340..291B} {340, 291}

\bibitem[\protect\citeauthoryear{{Becker} \& {Rozo}}{{Becker} \& {Rozo}}{2016}]{Becker2016}
{Becker} M.~R.,  {Rozo} E.,  2016, \mn@doi [\mnras] {10.1093/mnras/stv3018}, \href {https://ui.adsabs.harvard.edu/abs/2016MNRAS.457..304B} {457, 304}

\bibitem[\protect\citeauthoryear{{Bertin}}{{Bertin}}{2011}]{Bertin2011}
{Bertin} E.,  2011, in {Evans} I.~N.,  {Accomazzi} A.,  {Mink} D.~J.,   {Rots} A.~H.,  eds,  Astronomical Society of the Pacific Conference Series Vol. 442, Astronomical Data Analysis Software and Systems XX. p.~435

\bibitem[\protect\citeauthoryear{Bosch et~al.,}{Bosch et~al.}{2017}]{Bosch2018}
Bosch J.,  et~al., 2017, \mn@doi [Publications of the Astronomical Society of Japan] {10.1093/pasj/psx080}, 70, S5

\bibitem[\protect\citeauthoryear{{Bridle} et~al.,}{{Bridle} et~al.}{2010}]{Bridle2010}
{Bridle} S.,  et~al., 2010, \mn@doi [\mnras] {10.1111/j.1365-2966.2010.16598.x}, \href {https://ui.adsabs.harvard.edu/abs/2010MNRAS.405.2044B} {405, 2044}

\bibitem[\protect\citeauthoryear{{Chang} et~al.,}{{Chang} et~al.}{2019}]{Chang2019}
{Chang} C.,  et~al., 2019, \mn@doi [\mnras] {10.1093/mnras/sty2902}, \href {https://ui.adsabs.harvard.edu/abs/2019MNRAS.482.3696C} {482, 3696}

\bibitem[\protect\citeauthoryear{{Chisari} et~al.,}{{Chisari} et~al.}{2019}]{Chisari2019}
{Chisari} N.~E.,  et~al., 2019, \mn@doi [\apjs] {10.3847/1538-4365/ab1658}, \href {https://ui.adsabs.harvard.edu/abs/2019ApJS..242....2C} {242, 2}

\bibitem[\protect\citeauthoryear{{Chon}, {Challinor}, {Prunet}, {Hivon}  \& {Szapudi}}{{Chon} et~al.}{2004}]{Chon2004}
{Chon} G.,  {Challinor} A.,  {Prunet} S.,  {Hivon} E.,   {Szapudi} I.,  2004, \mn@doi [\mnras] {10.1111/j.1365-2966.2004.07737.x}, \href {https://ui.adsabs.harvard.edu/abs/2004MNRAS.350..914C} {350, 914}

\bibitem[\protect\citeauthoryear{{DES Collaboration}}{{DES Collaboration}}{2005}]{DES:2005}
{DES Collaboration} 2005, arXiv e-prints, \href {https://ui.adsabs.harvard.edu/abs/2005astro.ph.10346T} {pp astro--ph/0510346}

\bibitem[\protect\citeauthoryear{{Dalal} et~al.,}{{Dalal} et~al.}{2023}]{Dalal2023}
{Dalal} R.,  et~al., 2023, \mn@doi [\prd] {10.1103/PhysRevD.108.123519}, \href {https://ui.adsabs.harvard.edu/abs/2023PhRvD.108l3519D} {108, 123519}

\bibitem[\protect\citeauthoryear{Dodelson \& Schneider}{Dodelson \& Schneider}{2013}]{Dodelson2013}
Dodelson S.,  Schneider M.~D.,  2013, \mn@doi [Phys. Rev. D] {10.1103/PhysRevD.88.063537}, 88, 063537

\bibitem[\protect\citeauthoryear{{Doux} et~al.,}{{Doux} et~al.}{2022}]{Doux2022}
{Doux} C.,  et~al., 2022, \mn@doi [\mnras] {10.1093/mnras/stac1826}, \href {https://ui.adsabs.harvard.edu/abs/2022MNRAS.515.1942D} {515, 1942}

\bibitem[\protect\citeauthoryear{{Fenech Conti}, {Herbonnet}, {Hoekstra}, {Merten}, {Miller}  \& {Viola}}{{Fenech Conti} et~al.}{2017}]{FenechConti2017}
{Fenech Conti} I.,  {Herbonnet} R.,  {Hoekstra} H.,  {Merten} J.,  {Miller} L.,   {Viola} M.,  2017, \mn@doi [\mnras] {10.1093/mnras/stx200}, \href {https://ui.adsabs.harvard.edu/abs/2017MNRAS.467.1627F} {467, 1627}

\bibitem[\protect\citeauthoryear{Flaugher et~al.,}{Flaugher et~al.}{2015}]{Flaugher2015}
Flaugher B.,  et~al., 2015, \mn@doi [The Astronomical Journal] {10.1088/0004-6256/150/5/150}, 150, 150

\bibitem[\protect\citeauthoryear{{Foreman-Mackey} et~al.,}{{Foreman-Mackey} et~al.}{2013}]{Foreman-Mackey2013}
{Foreman-Mackey} D.,  et~al., 2013, {emcee: The MCMC Hammer}, Astrophysics Source Code Library, record ascl:1303.002

\bibitem[\protect\citeauthoryear{{Gatti} et~al.,}{{Gatti} et~al.}{2021}]{Gatti_2021}
{Gatti} M.,  et~al., 2021, \mn@doi [Monthly Notices of the Royal Astronomical Society] {10.1093/mnras/stab918}, 504, 4312

\bibitem[\protect\citeauthoryear{{Giblin} et~al.,}{{Giblin} et~al.}{2021}]{Giblin_2021}
{Giblin} B.,  et~al., 2021, \mn@doi [Astronomy \& Astrophysics] {10.1051/0004-6361/202038850}, 645, A105

\bibitem[\protect\citeauthoryear{{G{\'o}rski}, {Hivon}, {Banday}, {Wandelt}, {Hansen}, {Reinecke}  \& {Bartelmann}}{{G{\'o}rski} et~al.}{2005}]{Healpix}
{G{\'o}rski} K.~M.,  {Hivon} E.,  {Banday} A.~J.,  {Wandelt} B.~D.,  {Hansen} F.~K.,  {Reinecke} M.,   {Bartelmann} M.,  2005, \mn@doi [\apj] {10.1086/427976}, \href {http://adsabs.harvard.edu/abs/2005ApJ...622..759G} {622, 759}

\bibitem[\protect\citeauthoryear{Harris et~al.,}{Harris et~al.}{2020}]{Numpy2020}
Harris C.~R.,  et~al., 2020, \mn@doi [Nature] {10.1038/s41586-020-2649-2}, 585, 357

\bibitem[\protect\citeauthoryear{Hartlap, Simon  \& Schneider}{Hartlap et~al.}{2006}]{Hartlap2006}
Hartlap J.,  Simon P.,   Schneider P.,  2006, \mn@doi [Astronomy and Astrophysics] {10.1051/0004-6361:20066170}, 464

\bibitem[\protect\citeauthoryear{{Heymans} et~al.,}{{Heymans} et~al.}{2006}]{Heymans2006}
{Heymans} C.,  et~al., 2006, \mn@doi [\mnras] {10.1111/j.1365-2966.2006.10198.x}, \href {https://ui.adsabs.harvard.edu/abs/2006MNRAS.368.1323H} {368, 1323}

\bibitem[\protect\citeauthoryear{{Heymans} et~al.,}{{Heymans} et~al.}{2012}]{Heymans2012}
{Heymans} C.,  et~al., 2012, \mn@doi [\mnras] {10.1111/j.1365-2966.2012.21952.x}, \href {https://ui.adsabs.harvard.edu/abs/2012MNRAS.427..146H} {427, 146}

\bibitem[\protect\citeauthoryear{Heymans et~al.,}{Heymans et~al.}{2021}]{Heymans2020}
Heymans C.,  et~al., 2021, \mn@doi [A&A] {10.1051/0004-6361/202039063}, 646, A140

\bibitem[\protect\citeauthoryear{{Hildebrandt} et~al.,}{{Hildebrandt} et~al.}{2017}]{Hildebrandt2017}
{Hildebrandt} H.,  et~al., 2017, \mn@doi [\mnras] {10.1093/mnras/stw2805}, \href {https://ui.adsabs.harvard.edu/abs/2017MNRAS.465.1454H} {465, 1454}

\bibitem[\protect\citeauthoryear{{Hirata} \& {Seljak}}{{Hirata} \& {Seljak}}{2003}]{Hirata2003}
{Hirata} C.,  {Seljak} U.,  2003, \mn@doi [\mnras] {10.1046/j.1365-8711.2003.06683.x}, \href {https://ui.adsabs.harvard.edu/abs/2003MNRAS.343..459H} {343, 459}

\bibitem[\protect\citeauthoryear{Huff \& Mandelbaum}{Huff \& Mandelbaum}{2017}]{HuffMandelbaum2017}
Huff E.,  Mandelbaum R.,  2017, Metacalibration: Direct Self-Calibration of Biases in Shear Measurement (\mn@eprint {arXiv} {1702.02600}), \url {https://arxiv.org/abs/1702.02600}

\bibitem[\protect\citeauthoryear{Hunter}{Hunter}{2007}]{Matplotlib2007}
Hunter J.~D.,  2007, \mn@doi [Computing in Science \& Engineering] {10.1109/MCSE.2007.55}, 9, 90

\bibitem[\protect\citeauthoryear{{Jarvis}}{{Jarvis}}{2015}]{Jarvis2015}
{Jarvis} M.,  2015, {TreeCorr: Two-point correlation functions}, Astrophysics Source Code Library, record ascl:1508.007

\bibitem[\protect\citeauthoryear{Jarvis, Bernstein  \& Jain}{Jarvis et~al.}{2004}]{Jarvis2004}
Jarvis M.,  Bernstein G.,   Jain B.,  2004, \mn@doi [Monthly Notices of the Royal Astronomical Society] {10.1111/j.1365-2966.2004.07926.x}, 352, 338

\bibitem[\protect\citeauthoryear{{Jarvis} et~al.,}{{Jarvis} et~al.}{2016}]{Jarvis2016}
{Jarvis} M.,  et~al., 2016, \mn@doi [\mnras] {10.1093/mnras/stw990}, \href {https://ui.adsabs.harvard.edu/abs/2016MNRAS.460.2245J} {460, 2245}

\bibitem[\protect\citeauthoryear{{Jarvis} et~al.,}{{Jarvis} et~al.}{2021}]{JarvisPSFDES2021}
{Jarvis} M.,  et~al., 2021, \mn@doi [\mnras] {10.1093/mnras/staa3679}, \href {https://ui.adsabs.harvard.edu/abs/2021MNRAS.501.1282J} {501, 1282}

\bibitem[\protect\citeauthoryear{{Kaiser}}{{Kaiser}}{1992}]{Kaiser1992}
{Kaiser} N.,  1992, \mn@doi [\apj] {10.1086/171151}, \href {https://ui.adsabs.harvard.edu/abs/1992ApJ...388..272K} {388, 272}

\bibitem[\protect\citeauthoryear{{Kaiser}, {Wilson}  \& {Luppino}}{{Kaiser} et~al.}{2000}]{Kaiser2000}
{Kaiser} N.,  {Wilson} G.,   {Luppino} G.~A.,  2000, ArXiv Astrophysics e-prints, \href {http://adsabs.harvard.edu/abs/2000astro.ph..3338K} {}

\bibitem[\protect\citeauthoryear{{Kannawadi} et~al.,}{{Kannawadi} et~al.}{2019}]{Kannawadi2019}
{Kannawadi} A.,  et~al., 2019, \mn@doi [\aap] {10.1051/0004-6361/201834819}, \href {https://ui.adsabs.harvard.edu/abs/2019A&A...624A..92K} {624, A92}

\bibitem[\protect\citeauthoryear{{Kilbinger}}{{Kilbinger}}{2015}]{kilbinger2015}
{Kilbinger} M.,  2015, \mn@doi [Reports on Progress in Physics] {10.1088/0034-4885/78/8/086901}, \href {https://ui.adsabs.harvard.edu/abs/2015RPPh...78h6901K} {78, 086901}

\bibitem[\protect\citeauthoryear{{Kitching} et~al.,}{{Kitching} et~al.}{2012}]{Kitching2012}
{Kitching} T.~D.,  et~al., 2012, \mn@doi [\mnras] {10.1111/j.1365-2966.2012.21095.x}, \href {https://ui.adsabs.harvard.edu/abs/2012MNRAS.423.3163K} {423, 3163}

\bibitem[\protect\citeauthoryear{{Kitching} et~al.,}{{Kitching} et~al.}{2013}]{Kitching2013}
{Kitching} T.~D.,  et~al., 2013, \mn@doi [\apjs] {10.1088/0067-0049/205/2/12}, \href {https://ui.adsabs.harvard.edu/abs/2013ApJS..205...12K} {205, 12}

\bibitem[\protect\citeauthoryear{Krause et~al.,}{Krause et~al.}{2021}]{Krause2021}
Krause E.,  et~al., 2021, Dark Energy Survey Year 3 Results: Multi-Probe Modeling Strategy and Validation (\mn@eprint {arXiv} {2105.13548}), \url {https://arxiv.org/abs/2105.13548}

\bibitem[\protect\citeauthoryear{Kuijken et~al.,}{Kuijken et~al.}{2015}]{Kuijken2015}
Kuijken K.,  et~al., 2015, Monthly Notices of the Royal Astronomical Society, 454, 3500

\bibitem[\protect\citeauthoryear{{Kuijken} et~al.,}{{Kuijken} et~al.}{2019}]{Kuijken2019}
{Kuijken} K.,  et~al., 2019, \mn@doi [\aap] {10.1051/0004-6361/201834918}, \href {https://ui.adsabs.harvard.edu/abs/2019A\%26A...625A...2K} {625, A2}

\bibitem[\protect\citeauthoryear{{Li} et~al.,}{{Li} et~al.}{2022}]{Li2022}
{Li} X.,  et~al., 2022, \mn@doi [\pasj] {10.1093/pasj/psac006}, \href {https://ui.adsabs.harvard.edu/abs/2022PASJ...74..421L} {74, 421}

\bibitem[\protect\citeauthoryear{Li et~al.,}{Li et~al.}{2023a}]{Li2pt_2023}
Li X.,  et~al., 2023a, \mn@doi [Phys. Rev. D] {10.1103/PhysRevD.108.123518}, 108, 123518

\bibitem[\protect\citeauthoryear{{Li} et~al.,}{{Li} et~al.}{2023b}]{Li2023}
{Li} S.-S.,  et~al., 2023b, \mn@doi [\aap] {10.1051/0004-6361/202245210}, \href {https://ui.adsabs.harvard.edu/abs/2023A&A...670A.100L} {670, A100}

\bibitem[\protect\citeauthoryear{{Limber}}{{Limber}}{1953}]{Limber1953}
{Limber} D.~N.,  1953, \mn@doi [\apj] {10.1086/145672}, \href {http://adsabs.harvard.edu/abs/1953ApJ...117..134L} {117, 134}

\bibitem[\protect\citeauthoryear{LoVerde \& Afshordi}{LoVerde \& Afshordi}{2008}]{Loeverde2008}
LoVerde M.,  Afshordi N.,  2008, \mn@doi [Phys. Rev. D] {10.1103/PhysRevD.78.123506}, 78, 123506

\bibitem[\protect\citeauthoryear{{Longley} et~al.,}{{Longley} et~al.}{2023}]{Longley2023}
{Longley} E.~P.,  et~al., 2023, \mn@doi [\mnras] {10.1093/mnras/stad246}, \href {https://ui.adsabs.harvard.edu/abs/2023MNRAS.520.5016L} {520, 5016}

\bibitem[\protect\citeauthoryear{{MacCrann} et~al.,}{{MacCrann} et~al.}{2022}]{MacCrann2022}
{MacCrann} N.,  et~al., 2022, \mn@doi [\mnras] {10.1093/mnras/stab2870}, \href {https://ui.adsabs.harvard.edu/abs/2022MNRAS.509.3371M} {509, 3371}

\bibitem[\protect\citeauthoryear{{Mandelbaum} et~al.,}{{Mandelbaum} et~al.}{2015}]{Mandelbaum2015}
{Mandelbaum} R.,  et~al., 2015, \mn@doi [\mnras] {10.1093/mnras/stv781}, \href {https://ui.adsabs.harvard.edu/abs/2015MNRAS.450.2963M} {450, 2963}

\bibitem[\protect\citeauthoryear{{Mandelbaum} et~al.,}{{Mandelbaum} et~al.}{2018a}]{Mandelbaum2018b}
{Mandelbaum} R.,  et~al., 2018a, \mn@doi [\pasj] {10.1093/pasj/psx130}, \href {https://ui.adsabs.harvard.edu/abs/2018PASJ...70S..25M} {70, S25}

\bibitem[\protect\citeauthoryear{{Mandelbaum} et~al.,}{{Mandelbaum} et~al.}{2018b}]{Mandelbaum2018a}
{Mandelbaum} R.,  et~al., 2018b, \mn@doi [\mnras] {10.1093/mnras/sty2420}, \href {https://ui.adsabs.harvard.edu/abs/2018MNRAS.481.3170M} {481, 3170}

\bibitem[\protect\citeauthoryear{{Massey} et~al.,}{{Massey} et~al.}{2007}]{Massey2007}
{Massey} R.,  et~al., 2007, \mn@doi [\mnras] {10.1111/j.1365-2966.2006.11315.x}, \href {https://ui.adsabs.harvard.edu/abs/2007MNRAS.376...13M} {376, 13}

\bibitem[\protect\citeauthoryear{{McMahon}, {Banerji}, {Gonzalez}, {Koposov}, {Bejar}, {Lodieu}, {Rebolo}  \& {VHS Collaboration}}{{McMahon} et~al.}{2013}]{McMahon2013}
{McMahon} R.~G.,  {Banerji} M.,  {Gonzalez} E.,  {Koposov} S.~E.,  {Bejar} V.~J.,  {Lodieu} N.,  {Rebolo} R.,   {VHS Collaboration} 2013, The Messenger, \href {https://ui.adsabs.harvard.edu/abs/2013Msngr.154...35M} {154, 35}

\bibitem[\protect\citeauthoryear{Miller, Kitching, Heymans, Heavens  \& Van~Waerbeke}{Miller et~al.}{2007}]{Miller_2007}
Miller L.,  Kitching T.~D.,  Heymans C.,  Heavens A.~F.,   Van~Waerbeke L.,  2007, \mn@doi [Monthly Notices of the Royal Astronomical Society] {10.1111/j.1365-2966.2007.12363.x}, 382, 315–324

\bibitem[\protect\citeauthoryear{More et~al.,}{More et~al.}{2023}]{More2023}
More S.,  et~al., 2023, \mn@doi [Phys. Rev. D] {10.1103/PhysRevD.108.123520}, 108, 123520

\bibitem[\protect\citeauthoryear{Nelder \& Mead}{Nelder \& Mead}{1965}]{NelderMead}
Nelder J.~A.,  Mead R.,  1965, \mn@doi [The Computer Journal] {10.1093/comjnl/7.4.308}, 7, 308

\bibitem[\protect\citeauthoryear{{Paulin-Henriksson}, {Amara}, {Voigt}, {Refregier}  \& {Bridle}}{{Paulin-Henriksson} et~al.}{2008}]{PaulinHenriksson2008}
{Paulin-Henriksson} S.,  {Amara} A.,  {Voigt} L.,  {Refregier} A.,   {Bridle} S.~L.,  2008, \mn@doi [\aap] {10.1051/0004-6361:20079150}, \href {https://ui.adsabs.harvard.edu/abs/2008A&A...484...67P} {484, 67}

\bibitem[\protect\citeauthoryear{Prat et~al.,}{Prat et~al.}{2022}]{Prat2022}
Prat J.,  et~al., 2022, \mn@doi [Phys. Rev. D] {10.1103/PhysRevD.105.083528}, 105, 083528

\bibitem[\protect\citeauthoryear{{Prat} et~al.,}{{Prat} et~al.}{2023}]{Prat2023}
{Prat} J.,  et~al., 2023, \mn@doi [The Open Journal of Astrophysics] {10.21105/astro.2212.09345}, \href {https://ui.adsabs.harvard.edu/abs/2023OJAp....6E..13P} {6, 13}

\bibitem[\protect\citeauthoryear{{Rowe}}{{Rowe}}{2010}]{rowe2010}
{Rowe} B.,  2010, \mn@doi [\mnras] {10.1111/j.1365-2966.2010.16277.x}, \href {https://ui.adsabs.harvard.edu/abs/2010MNRAS.404..350R} {404, 350}

\bibitem[\protect\citeauthoryear{{Rowe} et~al.,}{{Rowe} et~al.}{2015}]{galsim}
{Rowe} B.~T.~P.,  et~al., 2015, \mn@doi [Astronomy and Computing] {10.1016/j.ascom.2015.02.002}, \href {https://ui.adsabs.harvard.edu/abs/2015A&C....10..121R} {10, 121}

\bibitem[\protect\citeauthoryear{{Schneider} \& {Seitz}}{{Schneider} \& {Seitz}}{1995}]{Seitz1991}
{Schneider} P.,  {Seitz} C.,  1995, \mn@doi [\aap] {10.48550/arXiv.astro-ph/9407032}, \href {https://ui.adsabs.harvard.edu/abs/1995A&A...294..411S} {294, 411}

\bibitem[\protect\citeauthoryear{{Schneider}, {Eifler}  \& {Krause}}{{Schneider} et~al.}{2010}]{Schneider2010}
{Schneider} P.,  {Eifler} T.,   {Krause} E.,  2010, \mn@doi [\aap] {10.1051/0004-6361/201014235}, \href {https://ui.adsabs.harvard.edu/abs/2010A&A...520A.116S} {520, A116}

\bibitem[\protect\citeauthoryear{Schutt et~al.,}{Schutt et~al.}{2025}]{Schutt2025}
Schutt T.,  et~al., 2025, \mn@doi [The Open Journal of Astrophysics] {https://doi.org/10.33232/001c.132299}, 8

\bibitem[\protect\citeauthoryear{{Secco} et~al.,}{{Secco} et~al.}{2022}]{Secco2022}
{Secco} L.~F.,  et~al., 2022, \mn@doi [\prd] {10.1103/PhysRevD.105.023515}, \href {https://ui.adsabs.harvard.edu/abs/2022PhRvD.105b3515S} {105, 023515}

\bibitem[\protect\citeauthoryear{{Sheldon}}{{Sheldon}}{2014}]{Sheldon:2014}
{Sheldon} E.~S.,  2014, \mn@doi [\mnras] {10.1093/mnrasl/slu104}, \href {https://ui.adsabs.harvard.edu/abs/2014MNRAS.444L..25S} {444, L25}

\bibitem[\protect\citeauthoryear{{Sheldon} \& {Huff}}{{Sheldon} \& {Huff}}{2017}]{Sheldon2017}
{Sheldon} E.~S.,  {Huff} E.~M.,  2017, \mn@doi [\apj] {10.3847/1538-4357/aa704b}, \href {https://ui.adsabs.harvard.edu/abs/2017ApJ...841...24S} {841, 24}

\bibitem[\protect\citeauthoryear{{Szapudi}, {Prunet}, {Pogosyan}, {Szalay}  \& {Bond}}{{Szapudi} et~al.}{2001}]{Szapudi2001}
{Szapudi} I.,  {Prunet} S.,  {Pogosyan} D.,  {Szalay} A.~S.,   {Bond} J.~R.,  2001, \mn@doi [\apjl] {10.1086/319105}, \href {https://ui.adsabs.harvard.edu/abs/2001ApJ...548L.115S} {548, L115}

\bibitem[\protect\citeauthoryear{Virtanen et~al.,}{Virtanen et~al.}{2020}]{SciPy2020}
Virtanen P.,  et~al., 2020, \mn@doi [Nature Methods] {10.1038/s41592-019-0686-2}, \href {https://rdcu.be/b08Wh} {17, 261}

\bibitem[\protect\citeauthoryear{{Wittman}, {Tyson}, {Kirkman}, {Dell'Antonio}  \& {Bernstein}}{{Wittman} et~al.}{2000}]{Wittman2000}
{Wittman} D.~M.,  {Tyson} J.~A.,  {Kirkman} D.,  {Dell'Antonio} I.,   {Bernstein} G.,  2000, \mn@doi [\nat] {10.1038/35012001}, \href {http://adsabs.harvard.edu/abs/2000Natur.405..143W} {405, 143}

\bibitem[\protect\citeauthoryear{Wright et~al.,}{Wright et~al.}{2010}]{Wright2010}
Wright E.~L.,  et~al., 2010, \mn@doi [The Astronomical Journal] {10.1088/0004-6256/140/6/1868}, 140, 1868

\bibitem[\protect\citeauthoryear{Zhang et~al.,}{Zhang et~al.}{2023}]{Zhang23}
Zhang T.,  et~al., 2023, \mn@doi [Monthly Notices of the Royal Astronomical Society] {10.1093/mnras/stad1801}, 525

\bibitem[\protect\citeauthoryear{Zonca, Singer, Lenz, Reinecke, Rosset, Hivon  \& Gorski}{Zonca et~al.}{2019}]{Healpy}
Zonca A.,  Singer L.,  Lenz D.,  Reinecke M.,  Rosset C.,  Hivon E.,   Gorski K.,  2019, \mn@doi [Journal of Open Source Software] {10.21105/joss.01298}, 4, 1298

\bibitem[\protect\citeauthoryear{{Zuntz}, {Kacprzak}, {Voigt}, {Hirsch}, {Rowe}  \& {Bridle}}{{Zuntz} et~al.}{2013}]{Zuntz2013}
{Zuntz} J.,  {Kacprzak} T.,  {Voigt} L.,  {Hirsch} M.,  {Rowe} B.,   {Bridle} S.,  2013, \mn@doi [\mnras] {10.1093/mnras/stt1125}, \href {https://ui.adsabs.harvard.edu/abs/2013MNRAS.434.1604Z} {434, 1604}

\bibitem[\protect\citeauthoryear{Zuntz et~al.,}{Zuntz et~al.}{2015}]{Cosmosis}
Zuntz J.,  et~al., 2015, \mn@doi [Astronomy and Computing] {https://doi.org/10.1016/j.ascom.2015.05.005}, 12, 45

\bibitem[\protect\citeauthoryear{Zuntz et~al.,}{Zuntz et~al.}{2018}]{Zuntz_2018}
Zuntz J.,  et~al., 2018, \mn@doi [Monthly Notices of the Royal Astronomical Society] {10.1093/mnras/sty2219}, 481, 1149

\bibitem[\protect\citeauthoryear{{de Jong} et~al.,}{{de Jong} et~al.}{2015}]{deJong2015}
{de Jong} J.~T.~A.,  et~al., 2015, \mn@doi [\aap] {10.1051/0004-6361/201526601}, \href {http://adsabs.harvard.edu/abs/2015A%26A...582A..62D} {582, A62}

\makeatother
\end{thebibliography}

\begin{appendix}

\section{Complete list of tests}
\label{sec:all_tests}
In this appendix we summarize the list of all tests that have ever been done in the three main papers as a reference. \\
\begin{itemize}
\item KiDS-1000:
\begin{itemize}
\item PSF ellipticity, $\sigma$, residual on focal plane
\item PH statistics
\item $\rho_1$ 
\item PSF ellipticity vs r-band mag, CCD chip ID
\item star-galaxy cross correlation
\item tangential ellipticity around BOSS galaxies
\item COSEBIs B-modes
\end{itemize}

\item DES-Y3:
\begin{itemize}
\item mean shape vs. PSF ellipticity/size, galaxy size/signal-to-noise ratio, CCD position, survey properties
\item PSF ellipticity/size vs. focal plane location, star magnitude/color
\item $\rho$ statistics and $\tau$ statistics
\item b-modes  (COSEBIs and NaMaster)
\item tangential ellipticity around stars, random points, field/chip centers
\end{itemize}

\item HSC-Y3:
\begin{itemize}
\item additive, multiplicative bias and $\sigma$ vs SNR, resolution (SNR,$R_2$) plane 
\item additive, multiplicative \textit{weight} bias and $\sigma$ vs (SNR,$R_2$)
\item additive, multiplicative bias vs photo-z (dNNz, DEmP, mizuki)
\item additive, multiplicative bias residuals vs SNR, $R_2$, PSF FWHM, mag
\item additive, multiplicative bias vs. $R_2$, aperture mag cuts
\item averaged fractional size residual vs PSF FWHM, mag
\item fractional size residual auto-correlation
\item $\rho$ statistics
\item weighted mean shape vs SNR, mag, $R_2$, PSF FWHM
\item tangential and cross shear around CMASS galaxy, stars, randoms
\item star-galaxy cross correlation
\item PDFs of gaussian-smoothed B-mode convergence maps
\end{itemize}
\end{itemize}

\section{Reproduction of published results}
\label{sec:reproduce}

To check our pipeline we attempt to reproduce published results in a subset of the tests. Below we list the tests that we have checked and whether our results were consistent with that of the published. Note that some of the tests were done with the first year (Y1) of the DES-Y3 catalog \citep{Zuntz_2018} and the first year (Y1) of the HSC-Y3 catalog \citep{Mandelbaum2018b}, which were not used in the main part of this paper but was useful in the early stages of the paper for developing the tests.

\begin{itemize}
\item DES-Y1
\begin{itemize}
\item PSF quantities as a function of focal plane position
\item mean shape as a function of PSF and galaxy quantities
\item rho statistics
\item tangential ellipticity around stars
\end{itemize}

\item DES-Y3
\begin{itemize}
\item mean shape as a function of PSF and galaxy quantities
\item rho statistics
\item tau statistics
\item tangential ellipticity around stars
\end{itemize}

\item KiDS-1000
\begin{itemize}
\item PH statistics
\end{itemize}

\item HSC-Y1
\begin{itemize}
\item tau statistics
\end{itemize}
\end{itemize}
\begin{figure*}
    \centering
    \includegraphics[width=0.95\textwidth]{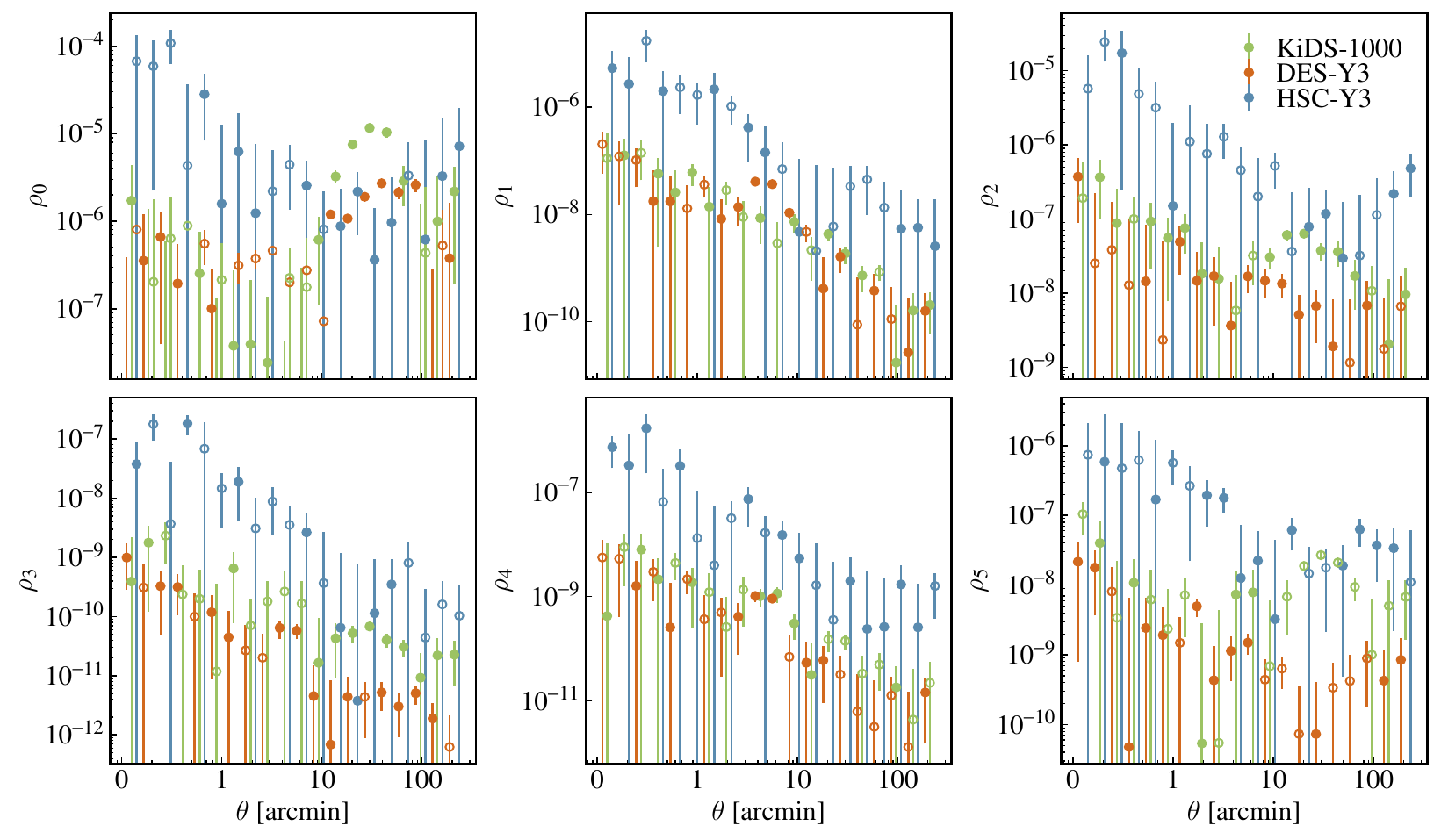}
    \caption{Rho statistics as described in Equation 12 for each survey, displaying only the components corresponding to $\xi_-$. Negative correlations are shown in absolute value with an open circle. Theta ranges are plotted with slight offsets for clearer visibility. The $\rho$s are evaluated for 20 angular bins between 0.1 to 250 arcminutes and error bars are estimated using jackknife resamplings for 250, 1000, and 150 patches for KiDS-1000, DES-Y3, and HSC-Y3.}
    \label{fig:rowesm}
\end{figure*}

\section{Rho Statistics, $\rho_-$ counterparts}
The rho statistics measurements corresponding to the $\xi_-$ component are included in Figure \ref{fig:rowesm}. These are combined with the $\rho_+$ displayed Figure \ref{fig:rowes} to model and fit the tau statistics as described Section \ref{sec:tau}. We see that overall, the measurements for the $\rho_-$ components are both smaller in amplitude and are estimated to have larger errors. We also find that some behavior shown in the $\rho_+$ measurements is repeated for the $\rho_-$ measurements i.e., HSC-Y3 consistently measures the highest amplitudes of the three surveys. However, we see some peculiarities arise where DES-Y3 and KiDS-1000 measure some signals with greater significance, as shown in $\rho_0$ for separations between 10 to approximately 100 arcminutes.

\section{Tomographic results for Tangential Ellipticity, Unweighted Stars}
\label{sec: GTStarsUW}
\begin{table*}
\center
    \caption{$\chi^2$/dof and $p$-value in parenthesis for tangential ellipticity around bright and faint stars for all tomographic bins. This quantity is computed for the full range of 0.1-250 arcminute separation (top) and for a subset within the scale cuts quoted in Table \ref{table:cosmology} (bottom).}
    \label{table:GTStars}
    \begin{tabular}{ccccccc} \toprule
    \multicolumn{2}{l}{{\bf Full range}}&&&&&\\[0.2cm]
    &{\textbf{KiDS-1000}}&Bin 1 & Bin 2 & Bin 3 & Bin 4 & Bin5 \\
     &Bright&8.4/20 (9.9e-1)&50.0/20 (2.2e-4)&35.8/20 (1.6e-2)&54.2/20 (5.5e-5)&77.3/20 (1.1e-8)\\
     &Faint&12.4/20 (9.0e-1)&9.0/20 (9.8e-1)&34.5/20 (2.3e-2)&136.0/20 (--)&261.720 (--)\\[0.2cm]
     &{\textbf{DES-Y3}}&Bin 1 & Bin 2 & Bin 3 & Bin 4&\\
     &Bright&418.3/20 (--)&168.3/20 (--)&211.4/20 (--)&152.2/20 (--)&\\
     &Faint&22.8/20 (3.0e-1)&88.1/20 (1.6e-10)&126.5/20 (--)&201.8/20 (--)&\\[0.2cm]
     &{\textbf{HSC-Y3}}&Bin 1 & Bin 2 & Bin 3 & Bin 4 &\\
     &Bright&12.5/20 (9.0e-1)&18.9/20 (5.3e-1)&9.4/20 (9.8e-1)&10.5/20 (9.6e-1)&\\
     &Faint&19.1/20 (5.1e-1)&16.0/20 (7.2e-1)&16.5/20 (6.9e-1)&11.7/20 (9.3e-1))&\\[0.2cm]
     \multicolumn{2}{l}{{\bf Scale cuts}}&&&&&\\[0.2cm]
     &{\textbf{KiDS-1000}}&Bin 1 & Bin 2 & Bin 3 & Bin 4 & Bin5 \\
     &Bright&6.6/16 (9.8e-1)& 6.1/16 (9.9e-1)&7.8/16 (9.6e-1)&36.7/16 (1.2e-3)&60.9/16 (3.7e-7)\\
     &Faint&9.0/16 (8.7e-1)& 7.3/16 (9.7e-1)&26.1/16 (5.3e-2)&121.1/16 (--)&257.7/16 (--)\\[0.2cm]
     &{\textbf{DES-Y3}}&Bin 1 & Bin 2 & Bin 3 & Bin 4 &\\
     &Bright&9.2/12 (6.9e-1)&7.8/12 (8.0e-1)&9.9/12 (6.3e-1)&4.8/12 (9.6e-1)&\\
     &Faint&21.5/12 (4.4e-2)&11.1/12 (5.2e-1)&15.8/12 (2.0e-1)&8.6/12 (7.4e-1)&\\[0.2cm]
     &{\textbf{HSC-Y3}}&Bin 1 & Bin 2 & Bin 3 & Bin 4 &\\
     &Bright&3.4/5 (6.3e-1)&9.7/5 (8.5e-2)&3.6/5 (6.1e-1)& 3.8/5 (5.8e-1)&\\
     &Faint&5.5/5 (3.6e-1)&2.4/5 (7.9e-1)&1.5/5 (9.1e-1)& 2.7/5 (7.4e-1)&\\
    \hline
    \end{tabular}
\end{table*}

\begin{figure}
\centering
\includegraphics[width=\columnwidth]{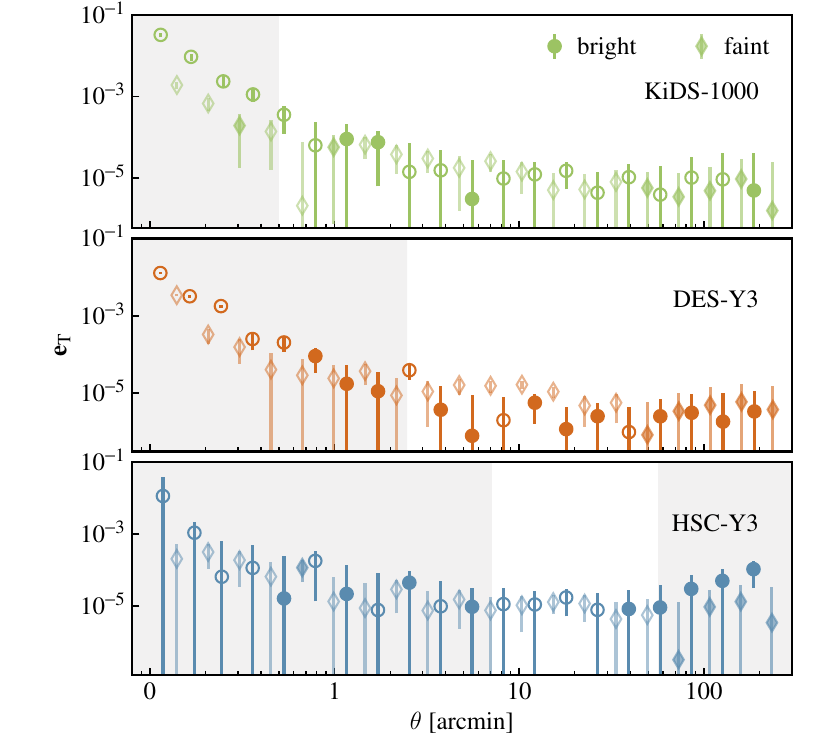}
\caption{Tangential component of the observed galaxy shape measured around two samples of stars. Results are shown for unweighted star samples (in contrast to an inverse stellar density weighting considered in Section~\ref{sec:tangential}). Measurements around bright stars are marked by opaque circles while measurements around faint stars are marked by semitransparent diamonds. Negative correlations are shown in absolute value with open markers. Gray regions denote the scale cuts for each survey, quoted in Table \ref{table:cosmology}. Error bars are estimated with jackknife resamplings for 250, 1000, and 150 patches for KiDS-1000, DES-Y3, and HSC-Y3. 
}
\label{fig:GTStarsuw}
\end{figure}

\begin{table*}
\center
    \caption{$\chi^2$/dof and $p$-value in parenthesis for tangential ellipticity around unweighted samples of bright and faint stars for all tomographic bins. This quantity is computed for the full range of 0.1-250 arcminute separation (top) and for a subset within the scale cuts quoted in Table \ref{table:cosmology} (bottom).}
    \label{table:GTStarsuw}
    \begin{tabular}{ccccccc} \toprule
    \multicolumn{2}{l}{{\bf Full range}}&&&&&\\[0.2cm]
    &{\textbf{KiDS-1000}}&Bin 1 & Bin 2 & Bin 3 & Bin 4 & Bin5 \\
     &Bright&9.6/20 (9.7e-1)&57.4/20 (1.8e-5)&41.1/20 (3.6e-3)&29.6/20 (7.7e-2)&36.0/20 (1.5e-2)\\
     &Faint &12.9/20 (8.9e-1)&11.0/20 (9.5e-1)&19.6/20 (4.8e-1)&35.2/20 (1.9e-2)&33.5/20 (3.0e-2)\\[0.2cm]
     &{\textbf{DES-Y3}}&Bin 1 & Bin 2 & Bin 3 & Bin 4&\\
     &Bright&510.8/20 (--)&232.9/20 (--)&264.6/20 (--)&212.3/20 (--)&\\
     &Faint &8.5/20 (9.9e-1)&100.4/20 (1.1e-12)&178.4 (--)&327.2/20 (--)&\\[0.2cm]
     &{\textbf{HSC-Y3}}&Bin 1 & Bin 2 & Bin 3 & Bin 4 &\\
     &Bright&12.6/20 (9.0e-1)&18.9/20 (5.3e-1)&9.4/20 (9.8e-1)&10.5/20 (9.6e-1)&\\
     &Faint &19.1/20 (5.1e-1)&16.0/20 (7.2e-1)&16.5/20 (6.9e-1)&11.7/20 (9.3e-1)&\\[0.2cm]
     \multicolumn{2}{l}{{\bf Scale cuts}}&&&&&\\[0.2cm]
     &{\textbf{KiDS-1000}}&Bin 1 & Bin 2 & Bin 3 & Bin 4 & Bin5 \\
     &Bright&6.5/16 (9.8e-1)&4.0/16 (1.0e0)&8.7/16 (9.3e-1)&10.4/16 (8.4e-1)&14.7/16 (5.5e-1)\\
     &Faint &8.5/16 (9.3e-1)&10.0/16 (8.7e-1)&10.4/16 (8.4e-1)&11.1/16 (8.0e-1)&17.3/16 (3.6e-1)\\[0.2cm]
     &{\textbf{DES-Y3}}&Bin 1& Bin 2 & Bin 3 & Bin 4 &\\
     &Bright&1.1/12 (8.1e-1)&8.0/12 (7.9e-1)&12.6/12 (4.0e-1)&6.7/12 (8.7e-1)&\\
     &Faint &5.8/12 (9.3e-1)&19.5/12 (7.7e-2)&35.4/12 (4.0e-4)&32.5/12 (1.2e-3)&\\[0.2cm]
     &{\textbf{HSC-Y3}}&Bin 1& Bin 2 & Bin 3 & Bin 4 &\\
     &Bright&3.4/5 (6.3e-1)&9.7/5 (8.5e-2)&3.6/5 (6.1e-1)&3.8/5 (5.8e-1)&\\
     &Faint &5.5/5 (3.6e-1)&2.4/5 (7.9e-1)&1.5/5 (9.1e-1)&2.7/5 (7.4e-1)&\\
    \hline
    \end{tabular}

\end{table*}

Table~\ref{table:GTStars} shows the tomographic reduced $\chi^2$ and $p$-value for the tangential ellipticity of galaxies around stars as discussed in Section~\ref{sec:tangential}. We find that both KiDS-1000 and DES-Y3 show significantly non-null signals for several redshift bins when considering the full range of angular separations. Within the scale cuts, KiDS-1000 also produces a significant non-null signal for higher redshift bins. We separately checked to ensure that the tangential ellipticity of galaxies around the uniform random sample were consistent with zero to ensure that this subtraction was not inducing a spurious signal. HSC-Y3 measures passing $p$-values for both the full range of angular separations and within the scale cuts.

We also explore the measured tangential ellipticity of galaxies around unweighted stars for each survey. Figure~\ref{fig:GTStarsuw} shows the non-tomographic measurement and Table~\ref{table:GTStarsuw} displays the tomographic reduced $\chi^2$ and $p$-value. We find that while KiDS-1000 and HSC-Y3 yield reasonable $p$-values within the science-ready scales, DES-Y3 yields a significantly non-null signal for the faint stars, which is comprised of stars used to model the PSF. \citet{Gatti_2021} cites that this is likely due to biasing effects from DES' star-finding algorithm, which motivated their choice to weight the measurements by the inverse star densities.\\

\section{Tomographic results for B-modes}
\label{sec:tomoBmodes}
\begin{table*}[!htbp]
\center
    \caption{$\chi^2$/dof and $p$-value in parenthesis for the pseudo-$C_{\ell}$ and \textsc{HybridEB} $B$-mode estimators used in this work, for all tomographic bin combinations.} 
    \label{table:bmodes}
\begin{tabular}{ccccccc} \toprule
    \multicolumn{2}{l}{{\bf Pseudo-C$\ell$}}&&&&\\[0.2cm]
   &{\bf KiDS-1000}       & Bin 1                      & Bin 2                      & Bin 3                      & Bin 4                      & Bin 5 \\ 
    &Bin 1 & 25.4/21 (2.3e-1) &                 &                 &                 &                \\
    &Bin 2 & 7.7/21 (1.0e0)  & 18.9/21 (5.9e-1) &                 &                 & \\
    &Bin 3 & 14.2/21 (8.6e-1) & 24.2/21 (2.8e-1) & 19.0/21 (5.9e-1) &                 & \\
    &Bin 4 & 23.2/21 (3.3e-1) & 14.9/21 (8.3e-1) & 36.0/21 (2.2e-2) & 16.7/21 (7.3e-1) & \\  
    &Bin 5 & 15.1/21 (8.2e-1) & 18.4/21 (6.2e-1) & 12.2/21 (9.3e-1) & 17.6/21 (6.7e-1) & 23.9/21 (3.0e-1) \\[0.2cm]
    &{\bf DES-Y3}         & Bin 1                      & Bin 2                      & Bin 3                      & Bin 4                      &  \\
    &Bin 1 & 32.1/21 (5.7e-2) &                 &                 &                 & \\
    &Bin 2 & 23.2/21 (3.4e-1) & 17.4/21 (6.9e-1) &                 &                 & \\
    &Bin 3 & 21.6/21 (4.2e-1) & 23.1/21 (3.4e-1) & 18.4/21 (6.2e-1) &                 & \\
    &Bin 4 & 24.9/21 (2.5e-1) & 19.0/21 (5.9e-1) & 34.5/21 (3.2e-2) & 17.4/21 (6.9e-1) & \\  [0.2cm]
    
    &{\bf HSC-Y3}         & Bin 1                      & Bin 2                      & Bin 3                      & Bin 4                      &  \\
    &Bin 1 & 7.5/13 (8.8e-1) &                  &                 &                 &  \\
    &Bin 2 & 23.9/13 (3.2e-2) & 9.0/13 (7.7e-1)  &                 &                 & \\
    &Bin 3 & 12.3/13 (5.0e-1) & 11.5/13 (5.7e-1) & 20.1/13 (9.4e-2) &                 & \\
    &Bin 4 & 15.8/13 (2.6e-1) & 29.9/13 ({4.8e-3}) & 13.9/13 (3.8e-1) & 11.6/13 (5.7e-1) & \\ [0.2cm]  
    \multicolumn{2}{l}{ \textsc{\textbf{HybridEB}}}&&&&\\[0.2cm]
   &{\bf KiDS-1000}       & Bin 1                      & Bin 2                      & Bin 3                      & Bin 4                      & Bin 5 \\ 
    &Bin 1 & 11.2/10 (3.4e-1) &  &  &  &  \\
    &Bin 2 & 1.1/10 (1.0e0) & 3.6/10 (9.7e-1) &  &  &  \\
    &Bin 3 & 2.1/10 (1.0e0) & 1.0/10 (1.0e0) & 10.6/10 (3.9e-1) &  &  \\
    &Bin 4 & 2.0/10 (1.0e0) & 3.9/10 (9.5e-1) & 6.1/10 (8.1e-1) & 7.7/10 (6.6e-1) &  \\
    &Bin 5 & 1.4/10 (1.0e0) & 6.7/10 (7.6e-1) & 10.8/10 (3.8e-1) & 7.2/10 (7.1e-1) & 16.0/8 (1.0e-2)  \\[0.2cm]
    &{\bf DES-Y3}         & Bin 1                      & Bin 2                      & Bin 3                      & Bin 4                      &  \\
    &Bin 1 & 7.1/8 (7.1e-1) &  &  &  \\
    &Bin 2 & 3.2/8 (9.7e-1) & 16.4/8 (8.8e-2) &  &  \\
    &Bin 3 & 5.6/8 (8.5e-1) & 7.9/8 (6.4e-1) & 10.0/8 (4.4e-1) &  \\
    &Bin 4 & 5.1/8 (8.8e-1) & 9.2/8 (5.1e-1) & 10.5/8 (4.0e-1) & 37.9/10 (4.0e-5) & \\  [0.2cm]
    &{\bf HSC-Y3}         & Bin 1                      & Bin 2                      & Bin 3                      & Bin 4                      &  \\
    &Bin 1 & 4.1/8 (9.4e-1) &  &  &  \\
    &Bin 2 & 2.1/8 (1.0e-0) & 12.7/8 (2.4e-1) &  &  \\
    &Bin 3 & 6.2/8 (8.0e-1) & 17.1/8 (7.3e-2) & 36.0/8 (8.6e-5) &  \\
    &Bin 4 & 11.3/8 (3.3e-1) & 27.6/8 (2.1e-3) & 44.7/8 (2.4e-6) & 16.0/8 (9.8e-1) & \\
    \hline
\end{tabular}

\end{table*}

Table~\ref{table:bmodes} shows the bin-by-bin $\chi^{2}$ values for each survey for the two B-mode estimators used in this work. We find that all surveys yield reasonable $p$-values with the pseudo-$C_{\ell}$ estimator, however DES-Y3 and HSC-Y3 both yield significantly non-null values for some of the higher bin pairs with the \textsc{HybridEB} estimator. \\

\end{appendix}
\end{document}